\documentclass[graybox]{svmult}

\usepackage{type1cm}        
\usepackage{makeidx}         
\usepackage{graphicx}        
\usepackage{multicol}        
\usepackage[bottom]{footmisc}

\usepackage{newtxtext}       %
\usepackage[varvw]{newtxmath}       

\usepackage{cite}
\usepackage{float}
\usepackage{scrextend} 
\usepackage{subcaption}
\usepackage{url}
\usepackage{verbatim}           
\usepackage{xcolor}
\usepackage{xspace}

\usepackage{lineno}

\usepackage[pagebackref=true,colorlinks=true]{hyperref}
\usepackage[colorlinks=true]{hyperref}
\hypersetup{
     unicode=true,          
     pdfnewwindow=true,      
     bookmarksnumbered=true,
     colorlinks=true,       
     linkcolor=blue,          
     citecolor=blue,        
     filecolor=magenta,      
     urlcolor=cyan,           
     menucolor=blue,
     linktocpage=true
}

\def\TeV{\ifmmode {\mathrm{\ Te\kern -0.1em V}}\else
                   \textrm{Te\kern -0.1em V}\fi}%
\def\GeV{\ifmmode {\mathrm{\ Ge\kern -0.1em V}}\else
                   \textrm{Ge\kern -0.1em V}\fi}%

\makeindex             

\begin{document}

\newcommand{\pbar}{\ensuremath{\overline{p}}\xspace}
\newcommand{\p}{\ensuremath{p}\xspace}
\newcommand{\nbar}{\ensuremath{\overline{n}}}
\newcommand{\dbar}{\ensuremath{\overline{d}}}
\newcommand{\pim}{\ensuremath{\pi^-}\xspace}
\newcommand{\pip}{\ensuremath{\pi^+}\xspace}
\newcommand{\km}{\ensuremath{K^-}\xspace}
\newcommand{\kp}{\ensuremath{K^+}\xspace}
\newcommand{\hm}{\ensuremath{h^-}\xspace}
\newcommand{\pt}{\ensuremath{p_{\rm T}}\xspace}
\newcommand{\pl}{\ensuremath{p_{\rm L}}\xspace}

\newcommand{\GeVc}{\ensuremath{\mbox{~Ge\kern-0.1em V}\!/\!c}\xspace}
\newcommand{\AGeVc}{\ensuremath{A\,\mbox{Ge\kern-0.1em V}\!/\!c}\xspace}
\newcommand{\AGeV}{\ensuremath{A\,\mbox{Ge\kern-0.1em V}}\xspace}

\newcommand{\mvl}[1]{\textcolor{blue}{MvL: #1}}
\newcommand{\mg}[1]{\textcolor{red}{MG: #1}}

\title*{Experiments at the CERN SPS: first signals of deconfinement}


%
%


\author{Federico Antinori, Marek Gazdzicki, Tapan K. Nayak, Guy Paic, 
Karel Šafařík, 
Enrico Scomparin, Itzhak Tserruya, Emanuele Quercigh, and Gianluca Usai}

\institute
{
Federico Antinori \at INFN Padova, Italy and CERN, Geneva, Switzerland; \email{Federico.Antinori@cern.ch} 
\and
Marek Gazdzicki \at Jan Kochanowski University, Kielce, Poland; \email{Marek.Gazdzicki@cern.ch} 
\and
Tapan K. Nayak \at CERN, Geneva, Switzerland; \email{Tapan.Nayak@cern.ch} 
\and
Guy Paic \at Universidad Nacional Autonoma de Mexico, Mexico; \email{Guy.Paic@cern.ch} 
\and
Karel Šafařík \at Czech Technical University (ČVUT), Prague, Czech Republic
\and
Enrico Scomparin \at INFN Torino, Italy; 
\email{scompar@to.infn.it} \and
Itzhak Tserruya \at  Weizmann Institute of Science, Rehovot, Israel; \email{itzhak.tserruya@weizmann.ac.il} \and
Emanuele Quercigh \at CERN, Geneva, Switzerland
\and
Gianluca Usai \at University of Cagliari and INFN Cagliari, Italy; \email{gianluca.usai@ca.infn.it}
}

\authorrunning{Antinori, Gazdzicki, Nayak, Paic, Scomparin, Tserruya, Usai }

\maketitle

\abstract{
Heavy-ion experiments at the CERN SPS began in the mid-1980s to study nuclear matter at extreme temperatures and densities.
The program started with light ions, such as oxygen and sulphur at energies of 60\AGeV and 200\AGeV, later advancing to lead ions at 158\AGeV. A series of experiments, employing novel detector technologies, 
explored various signatures of quark-gluon plasma (QGP) formation. 
In February 2000, these results
led CERN to announce evidence for the QGP formation~\cite{CERN2000NewState}. Subsequently, an energy scan was conducted with lead ions from 20\AGeV to 158\AGeV, to locate the threshold of QGP creation. 
}

\section{Heavy ions at CERN: a brief history}
\label{SPS_history}
\vspace{-0.3cm}

In its early years, CERN focused exclusively on accelerating protons. From 1964, more possibilities opened up, in particular the acceleration of deuterons and alpha particles, and antiprotons. 
A pivotal moment came in October 1980, when the GSI-LBL Collaboration submitted a Letter of Intent (LoI) to study neon-lead reactions at the CERN Proton Synchrotron (PS). This proposal led to a 1982 memorandum of understanding (MoU) between GSI, CERN, and LBL, aiming to get heavy ions into CERN's research program~\cite{Angert:1982rel}. With the initiative of Rudolf Bock (GSI), Herrmann Grunder (LBL), Reinhard Stock (University of Marburg), and Hans Gutbrod (GSI), this marked the birth of what would become a highly productive heavy-ion experimental program at CERN.
The ultimate physics goal was the observation of a new state of matter, the quark-gluon plasma (QGP)~\cite{Kapusta:1979fh,Shuryak:1980tp} and its characterization. 

\begin{figure}[b]
\begin{center}
\includegraphics[width=0.9\textwidth]{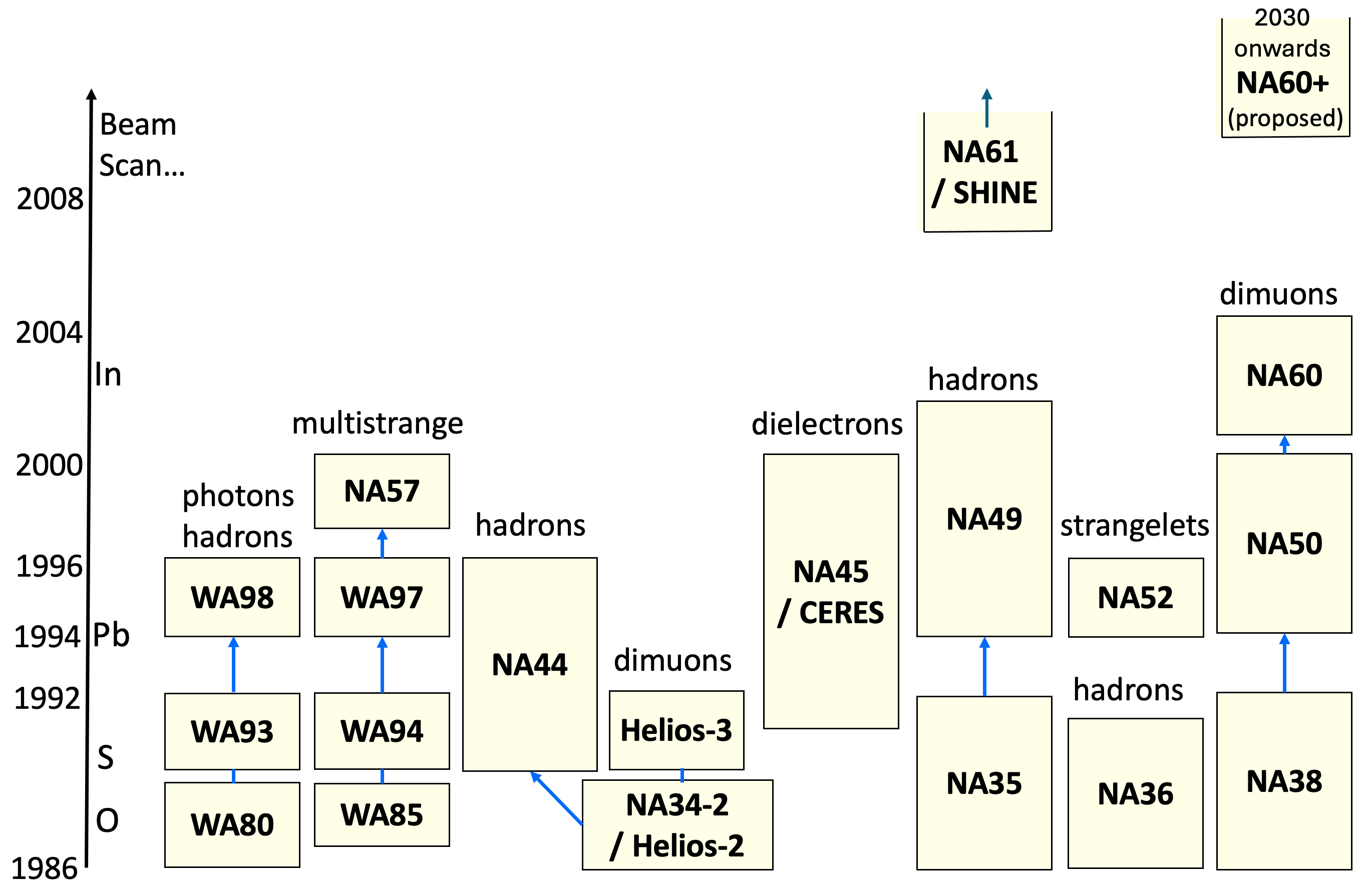}
\end{center}
\caption{Timeline of the heavy-ion physics experiments program at the CERN SPS.}
\vspace{-6mm}
\label{figures:SPS_1} 
\end{figure}

At the time, CERN was primarily focused on building the Large Electron-Positron Collider (LEP), which was being developed within a constant annual budget. Consequently, progress on heavy-ion experimentation was gradual. 
It was not until the fall of 1986 that the first oxygen ion beams became available at the Super Proton Synchrotron (SPS),
followed in 1987 by sulphur ion beams with an energy up to 200 AGeV. This program attracted
over 400 scientists participating in the 
first-generation of six heavy-ion experiments at the SPS. A timeline of these and subsequent heavy-ion experiments at CERN is shown in Fig.~\ref{figures:SPS_1}.

\begin{figure}[t]
\begin{center}
\includegraphics[width=0.95\textwidth]{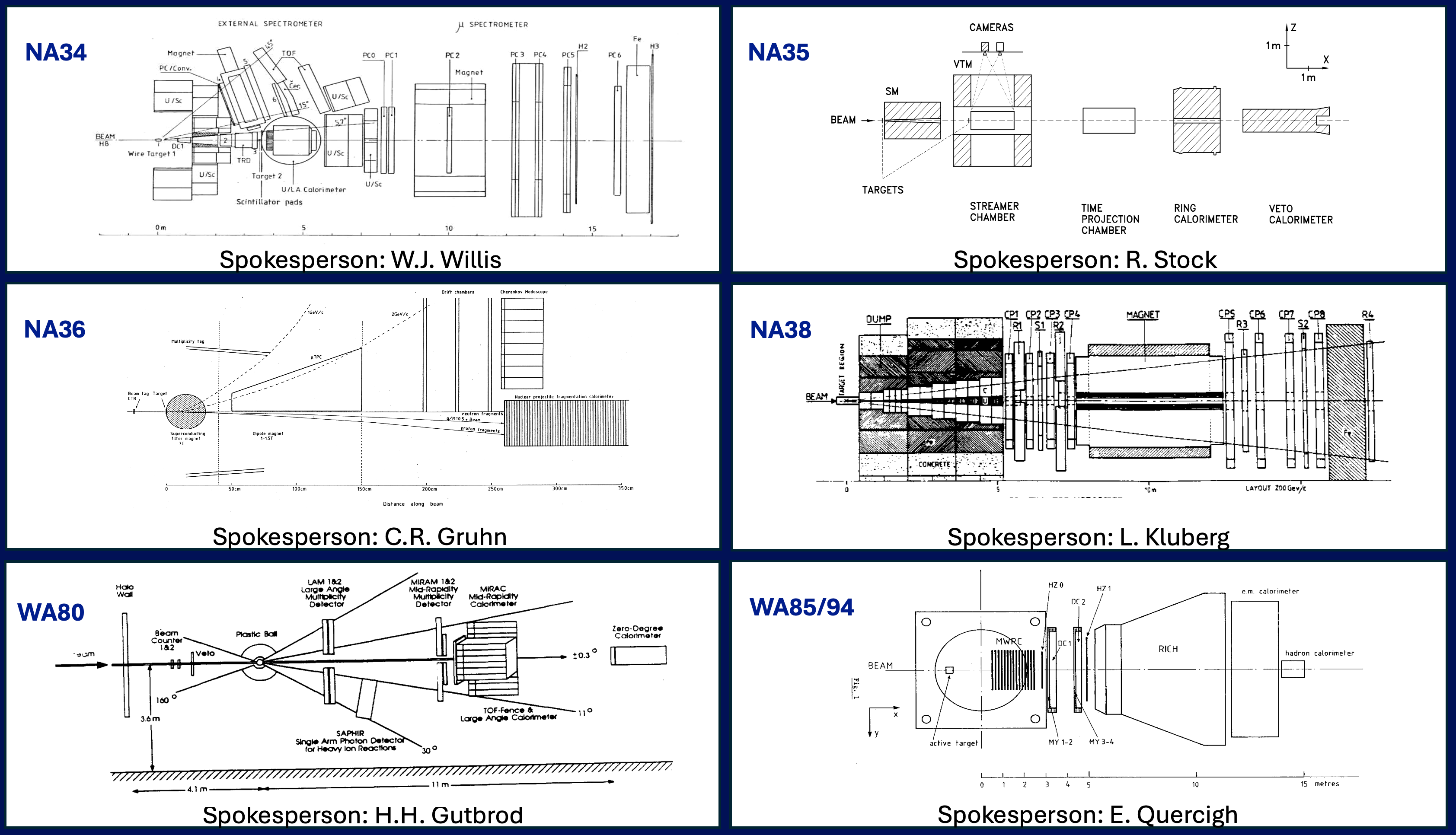}
\end{center}
\caption{Setup of the first-generation heavy-ion experiments at the CERN SPS.}
\vspace{-4mm}
\label{figures:SPS_2} 
\end{figure}

 The scientific potential of heavy-ion collisions spurred a range of theoretical predictions and experimental investigations. These included strangeness enhancement - studied primarily by NA35, NA36, and WA85/WA94; the detection of high temperatures through direct photon spectroscopy by WA80; and $J/\psi$ suppression, which NA38 sought to observe. Additionally, NA34 focused on dilepton spectra as indicators of dense hadronic matter. The setups of this first-generation experiments are shown in Fig.~\ref{figures:SPS_2}. 

From 1986 to 1992, the heavy-ion program at CERN evolved into a coherent and ambitious initiative. Every observation with oxygen and sulphur beams raised more questions than answers. Consequently, large-scale and multi-purpose heavy-ion experiments were proposed in 1991. 
This led the scientific community to develop a new ion source capable of producing lead beams, which would allow for more significant experiments.
A key development during this period was the construction of Linac 3, a linear accelerator specifically designed for accelerating lead beams. 
This project benefited from international collaboration, with contributions from the Czech Republic, France, India, Italy, Germany, Sweden, and Switzerland. 

The significant upgrades to the accelerator complex resulted in the acceleration of lead beams in 1994, achieving energies up to 158\AGeV. 
This marked the beginning of a nearly decade-long lead-beam program, including low-energy runs at 20, 30, 40, and 80\AGeV. 
The seven second-generation heavy-ion experiments, NA44, NA45, NA49, NA50, NA52, WA97/NA57, and WA98, specialized in measurement of charged hadrons, dielectrons, strange and multistrange hadrons, dimuons, strangelets and photons. 
Post 2000, the third-generation experiments started with NA60 serving as a follow-up to NA50, while NA61/SHINE (approved in 2007) evolved from the NA49 experiment. In this chapter, we discuss the historical development of the SPS experiments and their key findings.


\section{Experiment NA44 - particle correlations}
\label{SPS_NA44}
\vspace{-0.3cm}

The NA44 spectrometer~\cite{NA44Proposal1993} (Spokesperson: Hans Bøggild) was a unique experiment among the early SPS heavy-ion programs, which was specially designed for precision measurements of single particle spectra and the intensity interferometry in hadronic systems of high energy density using collisions of both hadrons and heavy ions. The spectrometer operated over a narrow central rapidity range. It featured excellent momentum resolution with the capability to identity ($\pi^{\pm}, K^{\pm}$, proton, anti-proton, deuteron, and anti-deuteron), and had an acceptance optimized for particle-pair measurements. 

NA44 was the only relativistic heavy-ion experiment to have both deuteron and anti-deuteron results in both 
p-Pb and Pb-Pb collisions and the first CERN experiment to study the physics topics addressed by deuteron and anti-deuteron production~\cite{NA44:1995lds}. The collaboration was the early producer of data that were explained with the coalescence mechanism\cite{Bearden2000Antideuteron}.

\begin{figure}[tbp]
\begin{center}
\includegraphics[width=0.8\textwidth]{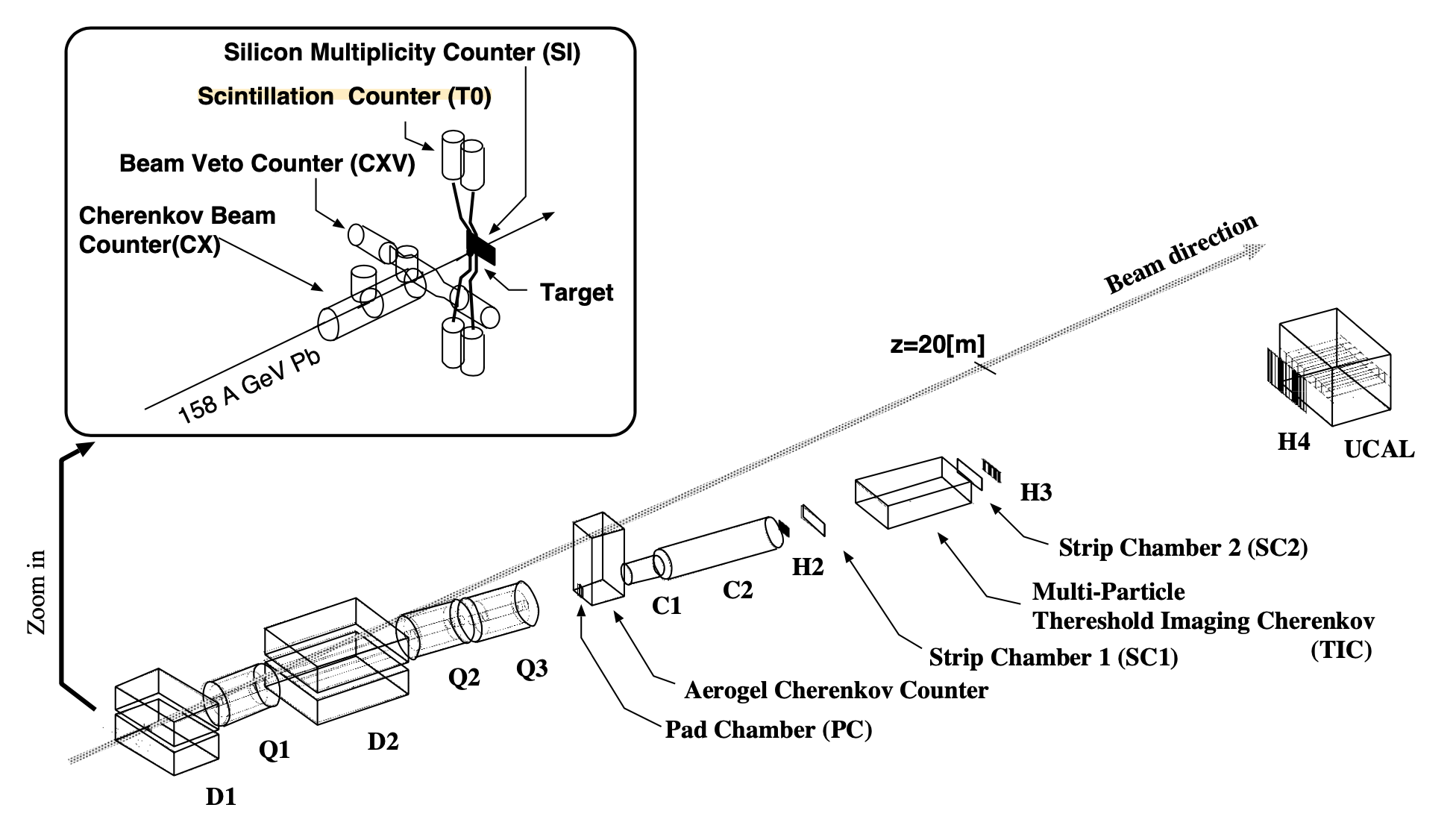}
\end{center}
\caption{The NA44 experimental setup for Pb-Pb running (taken from Ref.~\cite{Bearden:2000ex}).}
\label{fig:na44_1}
\end{figure}

\subsection{NA44 experimental setup}
\vspace{-0.3cm}
The experimental setup~\cite{Bearden:2000ex} in shown in Fig.~\ref{fig:na44_1}.
 The NA44 collaboration built a novel Threshold Imaging Cherenkov (TIC) detector for track-selective $\pi$-$K$ separation in the momentum range of 4-8 GeV/$c$ in combination with Time-Of-Flight (TOF) detector. Cherenkov light produced in an isobutane radiator is reflected by flat mirrors and detected in MWPCs (Multiwire proportional counters) with 2-dimensional cathode-pad readout and  TMAE vapour as photo-converter. Operating the detector in the experiment during the 1994 lead run, a best average of 20 pads hit per 4 GeV pion was obtained, giving excellent separation from the heavier particles (kaons, protons). However, to use the TIC in lead beams it was necessary to diminish the noise levels by diminishing the size of the photon conversion layer. The photocathodes that were just developed in the RD26 collaboration (led by Guy Paic and Francois Piuz) were tested for the first time in NA44. The main advantage of the CsI photocathode for the intended application was that the conversion layer of photons into photoelectrons was considerably smaller than using TMAE added to the gas of the MWPC. The neutron background was considerable making the detection very difficult.

The successful operation of the TIC demonstrated the feasibility to use CsI photocathodes in a complicated and noisy environment. Afterwards the CsI photocathodes have been used in many detectors in the world and prominently in the ALICE HMPID which is still functioning to this day. It is also interesting that the first test of a single HMPID prototype was tested in the STAR experiment at RHIC at the Brookhaven National Laboratory.

\subsection{Key findings from NA44}
\vspace{-0.3cm}

\begin{figure}[tbp]
\begin{center}
\includegraphics[width=0.4\textwidth]{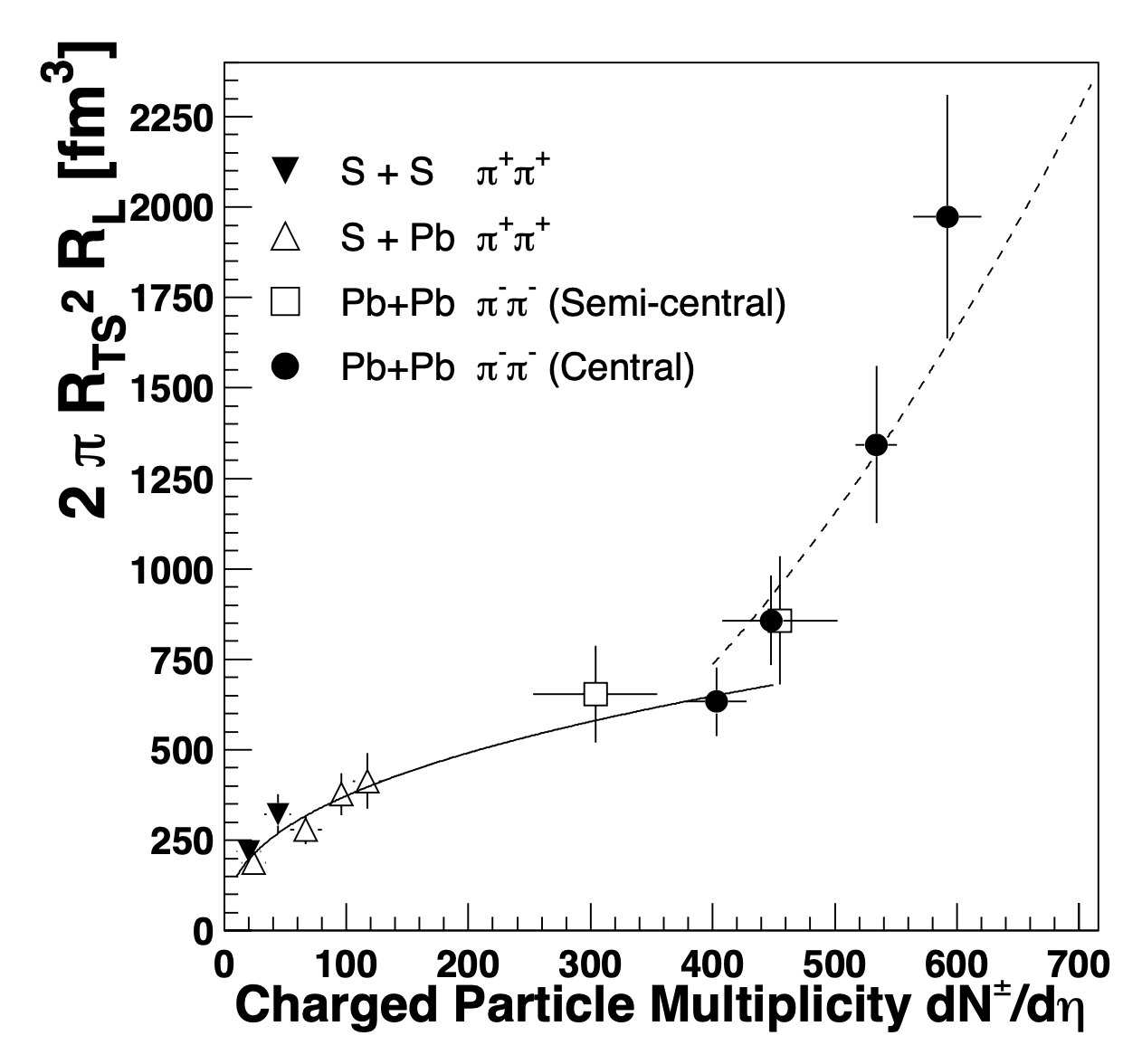}
\includegraphics[width=0.55\textwidth]{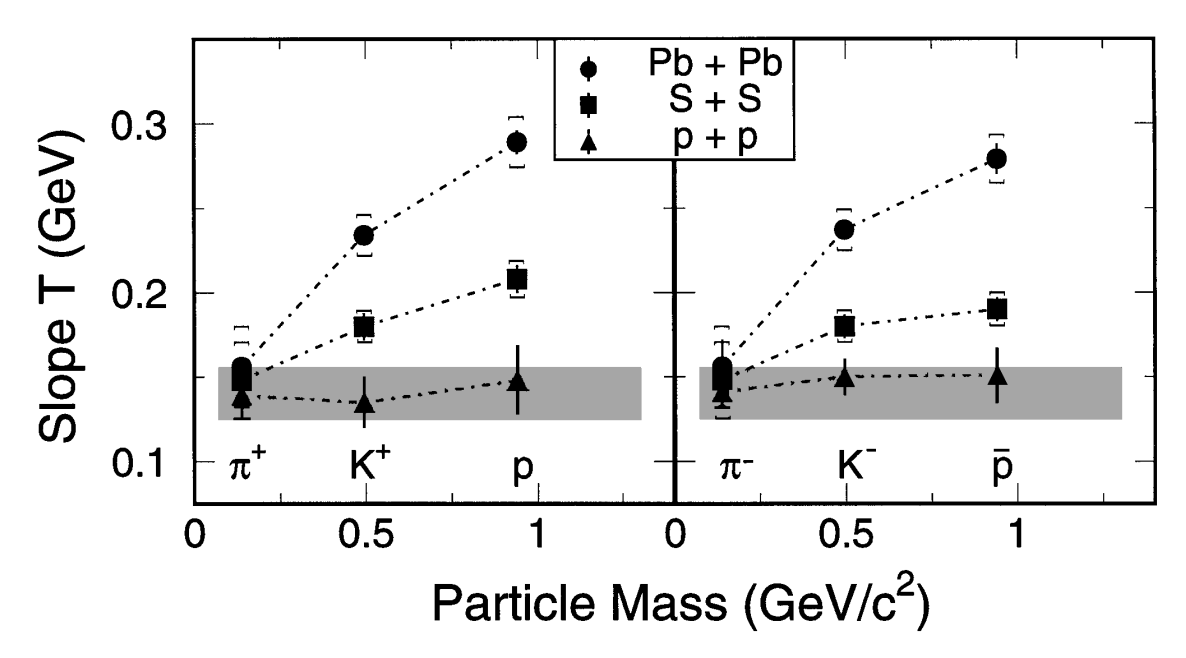}
\end{center}
\caption{({\it{Left}}) 
The source size at freeze-out as a function of charged particle multiplicity~\cite{Bearden:2000ex}.
({\it{Right}}) The inverse
slope parameter $T$ as a function of particle mass~\cite{NA44:1996xlh}. 
}
\label{fig:na44_2}
\end{figure}

One of the novelties of the NA44 experiment is the measurement of the two-particle correlations of negative pions as a function of the multiplicity of charged particles in 158\AGeV lead-lead collisions. The left panel of Fig.~\ref{fig:na44_2} shows the extracted source size as a function of multiplicity. The source size can be seen to increase with multiplicity, with a more rapid increase for larger systems \cite{Bearden:2000ex}. 

In retrospect, the most important contribution to the field was probably the detection of the radial flow in Pb-Pb collisions at SPS. The paper~\cite{NA44:1996xlh} presents results on the transverse mass spectra of pions, kaons, and protons measured 
in S-S  and Pb-Pb  collisions at 200\AGeV and 158\AGeV, respectively.

Fig.~\ref{fig:na44_2}~({\it{Right}}) shows the inverse slope parameter, $T$, of transverse mass distributions for pions, kaons, and protons as a function of particle mass in pp, S-S  and Pb-Pb  collisions at 200\AGeV and 158\AGeV~\cite{NA44:1996xlh}. 
The mass dependence provides evidence of collective transverse flow from expansion of the system in
heavy-ion induced central collisions. 
 While in pp collisions, no effect of radial flow was detected, the behaviour of the spectra of identified particles in heavy-ion collisions clearly pointed to the existence of an expansion in the created system. Radial flow and anisotropic flow became the stalwart of the study of ultra-relativistic collisions at SPS, RHIC, and LHC experiments and was an important part of the press release of CERN on the creation of a new state of matter 
in February 2000. 

\section{Experiment NA45/CERES - Dileptons}
\label{SPS_NA45}
\vspace{-0.3cm}
CERES pioneered the measurement of electron-positron pairs at the SPS. Electron pairs and dileptons in general, are powerful observables of the QGP. They are sensitive to the two fundamental properties of the QGP: the deconfinement of quarks and gluons and the restoration of chiral symmetry. 
The importance of dileptons resides in their mean-free-path being large compared to the size of the system formed in nuclear collisions. Therefore, once produced, they escape the interaction region unaffected by final state interactions, carrying information to the detectors about the conditions and properties of the medium at the time of their production.  
The primary interest is in identifying the thermal radiation emitted by the system in the form of virtual photons. The thermal radiation provides a measurement of the system's temperature. Furthermore, it carries a direct fingerprint of the matter 
formed\footnote{For theoretical and experimental reviews, see the papers of R. Rapp et al. (p. 4-1) and I. Tserruya (p. 4-43) in ”Relativistic Heavy Ion Physics”, published in Landolt-Boernstein, Vol. I-23 (2010).}: 
the QGP via the quark annihilation (${\it q} {\it \overline{q}} \rightarrow e^+e^-$) or the hadronic gas via the pion annihilation mediated by the $\rho$ meson ($\pi^+\pi^- \rightarrow \rho \rightarrow \gamma^* \rightarrow e^+e^-$).  This was the original goal of the CERES experiment.

\subsection{Brief history}
\vspace{-0.3cm}
The measurement of electron-positron pairs is notoriously challenging. The main difficulty is the need to detect a very weak source of $e^+e^-$ pairs in the presence of an overwhelming yield of pairs originating from $\pi^0$ Dalitz decays and $\gamma$ conversions, which lead to a large combinatorial background if they are not recognized and rejected.   
Therefore, it is no wonder that no electron measurement was proposed in the first round of heavy-ion experiments approved for running at the CERN SPS. 
Hans Specht initiated the measurement of electron pairs in heavy-ion collisions at the SPS. The original plan was to perform the measurement within the HELIOS-3 setup, adding a RICH detector in the field-free region downstream of the target to recognize $\pi^0$ Dalitz decays and  $\gamma$ conversions by their relatively small opening angle. After several meetings and discussions with Hans and his group, the Weizmann Institute heavy-ion group of Itzhak Tserruya joined the HELIOS-3 Collaboration with a declared interest in the measurement of electron pairs. The two groups started a comprehensive program of simulations and detector prototyping. The limited space available for the RICH detector resulted in a marginal performance. This, together with the lack of support from the HELIOS-3 management for such a measurement, convinced us to separate from HELIOS-3 and design a dedicated experiment for the measurement of electron-positron pairs. This was the birth of CERES.

A proposal was submitted to the SPSC in 1998, signed by the Heidelberg University and Weizmann Institute groups together with the group of Peter Wurm from the Max Planck Institut fur Kernphysik \cite{CERN-SPSC-88-25}. The three groups remained at the core of the collaboration. Other groups and individual collaborators joined soon after approval of the proposal, notably Pavel Rehak from BNL, who produced the Si radial drift chambers, and the JINR group of Yuri Panebrattsev. In 1995, Johanna Stachel and Peter Braun-Munzinger returned from Stony Brook University to the Heidelberg University and GSI, respectively, and both joined the CERES experiment. The CERES collaboration remained relatively small at all times, with typically 50 collaborators. 

Historically, the Heidelberg University and the Weizmann Institute groups have shared the responsibility for the development of the spectrometer concept and the RICH detector prototyping. Heidelberg was responsible for the general layout and integration, the mechanics, including the Ring Imaging Cherenkov (RICH) radiator tanks, mirrors, and UV windows, the 
Time Projection Chamber (TPC) upgrade, the readout electronics, and the data acquisition system. Peter Glassel designed the magnet system of the double RICH spectrometer and later the magnet system of the TPC. Weizmann was in charge of the design,  construction and operation of all the gas detectors, including the UV chambers of RICH-1 and RICH-2, the large pad chamber behind RICH-2, and the MWPC of the TPC. The Max Planck group took responsibility for the assembly, installation, and operation of the Si radial drift chambers produced at BNL  and the Si pad detector. JINR was in charge of the first-level trigger detectors, and the production of the magnet system for the TPC.  Software development and data analysis were shared by all the institutions. 

\noindent There were three distinct phases in the running history of the CERES experiment:
\begin{itemize}
\vspace{-0.2cm}
\item First phase (1988-1993, spokesperson Hans Specht): submission and approval of the proposal, construction and installation of the spectrometer, commissioning in 1990 with the challenging spark breakdown of the UV detectors with a S beam and the first successful S-Au run in 1992. 
\item Second phase (1993-1997, spokesperson Itzhak Tserruya): 
p-Be and p-Au runs in 1993, the first CERES upgrade with improved tracking for the Pb-Au runs in 1995 and 1996. During this period, CERES published its first impactful physics results, the enhancement of low-mass electron pairs in S-Au collisions \cite{CERES:1995vll}. 
\item Third phase (1997-2000, spokesperson Johanna Stachel): second CERES upgrade with the addition of a radial drift TPC to improve the mass resolution, data taking with Pb-Au collisions at 40\AGeV in 1999 and 158\AGeV in the year 2000. This was the last run of the CERES experiment.
\end{itemize}

\subsection{The CERES spectrometer}
\label{subsec:na45_exp}
\vspace{-0.3cm}
\subsubsection*{The original set-up}
\begin{figure}[btp]
 \begin{center}
     \includegraphics[width=0.8\textwidth]{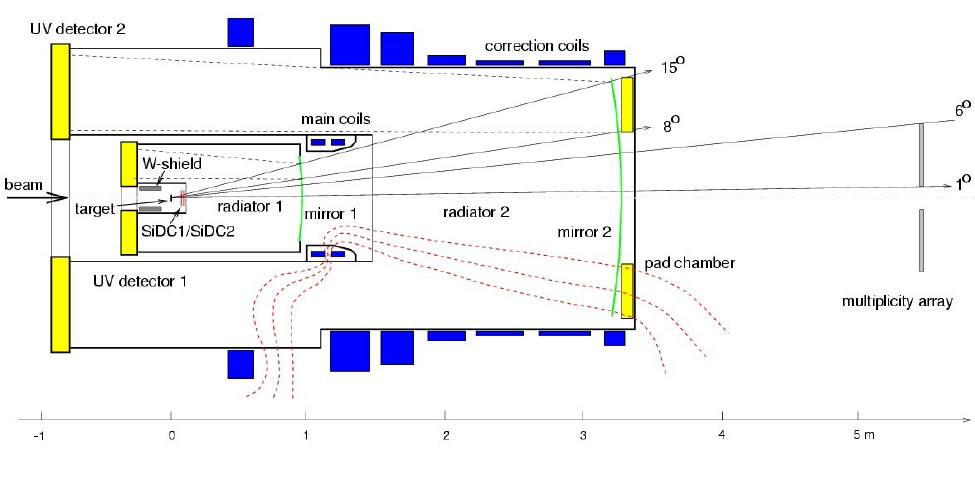}
    \caption{CERES setup for the 1995 and 1996 Pb beam runs.  The original setup had one SiDC and a Si pad chamber downstream of the target, and no pad chamber and multiplicity array downstream of mirror-2. In the last two runs (1999 and 2000), these were replaced by a radial drift TPC to improve the spectrometer's mass resolution.
    }
  \label{fig:ceres-setup}
 \end{center}
 \vspace{-6mm}
 \end{figure}

\vspace{-0.2cm}
The original CERES spectrometer consisted of two azimuthally symmetric RICH detectors, a Si pad detector, and a Si radial drift chamber (SiDC). A superconducting short ring-like double solenoid located between RICH-1 and RICH-2 provided the momentum and charge analysis (see Fig.~\ref{fig:ceres-setup}) \cite{Baur:1993mw}.
This was a rather unconventional design for a spectrometer, as the RICH detectors do not provide position tracking, only directional tracking. It was also a rather novel one. CERES was the first experiment using the concept of hadron blind tracking. Furthermore, although the RICH technology had been proposed long before, by the time of the CERES proposal, no experiment had published physics results based on a RICH detector, and CERES was, in fact, the first experiment that did it. CERES was also the first experiment to successfully use the novel silicon drift chamber concept. 
 
The spectrometer covers the pseudo-rapidity region close to mid-rapidity, 2.1 $< \eta < $ 2.65, with 2$\pi$ azimuthal coverage. 
The Cherenkov UV photons produced in the radiator gas are reflected by a spherical mirror and focused into a ring in the mirror's focal plane where the UV photon detectors are located. The radiator gas in both detectors is $CH_4$ at atmospheric pressure. Its high Cherenkov threshold of $\gamma_{th} \sim$ 32 ensures the hadron blindness of the spectrometer. 
The hadron blindness is further enhanced by the location of the UV detectors upstream of the target, such that they are not traversed by the large forward flux of charged particles produced in fixed-target central heavy-ion collisions.

The UV detectors of RICH-1 and RICH-2 are the crucial elements of the spectrometer. They consist of two stages of PPAC (Parallel-plate avalanche counter) followed by a MWPC. The initially planned running mode with only PPAC was abandoned due to a severe spark breakdown problem encountered in the 1990 commissioning run \cite{Baur:1993mz}. Adding a MWPC as a third amplification stage with a resistive layer cathode solved the sparking problem and allowed stable operation of the detectors with a very low sparking rate. 
The two PPACs were operated at very low gains of $<$ 100 each, whereas the entire structure with the MWPC reached full gain in the range of $2-4\times10^5$.

The UV detectors are separated from the radiators by high UV-transmission windows. RICH-2 uses a quartz window and is thus sensitive between the TMAE ionization threshold of 5.4~eV and the quartz window cut-off of 7.4~eV. On the other hand, RICH-1 with a $CaF_2$ window has a higher cut-off of 8.1 eV given by the detector gas, $C_2H_6$.

On the other hand, RICH-1 with a $\mathrm{CaF_2}$ window has a higher cut-off given by the detector gas, 8.1 eV in the case of $\mathrm{C_2H_6}$. Thus, both RICH-1 and RICH-2, despite having quite different radiator lengths, 86~cm and 175~cm, respectively, produce similar number of $\sim$12 photoelectrons per ring. For RICH-1, it was a significant achievement, corresponding to an effective figure of merit, $N_0$ of 131 cm$^{-1}$ \cite{Baur:1993mw},
a record for a gas Cherenkov detector at the time,  to be surpassed only much later by the PHENIX HBD \cite{Tserruya:2020vyt}.

The two superconducting coils carry current in opposite senses, deflecting the particles in the azimuthal direction while keeping the polar angle unchanged. The relative magnitude of the currents is adjusted such that, together with a warm correction coil, the magnetic field is compensated to nearly zero in the region of the RICH-1 radiator, thus preserving the original direction of the particles. This field-free region is essential for the identification of pairs from $\gamma$ conversions and $\pi^0$ Dalitz decays by their typically small opening angle. A set of warm correction coils shapes the field in the RICH-2 radiator such that it points back to the target (see the field lines in Fig.~\ref{fig:ceres-setup}). The particle trajectories are thus parallel to the field lines and not deflected within RICH-2, an essential requirement for sharp ring images.
 
Special care was taken to keep the amount of material within the acceptance of the spectrometer as low as possible to minimize the number of $e^+e^-$ pairs from conversions and the deterioration of the momentum resolution due to multiple scattering. The mirror of RICH-1 is thus very thin, made of 1.1 mm carbon fibre laminate. The volume between the two radiators, i.e., the region of the main superconducting coils, is filled with He gas. With these special features, the total amount of material downstream of the target could be kept at  $\sim$ 1.3\% of a radiation length.
 
The novel Si radial drift chamber~\cite{Chen:1993sh} and the Si pad detector are closely spaced downstream of the target. 
The SiDC fulfils several tasks: (i) provides the event centrality, (ii) locates the primary vertex within the segmented target by matching the RICH electron tracks to SiDC hits, (iii) helps the pattern recognition of RICH-1 by providing a priori information on the potential location of ring centres, (iv) further rejects conversions and Dalitz pairs by detecting a double pulse height signal or close tracks. 
The Si pad detector, segmented into 64 pads arranged in 8 rings and 8 azimuthal sectors, provides information about the charged-particle multiplicity used for first-level triggering.

In all measurements, CERES used a segmented Au target. For the Pb-Au run, the target consisted of 8 discs of 
25$~\mu$m thickness and 600~$\mu$m diameter, corresponding to a total interaction length of 0.83\% for Pb-Au collisions. The spacing of 3.2 mm between subsequent discs ensured that photons and electrons propagating inside the spectrometer acceptance see only a small fraction of the total radiation length.

\vspace{-0.4cm}
\subsubsection*{The first upgrade}
\vspace{-0.4cm}
The original CERES spectrometer proved to work very well in low-multiplicity collisions like p-Be and p-Au, where a pair reconstruction efficiency of 50\% and a signal-to-background ratio (S/B) of 1/2 could be achieved. However, increasing the collision multiplicity impaired the spectrometer’s performance. It was still acceptable for central 
S-Au collisions, where the efficiency was 9\% at a S/B of 1/6. But the setup was inadequate to cope with the much higher multiplicities expected in Pb-Au collisions. With more than 300 charged particles emitted within the CERES acceptance, the pattern recognition based solely on the RICH detectors breaks down. Therefore, in preparation for the Pb beam, the basic double RICH spectrometer was supplemented with new devices (see Fig.~\ref{fig:ceres-setup}): (i) a doublet of SiDC replaced the previously used single SiDC and Si pad chamber. The doublet of SiDC provides a precise two-point charged particle location, improving the tracking before RICH-1 and also the momentum resolution of the spectrometer \cite{Faschingbauer:1995qj}, (ii) a Pad Chamber was added behind the mirror of RICH-2 to provide external tracking downstream of RICH-2 and (iii) a plastic scintillator multiplicity array was added behind the pad chamber to provide a first-level centrality trigger. The performance of the upgraded CERES spectrometer with the Pb beam is described in \cite{CERES:1996bja}.

\vspace{-0.4cm} 
\subsubsection*{The second upgrade}
\vspace{-0.4cm}
The precision of the first di-electron results was insufficient to discriminate among the theoretical approaches and prompted the need to improve the mass resolution in the region of the $\phi$ meson. In 1998, the spectrometer was upgraded with the addition of a cylindrical TPC with radial electric drift field located downstream of RICH-2 and operated inside a new magnetic field provided by an anti-Helmholtz coil system surrounding the TPC \cite{CERES:2008asn}. The opposite currents in the two coils were unequal and tuned to make the fringe field negligible in the RICH-2 radiator volume. 
With the TPC, the magnetic field between the two RICHes was no longer needed (so turned off). The two detectors were used to redundantly identify electrons, resulting in an improved electron detection efficiency of $\sim$0.9 and improved rejection of close pairs. With the TPC upgrade, the mass resolution improved by almost a factor of 2, from $\Delta m/m$ = 7\% in the Pb runs of 1995 and 1996 to 3.8\%. 

\subsection{Main results from CERES/NA45}
\label{subsec:na45_res}
\begin{figure}[tbp]
 \begin{center}
     \includegraphics[width=0.5\textwidth]{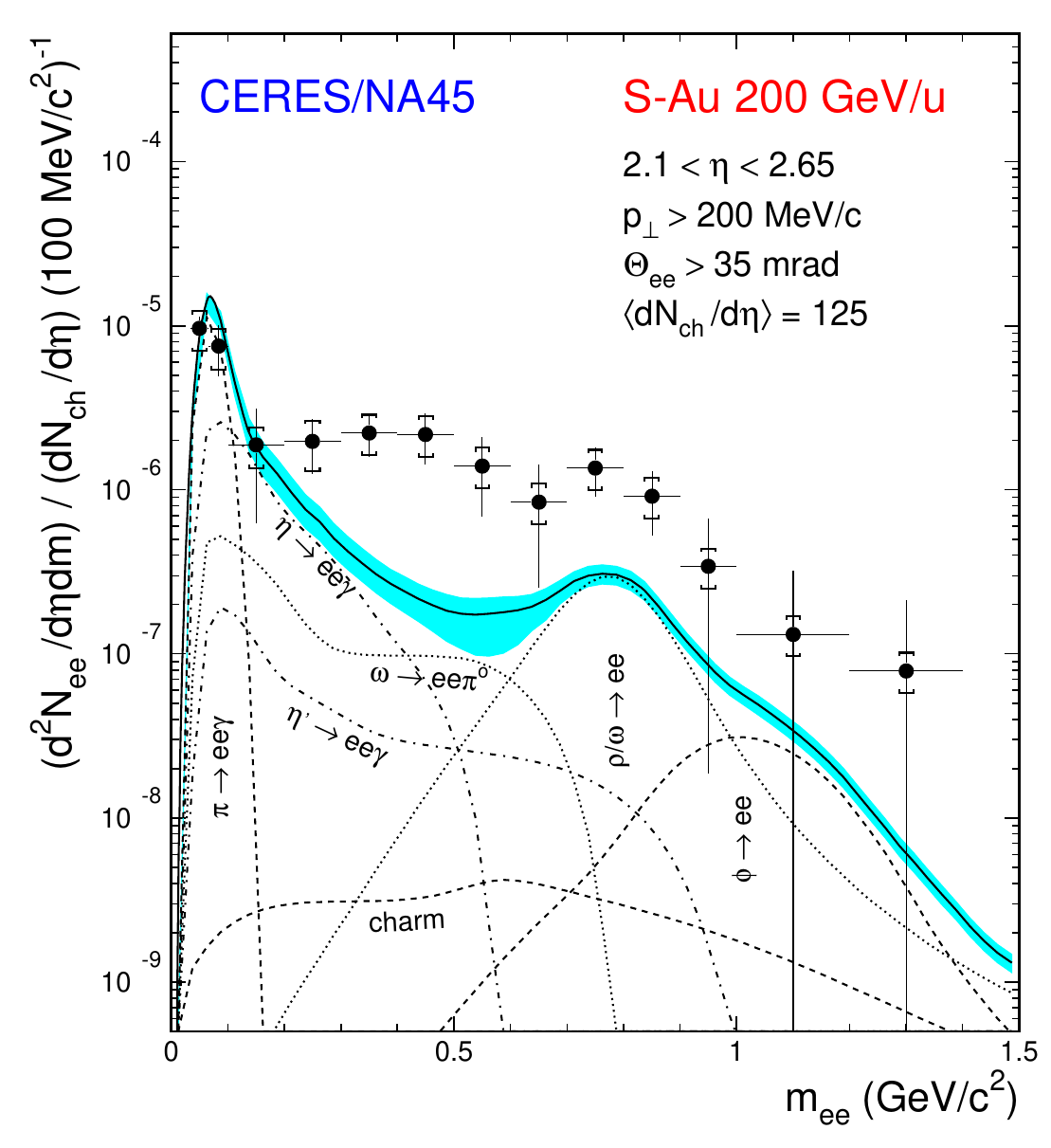}
     \vspace{-2mm}
    \caption{Invariant e$^+$e$^-$ mass spectrum in 200\AGeV S-Au collisions measured in the CERES acceptance. The bars and brackets in the data points reflect the statistical and systematic uncertainties, respectively. The figure also shows the cocktail of known hadron decays and its sum (solid line) with systematic uncertainty (band) \cite{CERES:1995vll}.}
  \label{fig:s-au}
 \end{center}
 \vspace{-6mm}
 \end{figure}

\vspace{-0.3cm}
In its first heavy-ion run of S-Au collisions at 200\AGeV, CERES discovered an enhancement of e$^+$e$^-$ pairs in the invariant-mass range $m_{ee}$ = 0.2 – 1.5 GeV/c$^2$, mostly below the $\rho$-meson mass (see Fig.~\ref{fig:s-au}), with respect to the expected yield from the cocktail of known hadronic sources decaying into $e^+e^-$ pairs \cite{CERES:1995vll}.
Similar enhancements were observed in all the subsequent heavy-ion measurements performed by CERES, in Pb-Au at 158\AGeV \cite{CERESNA45:1997tgc, CERES:2005uih} and 40\AGeV \cite{CERESNA45:2002gnc} (the strongest enhancement was observed at the lowest energy studied, 40\AGeV). But, no such enhancement was observed in p-Be (a proxy for \textit{pp}) or in p-Au \cite{Agakichiev:1998kip} collisions. It is interesting to note that the enhancement of low-mass pairs remains, up to now, one of the exceptional effects observed uniquely in heavy-ion collisions and not in pA or \textit{pp} collisions, contrary to many other observables like long-range near-side correlations, where effects, very similar to those observed in heavy-ion collisions, have also been reported in pA and \textit{pp} collisions \cite{CMS:2012qk}.

\begin{figure}[tbp]
 \begin{center}
 \includegraphics[width=0.44\textwidth]{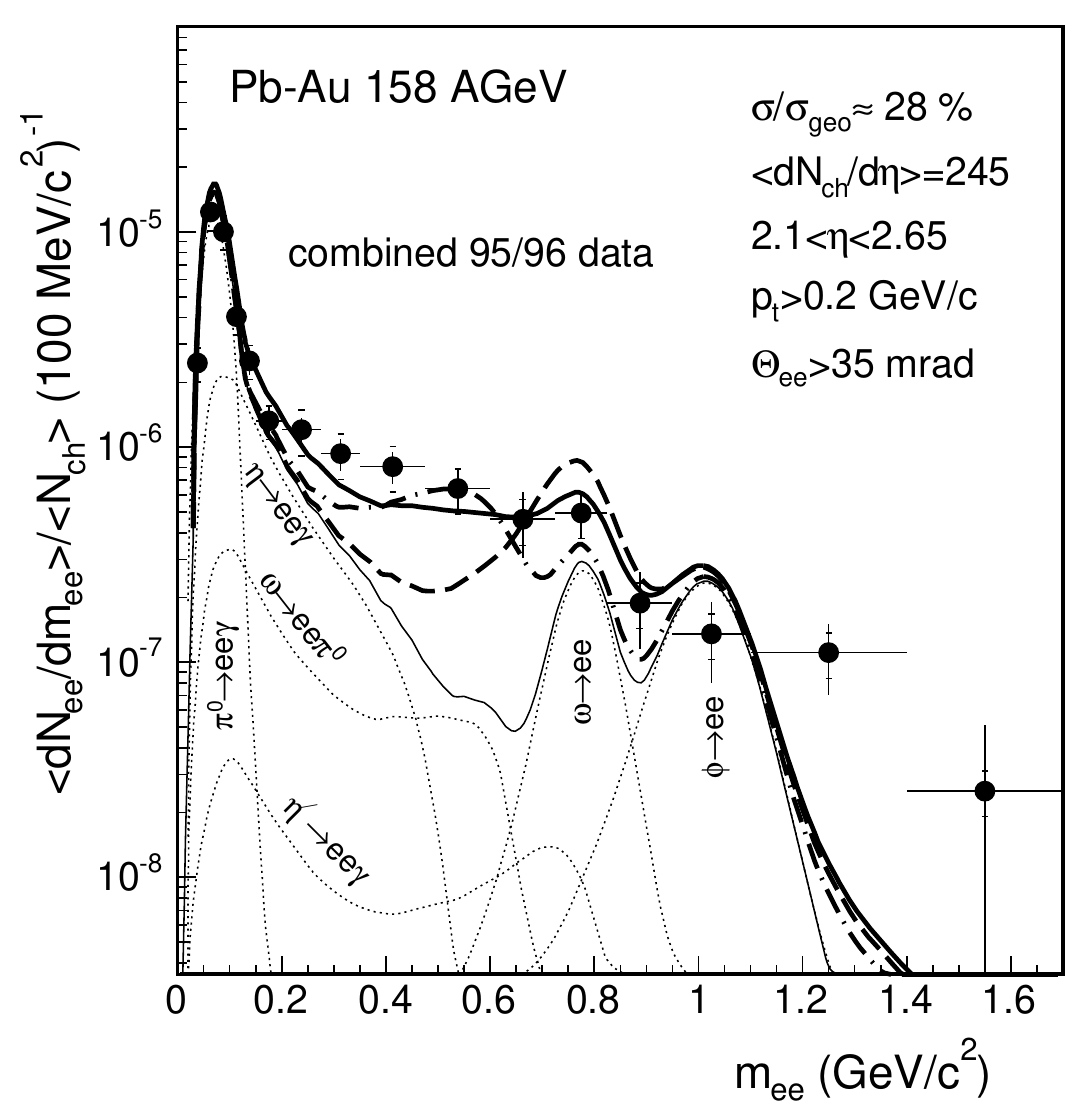}
 \includegraphics[width=0.47\textwidth]{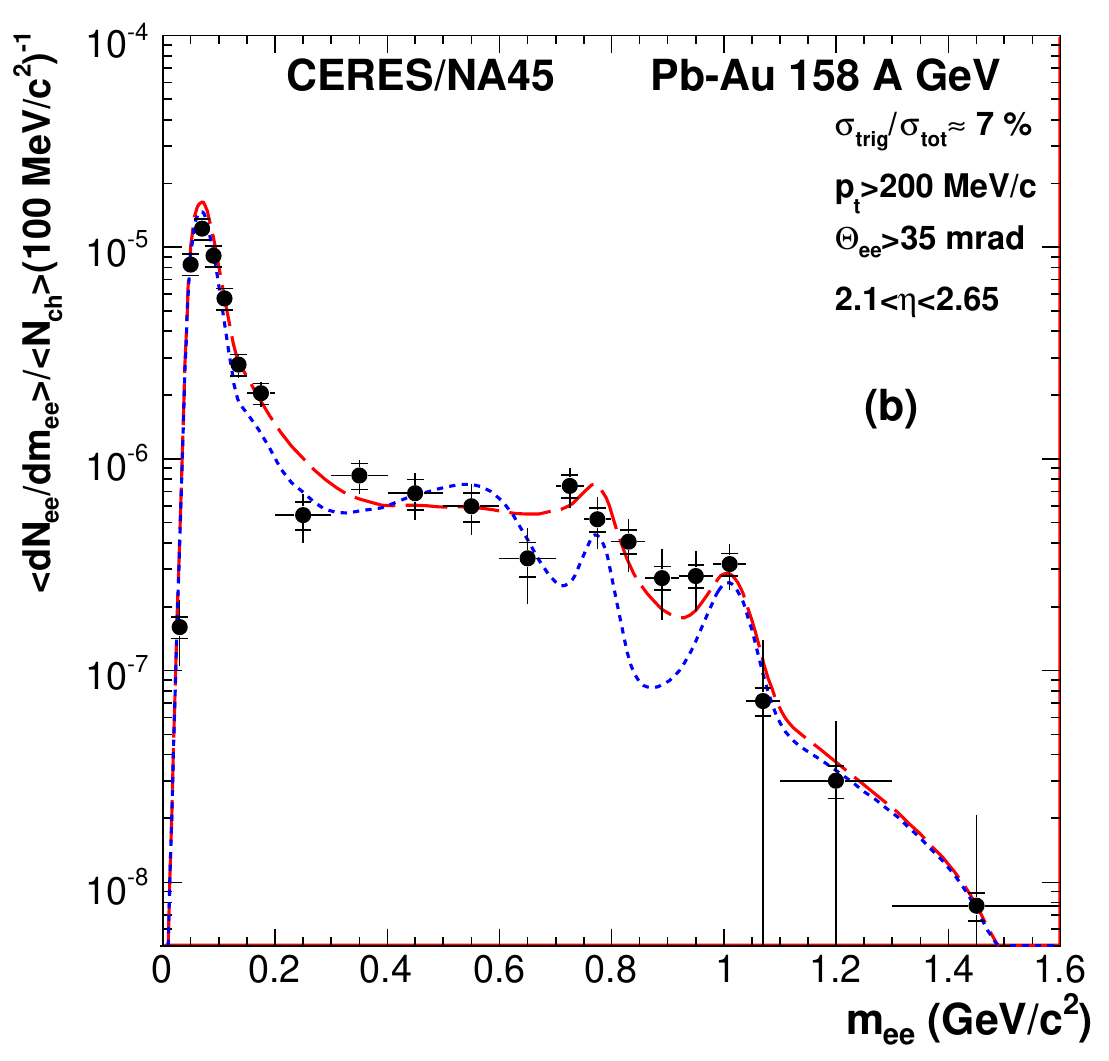}
    \caption{({\it{Left}}): Invariant e$^+$e$^-$ mass spectrum in 158\AGeV Pb-Au collisions measured by CERES, compared to calculations with cocktail of hadron decays (thin solid line), including pion annihilation with the vacuum $\rho$ (dashed line), dropping $\rho$ mass (dash-dot line) and broadened $\rho$ spectral function (thick solid line). ({\it{Right}}) improved measurements with TPC upgrade.  Calculations including a dropping $\rho$ mass (short dashed line) and a broadened $\rho$ spectral function 
    (long dashed line)are shown~\cite{CERES:2006wcq}.}
  \label{fig:pb-au1996}
 \end{center}
 \vspace{-6mm}
\end{figure}

The enhancement of low-mass electron pairs in S-Au collisions, first presented at the 1995 QM conference in Monterey \cite{Tserruya:2006ht}, was published in PRL (one of the most cited papers of the SPS heavy-ion program). The result attracted considerable interest in the community and triggered an intensive theoretical activity. The properties of the observed excess: its onset at $m_{ee} \sim 2m_{\pi}$, its extension into the $\rho$-mass region, and the possibility of a quadratic dependence with multiplicity, suggested that it originates from pion annihilation ($\pi^+\pi^- \rightarrow \rho \rightarrow \gamma^* \rightarrow e^+e^-$), a contribution practically non-existing in \textit{pp} or pA collisions, that had to be added to the hadronic cocktail for nucleus-nucleus collisions. This was the first experimental evidence of thermal radiation emitted by the dense hadronic matter formed in the late stage of the collisions.
The vacuum $\rho$ was insufficient to account for the magnitude of the enhancement and it was necessary to invoke in-medium modification of the $\rho$-meson spectral function. Two different models reproduced the experimental results equally well, as seen in Fig.~\ref{fig:pb-au1996}~({\it{Left}}). The first approach invokes a decrease of the $\rho$-meson mass in the dense medium following the conjectured Brown-Rho scaling. 
In this, the $\rho$ mass scales with the quark condensate, and the latter drops due to the high baryon density (rather than high temperature) of the medium as a precursor of chiral symmetry restoration. 
The second approach focuses on the calculation of the $\rho$-meson spectral function in the hot and dense hadronic medium, resulting in a broadening of the spectral function produced by the scattering of $\rho$ mesons mainly off baryons \cite{CERES:2005uih}.

The quality of the experimental results was insufficient to discriminate between the two models. However, it was realized that a good mass resolution would resolve the ambiguity. The dropping mass scenario shifts the dilepton yield to lower masses, depleting the yield between the $\omega$ and the $\phi$. In contrast, the broadening scenario adds yield below and above the $\rho$-meson mass, thus filling the gap between the $\omega$ and the $\phi$.  This prompted the upgrade of CERES with the addition of a radial drift TPC to improve the mass resolution. 
Fig.~\ref{fig:pb-au1996}~({\it{Right}}) shows the invariant dielectron mass spectra measured in 2000 with the upgraded spectrometer~\cite{CERES:2006wcq}. The data are compared with dropping mass and broadening calculations. Whereas the two models give very similar results for masses $m<0.8$~GeV/c$^2$, 
the data between the $\omega$ and the $\phi$ favor the broadening scenario. With this, CERES confirmed similar conclusion reached by NA60 in dimuon measurements in In-In collisions at 158\AGeV \cite{Hohler:2013eba}.
   
The deeper impact of the CERES results is in their relevance to chiral symmetry restoration. Combined with model calculations of the spectral shapes of both the $\rho$ and the $a_1$ resonances \cite{Hohler:2013eba}, the dilepton results tell us that the approach to chiral symmetry restoration proceeds through broadening of the $\rho$ together with a decreasing $a_1$ mass, resulting in the eventual degeneration of the two states. 

In addition to the dileptons, CERES performed various other studies, including the search for direct photons in S-Au collisions~\cite{CERES:1995rik}, the production of neutral mesons $\pi^0, \eta$, and $\omega$ in p-Be and p-Au collisions at 450 GeV \cite{Agakichiev:1998ign}. The TPC allowed for a broad range of measurements of hadronic observables, including pion interferometry \cite{CERES:2002rfr}, event-by-event fluctuations of 
the mean \pt~\cite{CERES:2003sap}, elliptic and triangular flow \cite{CERES:2012hbp, CERESNA45:2016rra}, two-particle correlations \cite{CERESNA45:2003pat}, jet-like correlations \cite{CERES:2009jng}, $\phi$ meson production via leptonic and charged kaon decay modes \cite{CERES:2005wwe}, and more. Twenty-five years since the last CERES run, the data are still being exploited. Thanks to the ingenuity and perseverance of Jovan Milosevic, the last CERES paper was published in 2024 \cite{CERESNA45:2024pkm}.

\section{Experiments NA35 to NA49 - strangeness and fluctuations
} 

\label{SPS_NA49}
\vspace{-0.2cm}

\subsection{A note on NA35}
\vspace{-0.3cm}
NA35 was the first experiment in the family of heavy-ion experiments conducted at the H2 beamline of the CERN North Area at the SPS. It collected data on collisions of light nuclear beams with various targets at 60\AGeV and 200\AGeV from 1986 to 1992. Initially, the collaboration was led by the experiment's initiator, Reinhard Stock. Later, when Reinhard shifted his focus to establish the NA49 experiment, Peter Seyboth, a former leader of the NA5 experiment, took over as the spokesperson.

The origin of the NA35 experiment can be traced back to the proposal~\cite{na35_roots} submitted by the GSI-LBL-Heidelberg-Marburg-Warsaw Collaboration to CERN's Proton Synchrotron and Synchro-Cyclotron Committee in January 1982. 
The proposal envisioned measurements of target fragmentation and hadron production in collisions of $^{16}$O with targets ranging from $^{40}$Ca to $^{206}$Pb at the CERN PS. The requested beam energy range was 9–13\AGeV, with an anticipated start of data taking in the spring of 1984, requiring 250 hours of dedicated PS running time. 
Although CERN accepted the proposal, accommodating the experiment in the East Hall PS proved challenging. As a result, CERN management suggested~\cite{Klapisch:1984yfi} transporting the ion beam from the PS to the SPS and distributing it to the North and West Halls of the SPS. 

\begin{figure}[btp]
\begin{center}
\includegraphics[width=0.7\textwidth]{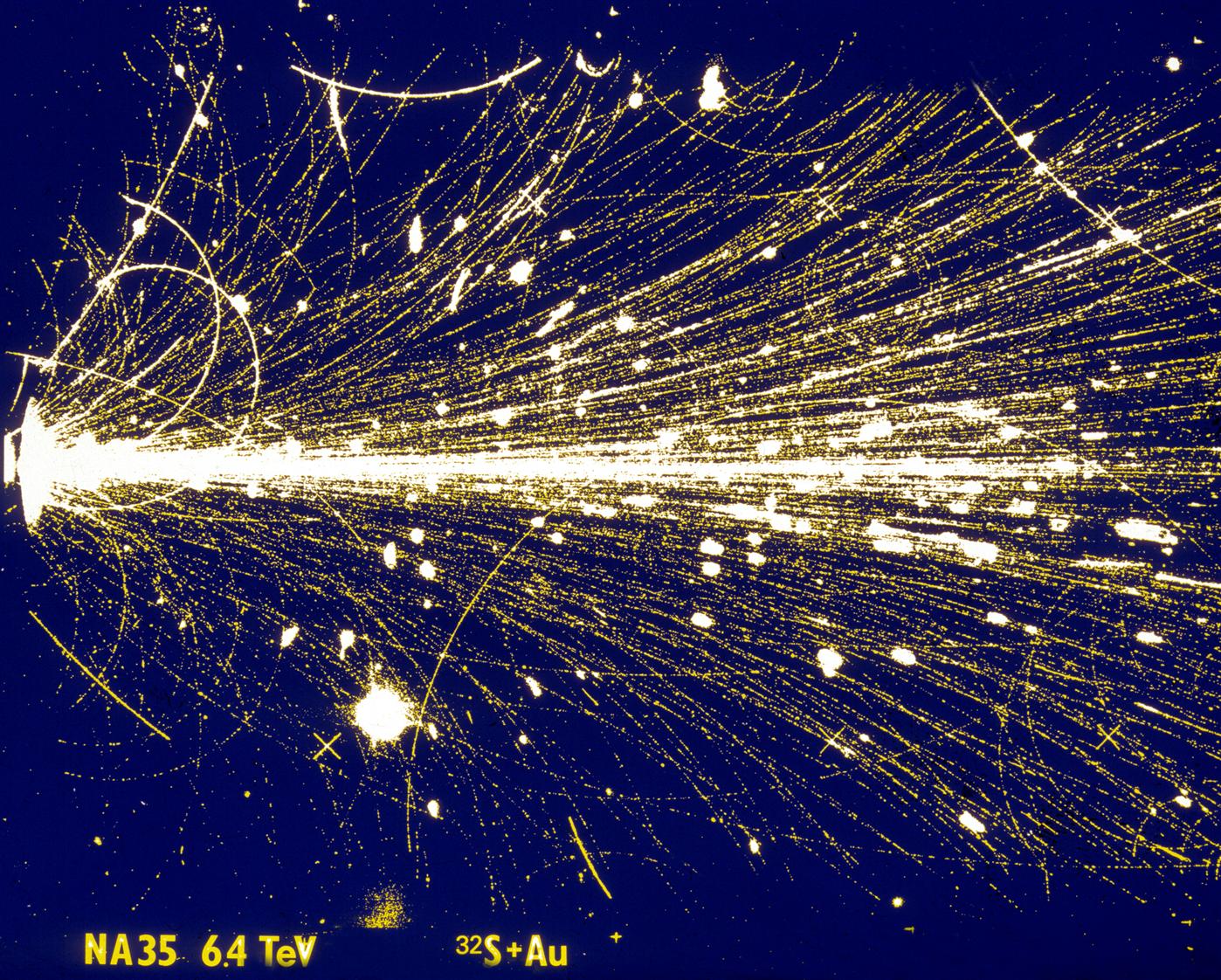}
\end{center}
\caption{An iconic image of a central S+Au collision at 158\AGeV recorded by the NA35 Streamer Chamber.
}
\vspace{-3mm}
\label{fig:na35_event}      
\end{figure}

The NA35 setup (Fig.~\ref{figures:SPS_2}) featured a visual tracking detector, a streamer chamber, with a sensitive volume of $200\times120\times70$~cm$^3$, placed in a 1.5~T vertical magnetic field. Downstream, the detector was followed by four calorimeters. Major components of the experiment were inherited from the NA5 and NA24 experiments at the CERN 
SPS~(See Ref.~\cite{Sandoval:170613}). NA35 recorded data on collisions of $p$, $d$, $^{16}$O, and $^{32}$S beams at 60\AGeV and 200\AGeV with S, Ag, and Au targets. The iconic image of a central S-Au collision at 158\AGeV is presented in Fig.~\ref{fig:na35_event}.
The main result of NA35 was the observation of enhanced production of strange hadrons in central S-S
collisions at 200\AGeV. This marked the first indication of QGP formation at the CERN SPS~\cite{NA35:1990teq}.

\subsection{NA49 program}
\label{SPS_NA49_program}
\vspace{-0.3cm}
The NA35 Collaboration initiated the NA49 experiment with a proposal submitted~\cite{Panagiotou:295042} to the SPS and LEAR Committee, which was approved in September 1991. The proposing institutions included groups from Athens, Berkeley, Darmstadt, Frankfurt am Main, Freiburg, Marburg, Munich, Seattle, Warsaw, and Zagreb. Reinhard Stock served as the spokesperson. 
The experiment encompassed two distinct physics programs: 

\begin{itemize}
\item 
{First phase (NA49-1: 1994-1997): focused on searching for anomalies in event-by-event fluctuations in central Pb-Pb collisions at 158\AGeV.}
\item 
{Second phase (NA49-2: 1998-2002): investigated the onset of QGP creation in Pb-Pb collisions by reducing the collision energy~\cite{Bächler:356588}. The primary goal was to study the collision energy dependence of hadron production properties.}
\end{itemize}

The primary aim of NA49-1 was to search for the deconfinement transition by studying event-by-event fluctuations in hadron production~\cite{Panagiotou:295042}. 
This goal necessitated a significant upgrade of NA35 to enable large acceptance measurements of hadron production in high-multiplicity events (up to 1000 charged particles) in Pb-Pb collisions. Building on NA35's strangeness enhancement results in S-S collisions~\cite{NA35:1990teq}, special emphasis was placed on identifying charged kaons and pions.

\begin{figure}[btp]
\begin{center}
\includegraphics[width=0.9\textwidth]{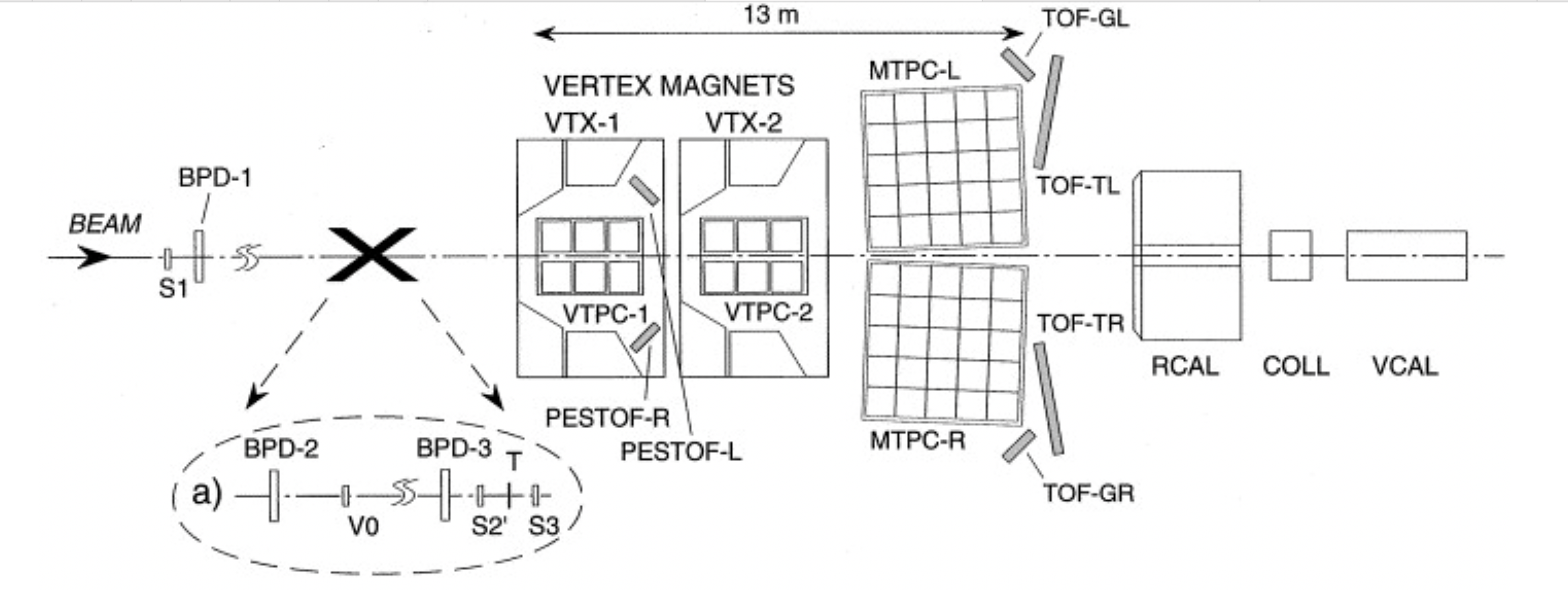}
\end{center}
\caption{
The experimental setup of the NA49 experiment~\cite{NA49:1999myq}. See text for details.
}
\vspace{-3mm}
\label{fig:na49_setup}      
\end{figure}

The experimental setup of the NA49 experiment is shown in Fig.~\ref{fig:na49_setup}~\cite{NA49:1999myq}. The main components included four large-volume Time Projection Chambers (TPCs) for tracking and identifying particles via energy loss in the TPC gas. TOF scintillator arrays extend charged kaon identification down to mid-rapidity. Calorimeters for transverse energy measurements and central collision selection using forward energy complete the downstream acceptance. The figure also shows the configuration of beam and trigger detectors used in data taking in Pb-Pb collisions.

In 1995 and 1996, approximately 1.5 million central Pb-Pb collisions at 158\AGeV were recorded, fulfilling the NA49-1 data-taking plan. Reinhard Stock transferred the role of spokesperson to Peter Seyboth in 1997.

\subsection{NA49-2: collision energy dependence of hadron production
}
\label{subsec:na492_exp}
\vspace{-0.3cm}
In the mid-1990s, numerous results were obtained from collisions of light nuclei at the BNL AGS (Si at 
14.6\AGeV) and the CERN SPS (O and S at 200\AGeV). Experiments with heavy nuclei (AGS: Au-Au at 11.6\AGeV; SPS: Pb-Pb at 158\AGeV) were just beginning. This period marked the first opportunity to examine the energy dependence of hadron production in nucleus–nucleus collisions at high energies. 

Two compilations, one on production of pions~\cite{Gazdzicki:1995zs} and another on 
strangeness~\cite{Gazdzicki:1996pk}, led to the conclusion that the energy dependence of mean hadron multiplicities measured in A-A collisions differs significantly from that observed in \textit{pp} interactions. Moreover, the 
A-A data suggested a notable change in the energy dependence of pion and strangeness yields at energies between the top AGS and top SPS energy. 

Using a statistical approach to strong interactions~\cite{Fermi:1950frz, Landau:1953wku}, it was speculated~\cite{Gazdzicki:1995ze} that this change is related to beginning of QGP creation with the increase of collision energy (onset of deconfinement)
in the early stage of A-A collisions. Following this conjecture, the Statistical Model of the Early Stage (SMES)~\cite{Gazdzicki:1998vd}, was developed by Marek Gazdzicki and Mark Gorenstein. SMES assumes the creation of early-stage matter according to the principle of maximum entropy. Depending on the collision energy, matter exists in the confined phase ($E \lesssim 30$\AGeV), the mixed phase between $30 \lesssim E \lesssim 60$\AGeV), or the deconfined phase ($E \gtrsim 60$\AGeV). The phase transition is assumed to be of the first order.

In 1997, based on those ideas, the collaboration proposed studying hadron production in Pb-Pb collisions at 40\AGeV~\cite{Bächler:356588}, marking the beginning of the second phase of the NA49 experiment.
The NA49-2 programme was conducted using essentially the same experimental setup as NA49-1, depicted in Fig.~\ref{fig:na49_setup}. The only modification was scaling the magnetic field in proportion to the beam momentum to ensure kaon identification at mid-rapidity via TOF and specific energy loss measurements in the TPCs.

The first 40\AGeV Pb beam was delivered to NA49-2 in 1998 as a test, followed by a five-week run in 1999. The success of these runs at low SPS energy, combined with the promising results by NA49-2, justified the continuation of the programme. In 2000, a beam at 80\AGeV was delivered to NA49 for a five-day run. 
The programme was completed in 2002 with two additional runs: one at 30\AGeV (seven days) and another at 20\AGeV (seven days).

\subsection{Key findings from NA35 and NA49}
\begin{figure}[bp]
\begin{center}
\includegraphics[width=0.8\textwidth]{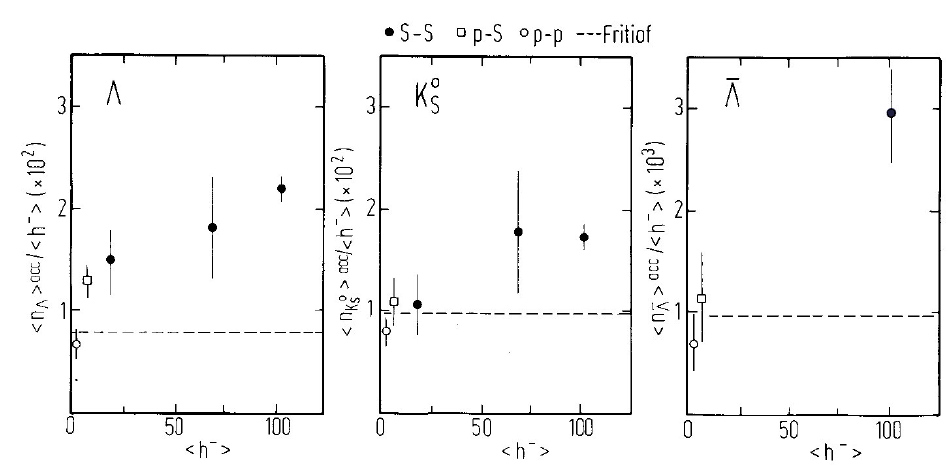}
\end{center}
\caption{
The ratio of the mean multiplicities of strange hadrons to the total negative hadron multiplicity in 
S-S collisions at 200\AGeV measured by NA35 for three centralities. 
The predictions of the simple superposition model, Fritiof, remain independent of the  
\hm mean multiplicity~\cite{NA35:1990teq}. 
}
\vspace{-6mm}
\label{fig:na35_strangeness}      
\end{figure}

\vspace{-0.3cm}
The most influential results of NA35 concerned neutral strange hadron production in central S-S collisions at 200\AGeV~\cite{NA35:1990teq}.  
Due to the reflection symmetry of the initial state, it was possible to calculate the total mean multiplicity of abundant hadrons carrying $s$ and $\bar{s}$ quarks, specifically $\Lambda$ hyperons and $K^0_S$ mesons. Their yields, relative to negatively charged hadrons (\hm, more than 90\% of which are \pim mesons), were approximately twice as large as the corresponding yields in \textit{pp} interactions. See Fig.~\ref{fig:na35_strangeness} for illustration. 
This result indicated substantial production of $s\bar{s}$ pairs, consistent with the predictions of Rafelski and Mueller regarding QGP formation in such 
collisions~\cite{Rafelski:1982pu, Koch:1986ud}.

NA35 also achieved other important results, particularly on two-pion correlations~\cite{NA35:1988eto}. It demonstrated that the pion emission source is significantly larger than the size of the colliding nuclei, indicating the expansion of the matter created in the early stages of the collisions. 

The NA49 phase-one results were summarised in 1998 as follows~\cite{Bächler:356588}:\\  
\textit{
The NA49 data analyses combines the study of particle yields with particle correlations on an event–by–event level. First results~\cite{NA49:1994hfj, NA49:1997qey, NA49:1997xvz, Roland:1997hs} shed new light on the salient questions concerned with the behaviour of matter at large energy density. In particular, the step from collisions of medium-size nuclei to large–size systems does not lead to a substantial change in the observables considered as possible signals of a phase transition like the pion/baryon and strangeness/pion ratios. A significant change in these observables is seen only
when comparing nucleon–nucleon interactions with collisions of medium– and large–size nuclei at the SPS. The event–by–event analysis of Pb-Pb data does not reveal the appearance of dynamically different new event classes down to the permille level of admixture. 
}

The basic NA49-2 results 
on search for
the onset of deconfinement were published in 2002~\cite{NA49:2002pzu} and 2008~\cite{NA49:2007stj}. These results, focusing on pion and kaon production in central Pb-Pb collisions at 40\AGeV, 80\AGeV, 158\AGeV, and later at 20\AGeV and 30\AGeV, have together accumulated approximately 1300 citations. 

The collaboration concluded~\cite{NA49:2007stj}:
\textit{
``A rapid change of the energy dependence is observed around 30\AGeV for the yields of pions and kaons as well as for the shape of the transverse mass spectra. The change is compatible with the prediction that the threshold for production of deconfined matter at the early stage of the collisions is located at low SPS energies." 
}
Here, we briefly report the two most striking results:  
The $E_S \equiv (\langle \Lambda \rangle + \langle K \rangle + \langle \bar{K} \rangle) / \langle \pi \rangle$ ratio is approximately proportional to the total strangeness-to-entropy ratio, which in the SMES model~\cite{Gazdzicki:1998vd} is assumed to remain constant from the early stage to freeze-out. At low collision energies, the strangeness-to-entropy ratio increases steeply with energy due to the low temperature at the early stage ($T \lesssim T_C$) and the high mass of the strangeness carriers in the confined state (e.g., the kaon mass of 500~MeV). 

\begin{figure}[btp]
\begin{center}
\includegraphics[width=0.45\textwidth]{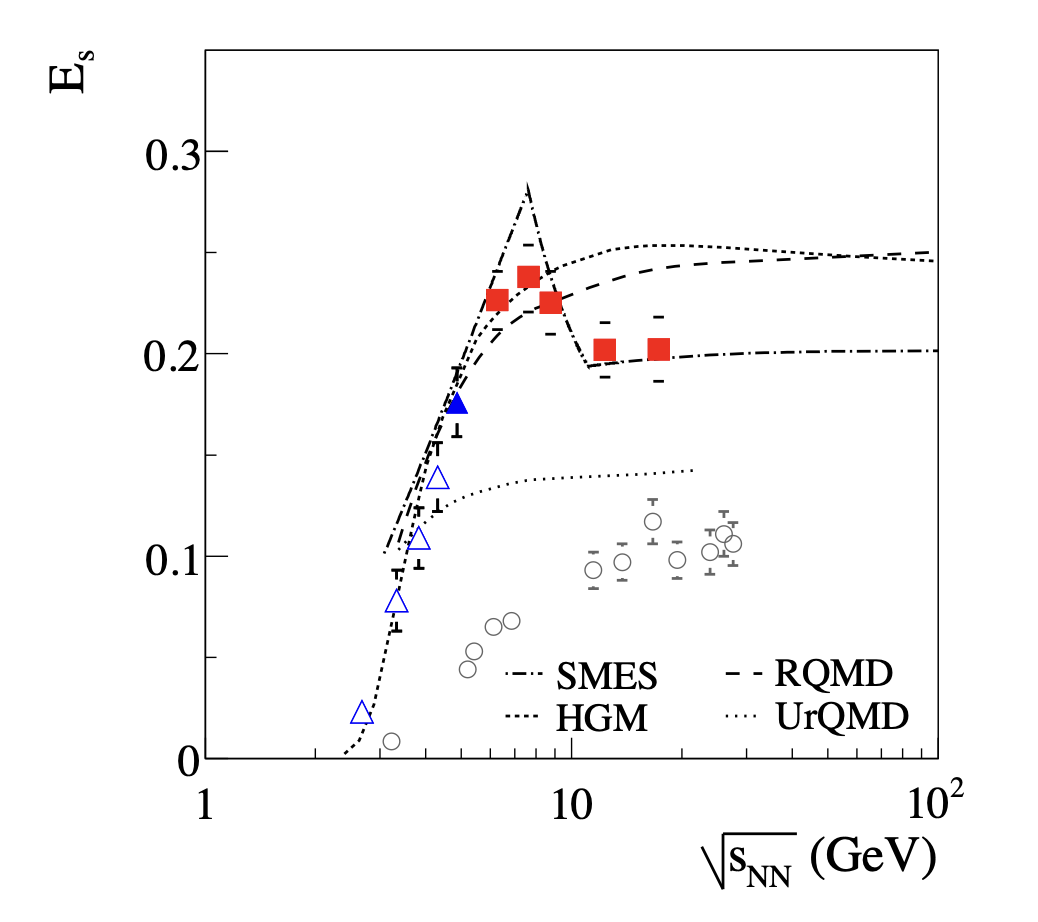}
\includegraphics[width=0.45\textwidth]{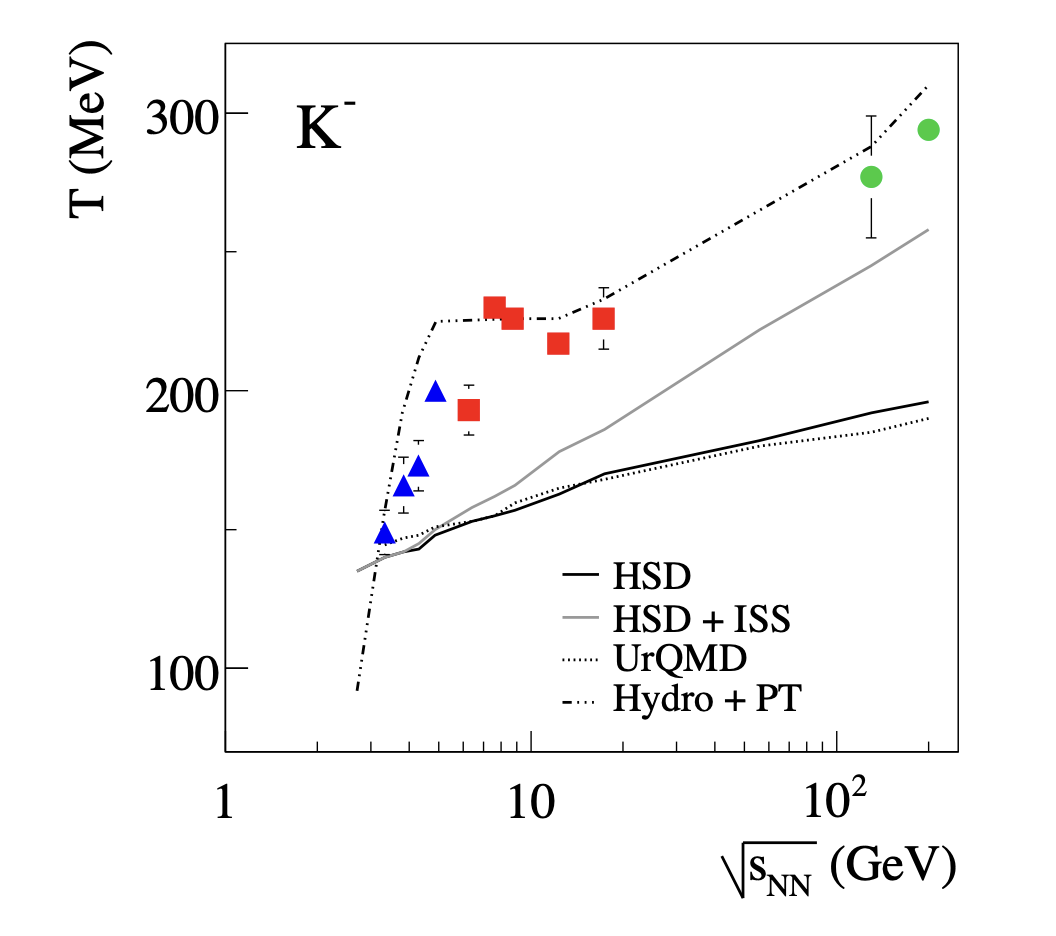}
\end{center}
\caption{
Collision energy dependence of (\textit{Left}) relative strangeness production ($E_S$) in central Pb-Pb and Au-Au collisions (full symbols), compared to  \textit{pp}($\bar{p}$) reactions (open circles),  (\textit{Right}) 
the inverse slope parameter $T$ of the transverse mass spectra for $K^-$ measured at mid-rapidity in central Pb-Pb and Au-Au collisions. The curves represent predictions from  SMES and Hydro+PT models which include the onset of deconfinement at the low SPS energies, and the other models those do not~\cite{NA49:2007stj}.
}
\vspace{-6mm}
\label{fig:na49_horn}      
\end{figure}

When the transition to a QGP is crossed ($T \gtrsim T_C$), the mass of the strangeness carriers drops significantly to the strange quark mass ($\approx$100~MeV). As a result, when $m < T$, the strangeness yield becomes approximately proportional to the entropy, and the strangeness-to-entropy (or pion) ratio becomes independent of energy. This transition leads to a decrease in the energy dependence of the ratio from the larger value for confined matter at $T_C$ to the QGP value. Consequently, the measured non-monotonic energy dependence of the strangeness-to-entropy ratio (Fig.~\ref{fig:na49_horn})
is followed by saturation at the QGP value. Within the SMES framework, this anomalous energy dependence, referred to as the ``horn",  is a direct consequence of 
the onset of deconfinement 
occurring at approximately 30\AGeV. 

In the mixed-phase region, the early-stage pressure and temperature are independent of the energy density~\cite{VanHove:1982vk}. Consequently, within the SMES model, this results in a weakening of the increase with energy of the inverse slope parameter $T$ (the ``step") or, equivalently, the mean transverse mass $\langle m_T \rangle$ in the SPS energy range~\cite{Gorenstein:2003cu}. 

This qualitative prediction is confirmed by the results shown in 
Fig.~\ref{fig:na49_horn}~(\textit{Left}). Furthermore, hydrodynamic calculations~\cite{Gazdzicki:2003dx} that incorporate both the deconfined and hadronic phases quantitatively describe the data, as demonstrated by the dashed-dotted curve (Hydro + PT) in 
Fig.~\ref{fig:na49_horn}~(\textit{Right}).

The observation of the predicted signals of deconfinement (the horn, step, and kink) and the markedly different collision energy dependence of hadron production properties in central 
Pb-Pb collisions compared to \textit{pp} interactions underscored the need to continue the heavy-ion programme at the CERN SPS and other facilities worldwide. This motivated several initiatives, including NA61/SHINE at the SPS, the Beam Energy Scan programme~\cite{STAR:2010vob} at RHIC, the CBM experiment~\cite{CBM:2016kpk} at 
the Facility for Antiproton and Ion Research (FAIR) in Darmstadt, and the MPD experiment~\cite{MPD:2022qhn} at NICA in Dubna.  

A series of workshops titled \textit{Critical Point and Onset of Deconfinement} was established in 2004 by Marek Gazdzicki, Peter Seyboth, and Edward Shuryak at the European Centre for Theoretical Studies in Nuclear Physics and Related Areas (ECT*), Trento~\cite{CPOD2024}. During the first workshop in Trento, the search for critical point of strongly interacting matter was identified as a key objective of these programmes.

\section{Experiments NA38 to NA50 -  quarkonium suppression}
\label{SPS_NA50}
\vspace{-0.3cm}

The first round of measurements carried out at the CERN SPS with heavy-ion collisions led to a number of exciting but also puzzling results. A number of experiments focused on different ``signatures'' of the QGP, with no real ``smoking gun'' being detected.  Among the various experiments, NA38, built as a modest upgrade of the older NA10 experiment, was specifically devoted to the study of muon pair production (Fig.~\ref{figures:SPS_2}. It had started its data taking with oxygen beams in 1986, with a timing remarkably coincident with the appearance of an incisive and bold paper by Helmut Satz and Tetsuo Matsui~\cite{Matsui:1986dk}, which stated in its abstract ``We conclude that $J/\psi$ suppression in nuclear collisions should provide an unambiguous signature of QGP formation". At the following Quark Matter conference, in August 1987, NA38 experiment  reported on the suppression of $J/\psi$~\cite{NA38:1988ahd}, this result being considered as a real highlight in the words of Mikolos Gyulassy at the conference summary: ``the most provocative observation, reported by NA38, was that $J/\psi$ production seems to be suppressed by $\sim$30\% in high \pt events``~\cite{Gyulassy:1988gg}.

However, it soon became clear that competing effects, as those due to $J/\psi$ dissociation in cold nuclear matter, could reproduce the observations to a good extent. It was also suspected that sulphur ions, the heaviest species technically feasible in the SPS at that time, could be too small to create a sufficiently extended region where the QGP could be formed.

In that situation, the availability of truly heavy ions (Pb), foreseen for 1994, was considered as a clear-cut occasion to distinguish QGP-related effects, expected to be strongly enhanced, from conventional mechanisms. This led to the formation of an extended collaboration, NA50, with Italian groups joining the historical mainly French-Portuguese collaboration. 

\subsection{NA50 experimental setup}
\vspace{-0.3cm}
\begin{figure}[tbp]
\begin{center}
\includegraphics[width=0.8\textwidth]{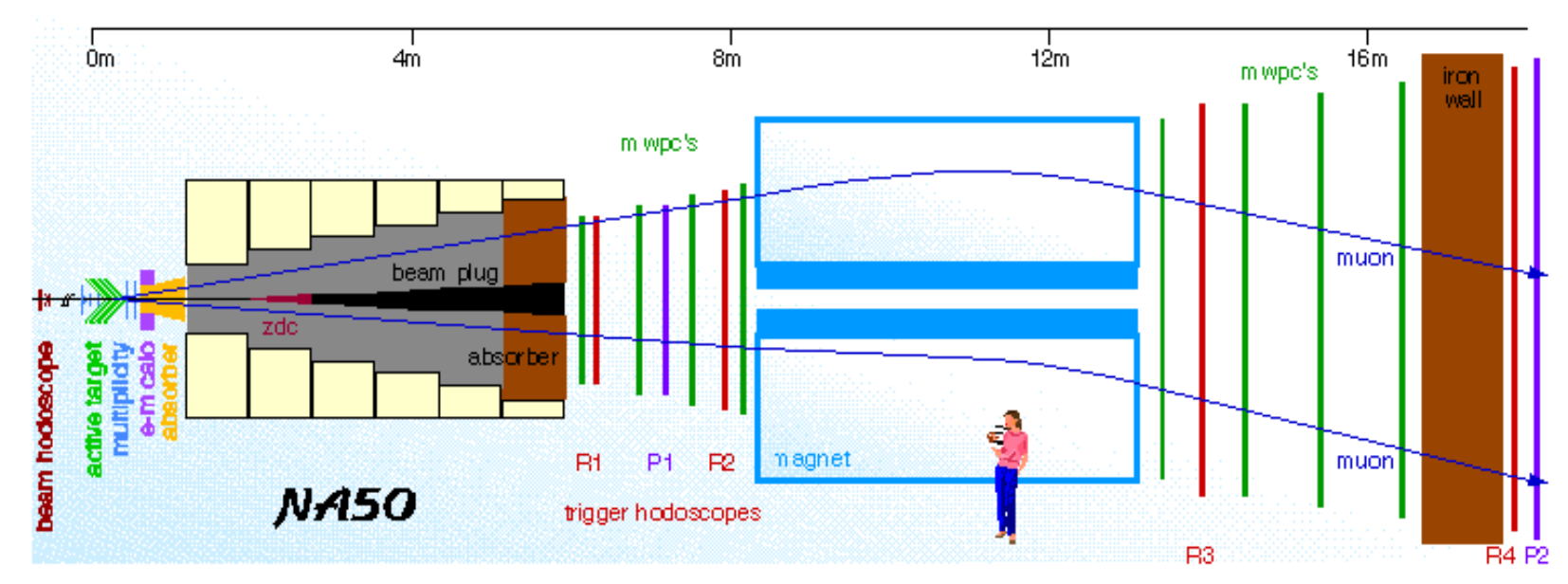}
\end{center}
\caption{The NA50 experimental apparatus~\cite{NA38:2000wlp}.}
\vspace{-6mm}
\label{fig:na50_1}
\end{figure}

The core of the apparatus remaining from the NA10 muon spectrometer, composed of trigger scintillators, MWPCs for tracking and featuring a large toroidal magnet. It was upgraded with several new detectors, particularly devoted to a precise determination of the centrality of the collisions. Among these ``entry tokens'' for the new groups, a quartz-fiber zero-degree calorimeter and a remarkably granular Si strip detector were added to the setup. The experimental setup of the NA50 experiment is shown in Fig.~\ref{fig:na50_1}.
One of its distinctive features was the possibility of working with beam intensities up to several 10$^7$ ions/spill, at last a factor 10 larger with respect to the other heavy-ion experiments. This necessity was dictated by the relatively low cross sections for charmonium production at SPS energies.

\subsection{Key findings of NA50}
\vspace{-0.3cm}
At that time, people were aware that the signals NA50 would be looking at in Pb-Pb collisions might represent the long-sought smoking gun for QGP discovery. Collaborations were relatively small at that time, with only a few tens of people active, but this small lot included several first-class physicists as Louis Kluberg from Ecole Polytechnique leading the experiment with a calm but authoritative approach, Claudie Gerschel from IPN Orsay who was keeping the contacts with theorists and was famous for her unconventional approach to the physics interpretation of the data, and Peter Sonderegger of CERN who was always unorthodox, provocative and brilliant in putting forward ideas for new measurements and physics signals. Another typical feature, probably shared by other collaborations at that time, was the coexistence of physicists coming from the  world of high-energy particle physics, together with lower-energy nuclear physicists, such as Emilio Chiavassa, the most senior researcher in the Torino group. Nobody was born, as it is the case today, as a ``heavy-ion physicist'', and that was an occasion for sharing the respective experience and learning on both sides. At the other end of the age span of the collaboration, a group of young post-docs were eager to analyze the first data, expected at the end of 1994. 

The integrated luminosity collected in that first year was in the end rather low, so it was with the second Pb-Pb data taking in 1995 that a first clear physics result was obtained. By studying the ratio $J/\psi$/Drell-Yan as a function of centrality and appropriately comparing with previously available p-A data, a clear signal of ``anomalous'' suppression was seen. 
\begin{figure}[btp]
\begin{center}
\includegraphics[width=0.5\textwidth]{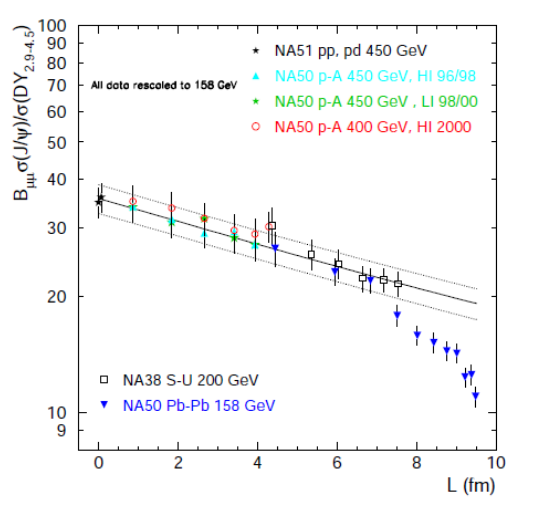}
\end{center}
\vspace{-6mm}
\caption{
($J/\psi$)/Drell-Yan cross-section ratio {\it vs.} $L$, the mean thickness of nuclear matter crossed by the $c\overline c$ pair on its way through nuclear matter~\cite{NA50:2000brc}. The black curve represents the $J/\psi$ suppression due to ordinary nuclear absorption, tuned on \textit{pp} and p-A data. The Pb-Pb data are shown to significantly deviate from the behaviour expected in case of absence of QGP effects. 
}
\vspace{-3mm}
\label{fig:na50_2}
\end{figure}

It was also a time of honest but heated competition between experiments and once this result was approved in a collaboration meeting, an appointment with the CERN Director of Research (Lorenzo Fo\`a at that time) was asked by Louis Kluberg for the same evening to present these findings,  
which were received with interest and courtesy by the CERN management. But of course the most exciting phase was the presentation and discussion of the new discovery in the heavy-ion community. This happened extensively at the Quark Matter 1996 conference in Heidelberg, where once more the $J/\psi$ results~\cite{NA50:1996lag} represented ``the most striking result'' as one can read in the preface to the proceedings of the conference. It was a time of heated discussions, where scientific facts, strong characters and also personal rivalry were meeting and very often clashing; a very exciting period to witness, when the heavy-ion field was still in its youth, a time  honestly difficult to compare with the more poised attitude of current experiments.
The anomalous $J/\psi$ suppression~\cite{NA50:2000brc} was scrutinized and slowly digested by the community and an alternative explanation involving the so-called ``hadronic comovers'', a kind of dense hadron gas that, as an alternative to QGP, was supposed to be also able to dissociate the $J/\psi$ was, sometimes quite vocally, put forward. 

\begin{figure}[btp]
\begin{center}
\includegraphics[width=0.5\textwidth]{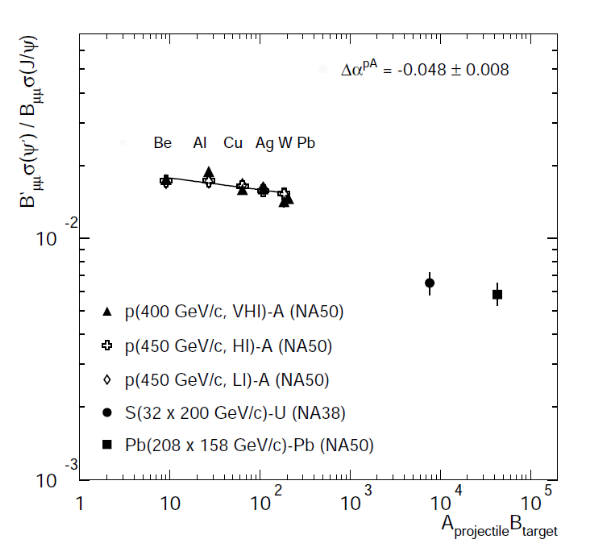}
\end{center}
\vspace{-4mm}
\caption{$\psi(2S)/$(J/$\psi$) cross-section ratio 
{\it vs.} the product of projectile and target mass numbers~\cite{NA50:2006yzz}. Data from \textit{pp} to pA, S-U and Pb-Pb collisions are shown. A strong decrease of the ratio is visible in both S-U and Pb-Pb collisions, indicating a stronger suppression for the more weakly bound $\psi(2S)$ state.}
\vspace{-4mm}
\label{fig:na50_3}
\end{figure}

All in all, the anomalous suppression (shown in Fig.~\Ref{fig:na50_2}) was able to withstand criticisms, a process that took time and found its natural completion a few years after, when in February 2000 all the SPS experiments collected their finding in a special seminar where ``evidence for colour deconfinement in the early collision stage and for a collective explosion of the collision fireball in its late stages'' was presented and a CERN press release~\cite{CERN2000NewState} on the discovery of a new state of matter was issued. 

Although the $J/\psi$ studies~\cite{NA50:2004sgj} remain its highlight, the NA50 physics program was also including other hot topics, as the observation of an enhancement in the muon pair production between the 
$\phi$ and the $J/\psi$, the so-called intermediate mass region~\cite{NA38:2000wlp}. This enhancement was suspected to be due to thermal production, although the evidence was reached years later by the NA60 experiment (see next section), a very successful spin-off of NA50. Another key results was the study of the $\psi(2S)$ production, a much more weakly bound $c\overline c$ state that was found to exhibit a stronger suppression with respect to the $J/\psi$~\cite{NA50:2006yzz} (see Fig.~\Ref{fig:na50_3}). This observation was seen as a confirmation of the ``sequential suppression'' scenario, corresponding to an increased level of suppression with decreasing binding energy of the charmonium state under study.
With the CERN press release of 2000 that marked the end of most SPS experiments, the energy frontier was moved to RHIC, where the story of $J/\psi$ production was about to continue with a new chapter. But that is still another story.

\section{Experiment NA60 - the dimuon specialist}
\label{SPS_NA60}
\vspace{-0.3cm}


The NA60 experiment was designed as a successor to the NA50 at the CERN SPS. The inherited NA50 spectrometer
 consisted of a thick hadron absorber followed by a toroidal magnet and a tracking system based on MWPCs. However, this configuration presented significant limitations. The multiple scattering in the hadron absorber substantially degraded the mass resolution, achieving only $70$ MeV at the $\omega$ meson mass. Additionally, the absorber severely constrained the transverse momentum (\pt) acceptance, particularly for muon pairs with invariant masses below $1$ GeV.

\begin{figure}[tbp]
\begin{center}
\includegraphics[width=0.8\textwidth]{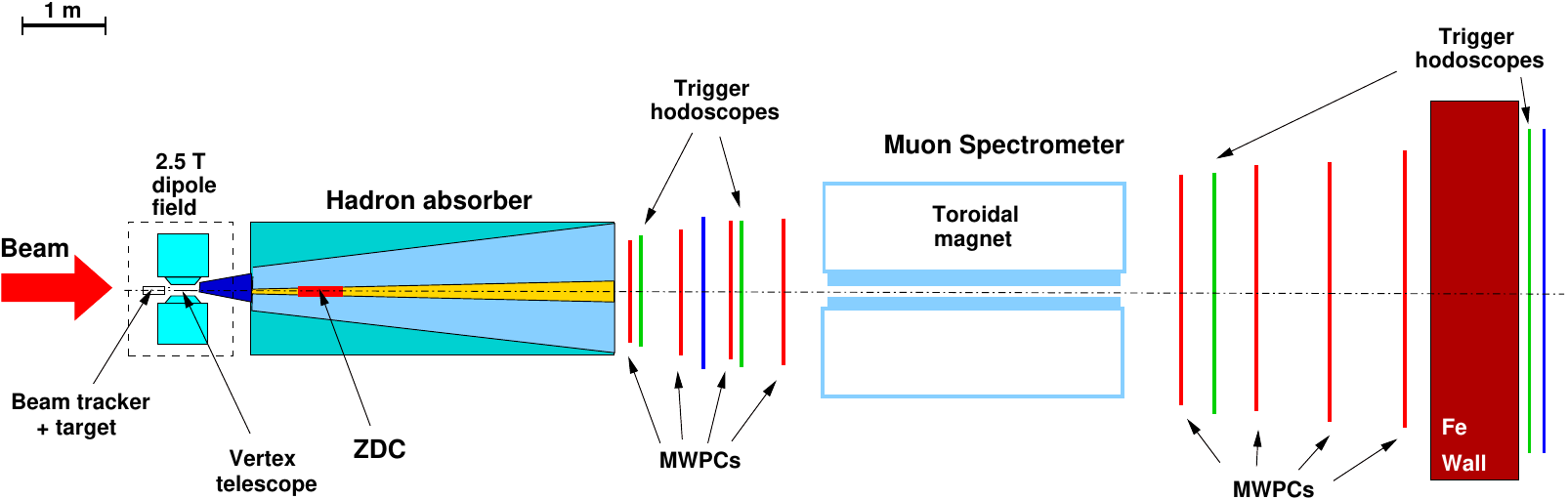}
\includegraphics[width=0.71
\textwidth]{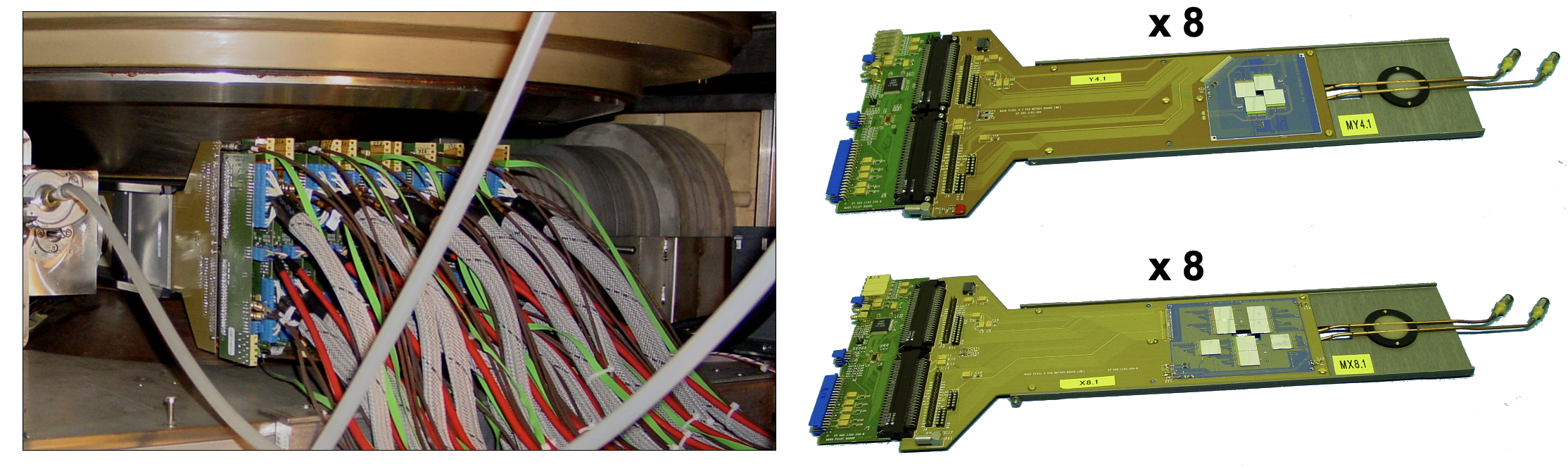}
\end{center}
\caption{Top: The NA60 apparatus: the NA50 muon spectrometer is complemented by a vertex spectrometer in front of the hadron absorber~\cite{NA60:2008dcb}. Bottom: stations of silicon pixel sensors installed inside the PT7 magnet. Details of the silicon stations are shown on the right.}
\vspace{-2mm}
\label{fig:na60_setup}
\end{figure}



In the late 1990s, radiation-hard silicon pixel detectors were being developed for LHC experiments. During this period, Peter Sonderegger conceived the innovative idea of complementing the muon spectrometer with a vertex detector. It consisted of a telescope with planes of pixel sensors operating within a magnetic field, capable of tracking all particles (including muons) before they entered the hadron absorber.

The complete NA60 apparatus is shown in Fig.~\ref{fig:na60_setup}(top), featuring the silicon tracker positioned inside a dipole magnet immediately upstream of the muon absorber~\cite{NA60:2008dcb}. Carlos Lourenco began collaborating with Peter during this time and secured a staff position at CERN specifically to manage NA60 operations. Gianluca Usai joined the experiment in 2000, taking responsibility for designing the silicon tracker readout system.
Despite being a small team, they successfully executed the experiment within an ambitious timeline: conducting the lead run in 2002 followed by the indium run in 2003. Fig.~\ref{fig:na60_setup}(bottom) shows the fully instrumented silicon tracker operating within the PT7 dipole magnet.




The silicon tracker required complete development from scratch, along with a significantly faster readout system for the muon spectrometer. A crucial collaboration was established with the ALICE experiment for the silicon sensors, which provided essential support in sensor procurement and offered valuable guidance on detector design.
A dedicated team of exceptional students and postdoctoral researchers was assembled. Under the leadership of Carlos Lourenco and Gianluca Usai, this team successfully designed and tested all components—including silicon planes, readout electronics, detector control systems (DCS), and data acquisition (DAQ)—within an impressive two-year timeframe. The group's efforts were further strengthened in 2001 when Ernest Radermacher joined as technical coordinator, bringing further organizational structure to the talented team of young researchers.


Among the original scientific motivations for NA60 was continuing the investigation of anomalous $J/\psi$ suppression in the QGP. This research direction was strongly supported by Helmut Satz, who specifically proposed studying quarkonium production in nuclei lighter than lead - a suggestion that directly led to the decision to conduct runs with indium. Helmut Satz maintained close collaboration with Carlos Lourenco and made several visits to the NA60 setup at CERN during the project's development.
 Louis Kluberg, the former spokesperson of NA50,
 was a key contributor to NA60.
 His extensive expertise in $J/\psi$ physics, gained through leadership of the predecessor experiment, proved invaluable for the NA60 research program.


The primary scientific motivation for NA60 was the measurement of muon pairs in the intermediate mass region (IMR) between the $\phi$ and J/$\psi$ mesons. While NA50 had observed an excess of muon pairs in this region compared to expectations from Drell-Yan production and semi-leptonic decays of $D\bar{D}$ mesons, the origin of this excess remained unclear. Two competing interpretations emerged: either an enhancement of the charm production cross-section or a prompt source such as thermal radiation from partonic and/or hadronic matter.
The vertex detector developed for NA60 provided the crucial capability to distinguish these scenarios through precise measurement of displaced vertices from $D$ meson decays. This technological advancement represented the key to resolving the longstanding puzzle.
However, significant questions remained about the experiment's ability to study the lower mass region. In particular, it was uncertain whether the $\rho$ meson - considered an important probe of chiral symmetry restoration in the hot medium - could be effectively measured given the \pt cuts imposed by the experimental setup.
In 2002, Hans Specht joined the NA60 collaboration. His involvement proved crucial when the team realized that the magnetic field in the vertex region enabled the detection of numerous soft muons that would otherwise have been lost. This breakthrough made it evident, for the first time, that precise measurements of the $\rho$ meson were indeed feasible with the NA60 apparatus. Hans Specht played a pivotal role in reshaping the experiment's scientific scope.

The planned lead run in 2002 was ultimately cancelled due to technical difficulties. Following a successful commissioning test beam for the silicon tracker in August 2003, the experiment proceeded with the indium run in October 2003. This run achieved remarkable success, accumulating a world-record statistics of $5\times 10^5$ reconstructed muon pairs - a milestone that remains unsurpassed more than two decades later.
The initial examination of mass spectra revealed extraordinary features. In the low-mass region, the $\omega$ and $\phi$ mesons appeared as exceptionally well-defined peaks with mass resolution of approximately 20 MeV, unprecedented in heavy-ion collision studies. These preliminary results were presented at major conferences including Quark Matter and Hard Probes in 2003 and 2004.
However, transforming these observations into quantitative scientific results required extensive analysis. The most demanding aspect involved the subtraction of combinatorial background from pion and kaon decays. Additional challenges arose from potential incorrect matches between muon tracks and corresponding tracks in the vertex tracker. Nearly two years of dedicated effort were needed to perfect the subtraction procedure and evaluate all systematic uncertainties.
Ruben Shahoyan played a pivotal role in this process, developing all reconstruction software through tireless efforts that often continued uninterrupted for days. The refined results were first presented internally during the collaboration meeting in Alghero, Italy in spring 2005, followed by public presentations at ECT* (Trento) and Quark Matter (Budapest) later that year. The community responded enthusiastically to these groundbreaking results, with widespread acclaim for the impressive volume of findings presented.
Gianluca Usai assumed the role of spokesperson in 2005, coinciding with the collaboration's readiness to publish its scientific results in peer-reviewed journals.

\subsection{Key findings of NA60}
\vspace{-0.3cm}

\begin{figure}[tbp]
\begin{center}
\includegraphics[width=1.0\textwidth]{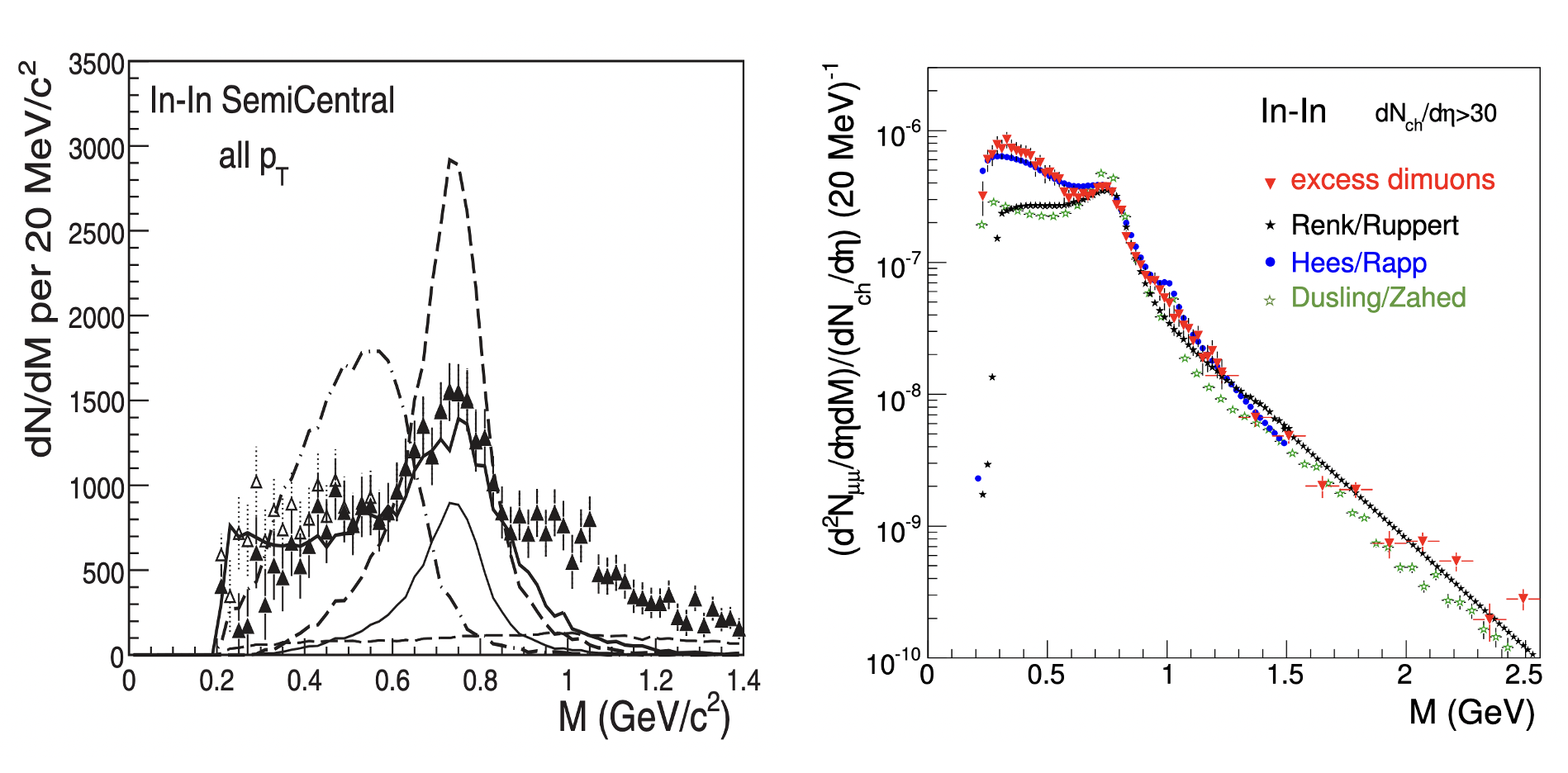}
\end{center}
\vspace{-3mm}
\caption{({\it Left}) NA60 results on excess mass spectrum in 
In-In at  $dN_{ch}/d\eta=140$ (semi-central) compared to model calculations: cocktail $\rho$(thin solid line), unmodified $\rho$(dashed line), in-medium broadening 
$\rho$~\cite{Rapp:1997fs,Rapp:1999ej} (thick solid line), 
in-medium moving $\rho$ 
(dashed-dotted ine)~~\cite{Brown:1991kk,Li:1995qm,Brown:2001nh}. ({\it Right}) Acceptance-corrected mass spectra of excess
dimuons for the combined  low-mass (LMR, $M<1$~GeV) and
intermediate-mass region (IMR). Results from theoretical calculations are shown~\cite{vanHees:2007th,Ruppert:2007cr,Dusling:2006yv}.}
\vspace{-2mm}
\label{fig:na60-thermal-radiation}
\end{figure}

The comprehensive analysis of thermal radiation represents the seminal contribution of Hans Specht and Sanja Damjanovic, the latter having completed her PhD research with the CERES experiment. The exceptional mass resolution achieved by NA60 enabled clear separation of the freeze-out contributions from $\omega$ and $\phi$ resonances, thereby isolating the thermal radiation component originating from $\pi\pi$ annihilation through the $\rho$ meson channel.
The extracted $\rho$ spectral function, representing an average over the space-time evolution of the collision fireball, exhibits two remarkable features: a nearly divergent width consistent with approaching chiral symmetry restoration, yet essentially no mass shift~\cite{NA60:2006ymb}. The excess mass spectrum relative to theoretical predictions is presented in Fig.~\ref{fig:na60-thermal-radiation} (Left).
These definitive measurements conclusively resolved a decades-long debate regarding the in-medium modifications of hadronic spectral functions near the QCD phase boundary.

Thanks to the vertex detector's precision, perfect consistency with unenhanced open charm production was demonstrated in the IMR. By systematically subtracting the Drell-Yan and open charm contributions from the total dimuon yield, the remaining excess was unambiguously identified as thermal radiation~\cite{NA60:2008dcb}, 
as shown in Fig.~\ref{fig:na60-thermal-radiation}~({\it{Right}}).
The excess follows an exponential distribution $dN/dM \sim M^{3/2}\exp(-M/T)$, characteristic of a nearly flat spectral function, where $T$ represents a space-time averaged temperature over the system's evolution. Fitting this distribution across the mass range $1.2 < M < 2$~GeV yielded $T = 205 \pm 12$~MeV~\cite{Specht:2010xu}. Crucially, as mass is a Lorentz-invariant quantity, this spectrum remains unaffected by source motion effects that typically complicate transverse mass analyses.
The extracted temperature value significantly exceeds the QCD pseudo-critical temperature $T_C = 155$~MeV, strongly suggesting dominant partonic emission in the IMR. This conclusion was further supported by two independent analyses: (1) the mass dependence of effective temperatures from transverse mass spectra, and (2) angular distribution studies revealing unpolarized thermal radiation~\cite{NA60:2007lzy,NA60:2008iqj}.


At the high-mass end of the spectrum, Enrico Scomparin, Roberta Arnaldi, and Pietro Cortese conducted detailed studies of $J/\psi$ production. The measurement of anomalous $J/\psi$ suppression requires careful calibration of cold nuclear matter effects, typically parametrized through an absorption cross section. Prior to NA60, this cross section had only been measured in pA collisions at 400~GeV, with the assumption of energy independence. Under this assumption, significant anomalous suppression was reported in In-In collisions~\cite{NA60:2006ncq}.
In fall 2004, NA60 performed dedicated pA runs at both 400~GeV and 158~GeV proton energies. These measurements presented substantial technical challenges due to the instantaneous rate of $10^9$ protons per second, which created overwhelming pile-up and would severely compromise event reconstruction. After evaluating potential mitigation strategies, the collaboration opted to employ ATLAS silicon pixel sensors, which offered four times faster readout compared to the ALICE sensors. Through collaboration with Leonardo Rossi, the ATLAS tracker project leader, two silicon planes were successfully integrated into the NA60 data acquisition within months.

The resulting absorption cross section measurements revealed a striking energy dependence, with larger values observed at lower energies~\cite{NA60:2010wey}. This finding fundamentally altered the interpretation of previous results: when properly accounting for the energy-dependent cross section, no anomalous suppression was observed in In-In collisions, while the suppression in central Pb-Pb collisions measured by NA50 remained significant. These results provide  evidence for the sequential melting scenario of charmonium states, where the $J/\psi$ yield depletion in central Pb-Pb collisions arises primarily from $\chi_c$ state melting~\cite{NA50:2004sgj}.


The NA60 experiment yielded many other significant results from both In-In and pA collisions. Alessandro De Falco, Michele Floris, and Gianluca Usai conducted comprehensive studies of $\phi$ meson production in In-In collisions, analyzing both muon and hadronic decay channels. This work resolved previous discrepancies between NA49 and NA50 measurements, demonstrating complete consistency in both yield and transverse mass spectra between the different decay channels. Additional precision measurements conducted by Gianluca Usai, Sanja Damjanovic, and Antonio Uras included $\phi$ production in pA collisions, the $\rho$ meson line-shape in cold nuclear matter, and Dalitz decay form factors of the $\eta$ and $\omega$ mesons, with detailed results published in~\cite{NA60:2009eih,NA60:2011zkq,NA60:2009una,NA60:2019tfy,NA60:2016nad}.


The scientific legacy of NA60 naturally evolved into the NA60+ proposal, which aims to perform precision measurements of thermal radiation and charm production at  high baryochemical potential  through a beam energy scan at the CERN SPS. Following the conclusion of NA60 operations, the conceptual framework for NA60+ was developed between 2010 and 2014. The collaboration expanded when Enrico Scomparin and other researchers joined the effort to transform this concept into a viable experimental program. Currently recommended by the CERN SPSC, the project anticipates beginning data collection in 2030.

\section{The Omega experiments and NA57: hyperon enhancements}
\label{SPS_WA97}
\vspace{-0.3cm}


The Omega Spectrometer, with its remarkable versatility (it had already been a workhorse for many experiments) and its high-field superconducting magnet, was an obvious choice for housing a suite of experiments focussing on the measurement of the production of multi-strange hyperons, whose production was predicted to be enhanced in a specific pattern, increasing with strangeness content in case of QGP formation ~\cite{Koch:1986ud}.
such studies.
The birth of Omega dates back to 1968~\cite{omega1968}. It
was conceived as an electronic ``bubble chamber'',
offering a large magnetic volume, which could be filled with spark chambers and operated under a variety of triggers and with a variety 
of incident beams. The spectrometer was later upgraded  (sometimes named as ``Omega Prime'') with 
Multi-Wire Proportional Chambers (MWPC)~\cite{Beusch:345021}, Cherenkov counters and large photon detectors. 

\begin{figure}[tbp]
    \centering
    \includegraphics[width=0.47\textwidth]{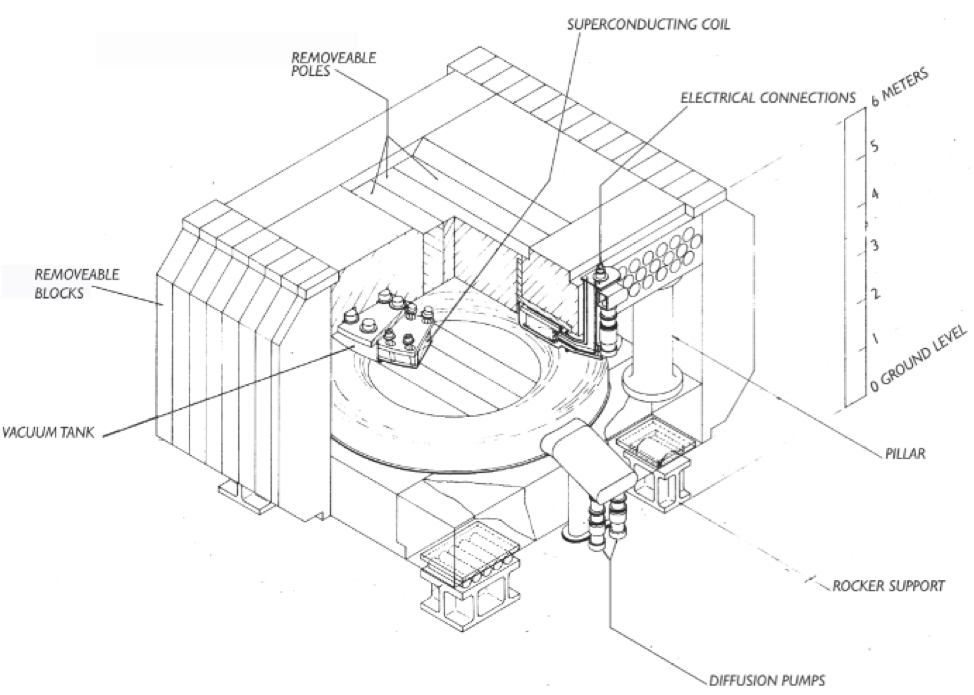}
    \includegraphics[width=0.35\textwidth]{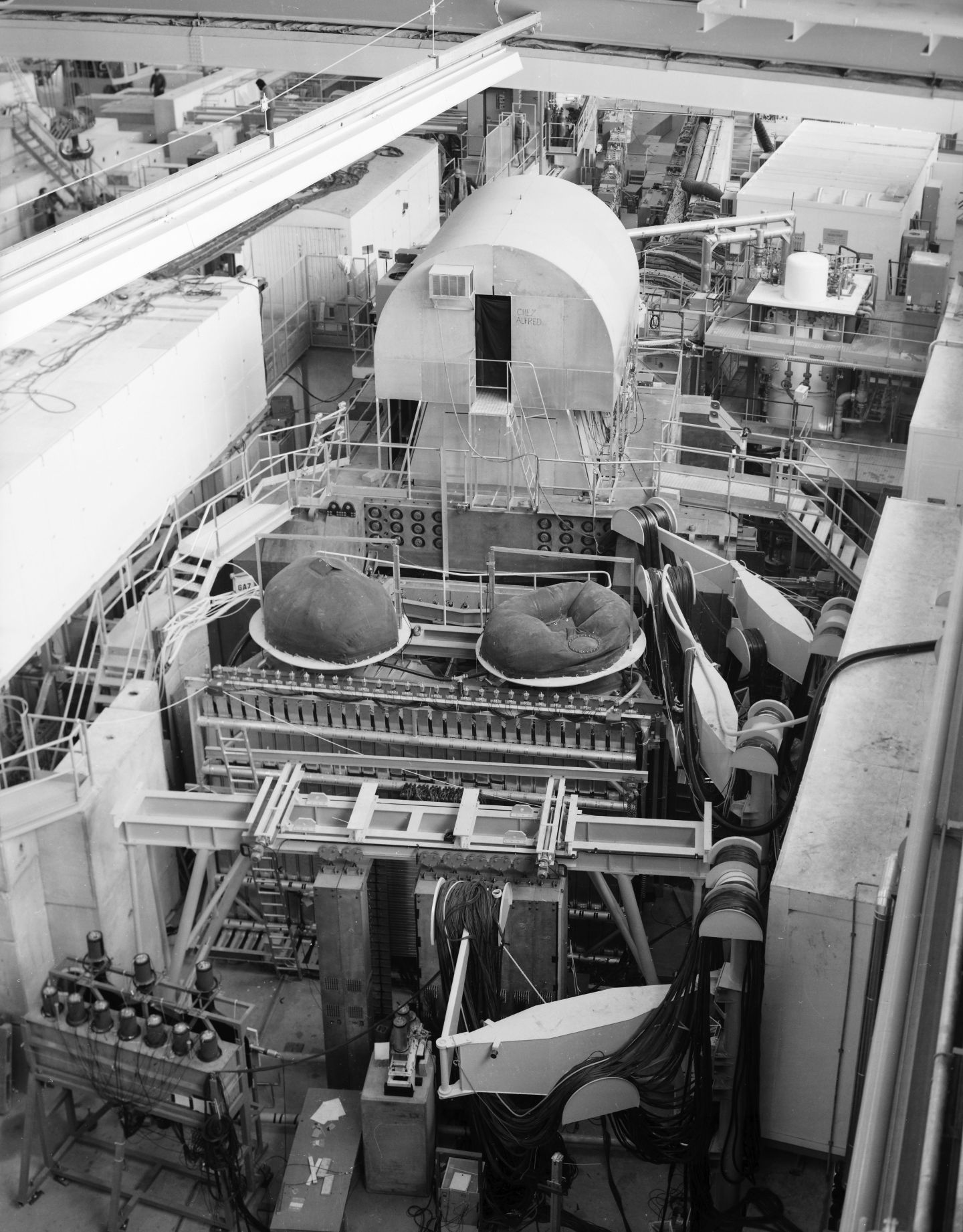}
    \caption{Sketch of the Omega spectrometer superconducting magnet and a photograph in the SPS West Hall~\cite{michelini1969omega,cern916772}.
   }
    \label{AQS:OmegaFig}
    \vspace{-3mm}
\end{figure}

Omega, located in the CERN West Area (WA), consisted of a large 1.8\,T superconducting magnet (Fig.~\ref{AQS:OmegaFig}) in which each approved experiment could install the detectors according to the need. The first ``nuclear'' experiment at Omega was WA72~\cite{WA72proposal}, proposed by the Warsaw group to study the production of fast protons and antiprotons in the interactions of 30\,GeV pions with various target nuclei. The use of nuclear targets had the specific purpose to study the differences in the $A$-behaviour of baryon and antibaryon yields.

\subsection{Experiments with sulphur beam: WA85 and WA94}
\label{AQS:WA85WA94}
\vspace{-0.3cm}
The first ion-beam experiment at the Omega Spectrometer, using a 200\AGeV SPS sulphur beam, 
was WA85 (1987-1990, Spokesperson: Emanuele Quercigh)
with the aim of investigating a possible strangeness enhancement in nucleus-nucleus 
collisions with respect to proton-nucleus,
by measuring the yields of strange baryons and antibaryons:
$\Lambda$, $\Xi^-$, and their antiparticles. 
A modification of the original Omega setup was necessary for high-energy ion collisions. This used modified MWPCs, reshaped with a butterfly geometry, a technique that had been developed by the WA77 collaboration~\cite{BEUSCH1986391}. 


In WA85 (Fig.~\ref{figures:SPS_2}), the beam was incident on a thin tungsten target (0.005 interaction lengths). Tracks were reconstructed using seven MWPCs, whose sensitive region had a butterfly shape,  
chosen so that charged particles from the target reach it through the magnetic field only for 
\pt $> 0.6$\,GeV/$c$, and laboratory rapidity $y_{\rm lab}$ within the interval $2.2 < y_{\rm lab} < 3.2$. The \pt threshold for particles coming from V$^0$ decays is somewhat lower, as they travel a shorter distance in the magnetic field. The two scintillator hodoscopes (HZ0 and HZ1) also have a butterfly shape and detect particles in the same \pt-rapidity range as the chambers. 
Multiplicity is obtained from the number of strips hit in the upper and lower microstrip arrays. A hadron calorimeter was placed 25\,m downstream of the target to veto non-central events.

    
    
\begin{figure}[tbp]
\includegraphics[width=0.47\textwidth]{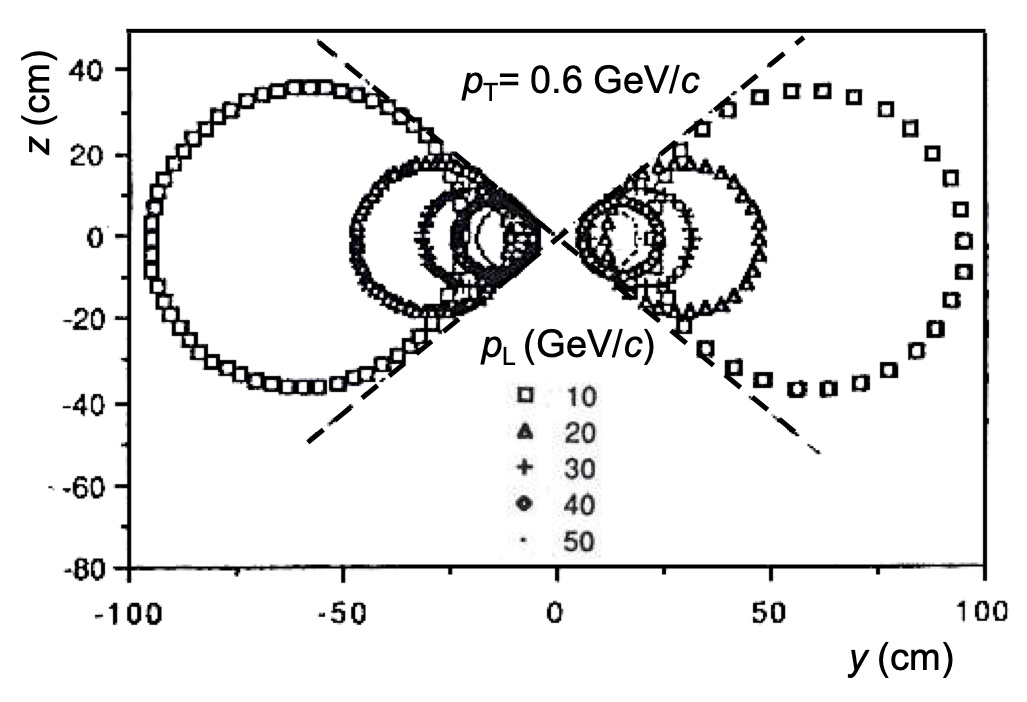} 
\includegraphics[width=0.47\textwidth]{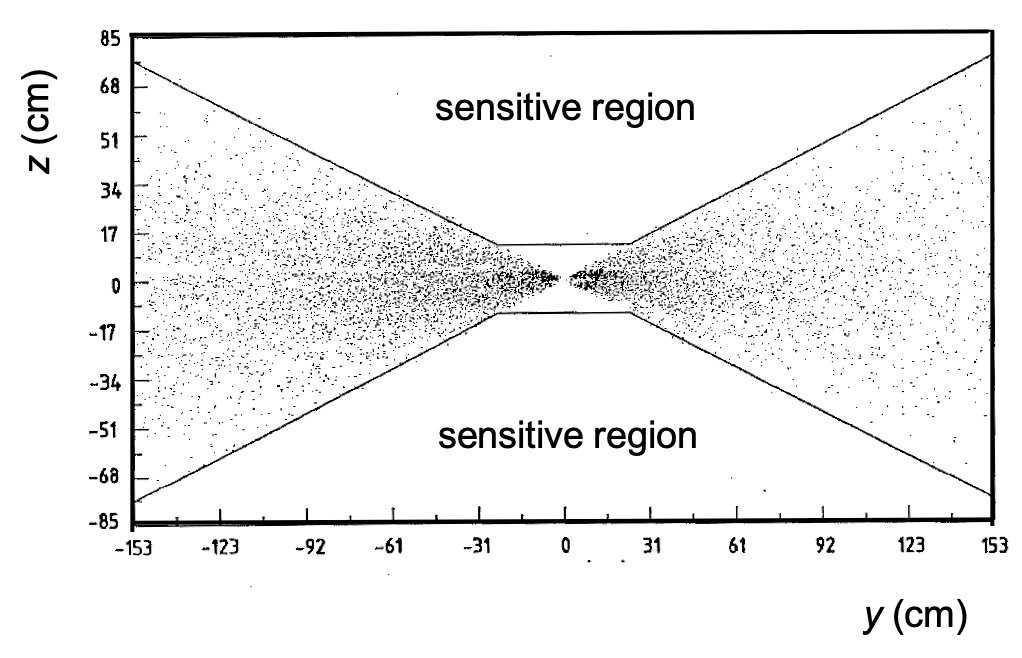}
    \caption{Butterfly principle~\cite{Abatzis:195599} in Omega:  
    ({\it{Left}}) Charged particle impact points with given \pt and \pl transported in the magnet from target to MWPC.
    ({\it{Right}}) 
        Hits of charged particles with \pt $> 0.6$\,GeV/$c$ showing the sensitive region.
        }
    \label{AQS:ButterflyFig}
    \vspace{-3mm}
\end{figure}
\begin{figure}[tbp]
    \centering
    \includegraphics[width=0.45\textwidth]{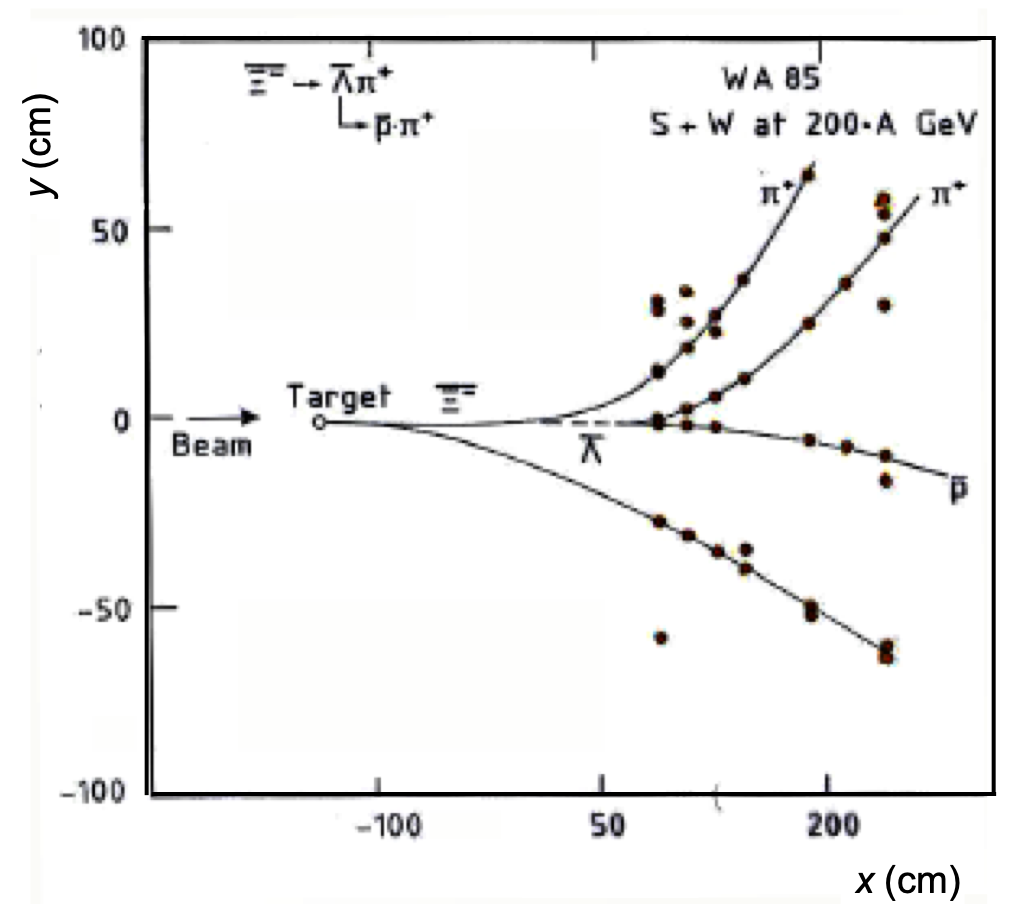}
    \vspace{-3mm}
    \caption{Decay of $\overline{\varXi}^{+}$ produced in central S-W collision at 200\AGeV.
    }
    \label{AQS:WA85Xi}
    \vspace{-3mm}
\end{figure}

The butterfly principle~\cite{Abatzis:195599} relies on how charged particles with the same \pt but varying longitudinal momentum (\pl) intersect a vertical detection plane in circular paths. In a magnetic field, these paths bend based on charge and momentum, with lower \pl causing more deflection. The combined effect forms a butterfly-shaped envelope, allowing only particles above a certain \pt to be detected in the active regions of the modified MWPCs. 
This is illustrated in Fig.~\ref{AQS:ButterflyFig}. Its effectiveness is demonstrated in Fig.~\ref{AQS:WA85Xi}, which shows a fully reconstructed decay of a $\overline{\Xi}^+$ produced in a central S-W collision at 200\AGeV, with very little background from other particles produced in the event, since most of their hits are located in the the inactive areas of the chambers. \\

\noindent WA85 was followed by the WA94 experiment~\cite{WA94Proposal} (Spokesperson: Emanuele Quercigh), which had two main objectives:
\begin{itemize}
\item{First run (1991): To measure strange and non-strange baryon and antibaryon spectra in S-S collisions.}
\item{Second run (1992): to test new detector elements as the first step in upgrading the tracking apparatus so as to cope with Pb-Pb collisions in the WA97 experiment.}
 \end{itemize}
During the WA94 second run, the main question was whether one could replace the existing array of butterfly MWPCs with a silicon pixel telescope. For this purpose, a cooperation 
with the RD19 collaboration at CERN~\cite{RD19Proposal} started,
which then became progressively intertwined with the WA94, WA97 and NA57 collaborations. 
 
\begin{figure}[tbp]
    \includegraphics[width=0.52\textwidth]{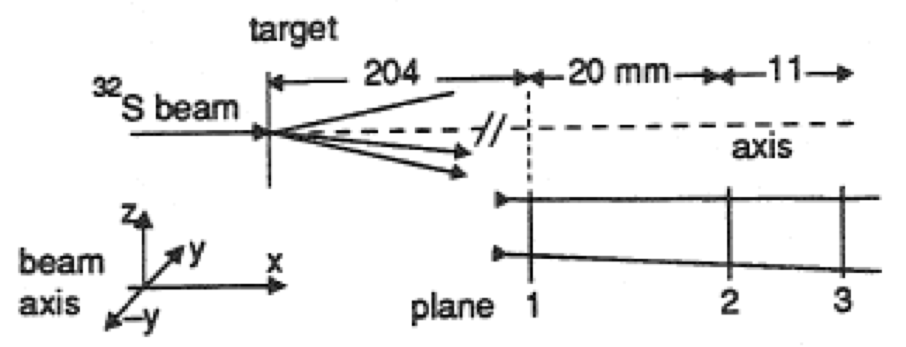} 
    \includegraphics[width=0.4\textwidth]{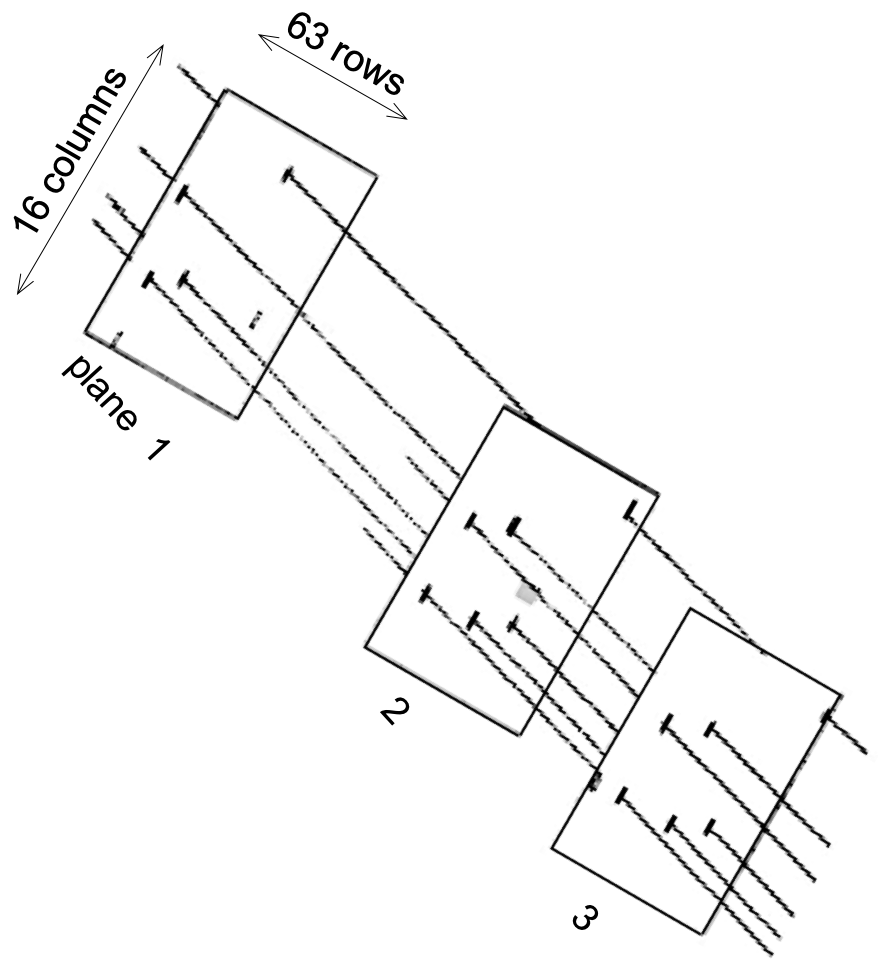}
    \caption{RD19-WA94 silicon-pixel test (1991): ({\it{Left}}) Side view with three Omega-Ion pixel chips below the beam axis. ({\it{Right}}) Hits in the three pixel chips showing six visible hand-drawn tracks from a high-multiplicity S-S event. 
    }
    \vspace{-3mm}
    \label{AQS:WA94pixeltest}
\end{figure}

In 1991, an array of silicon pixel detectors 
was successfully tested at CERN for the WA94 experiment - a first for particle physics~\cite{HEIJNE20011}. This array was made of three Omega-Ion  hybrid silicon pixel detectors~\cite{Cambell:240426}. Each detector plane consisted of 1006 active 
pixels ($500$\,$\umu$m\,$\times75$\,$\umu$m), bump-bonded to the readout chip, and arranged in 16 columns and 63 rows. 
The test setup and the hand-drawn tracks of an event are shown in Fig.~\ref{AQS:WA94pixeltest}.
It is clear that particles can be reliably tracked, even in such a hit density environment, with 
 virtually no spurious noise hits and easy pattern recognition. 
This test was an important step toward the use of pixel detectors in a heavy-ion experiment. 

\subsection{Lead beam and Silicon Pixels: the WA97 and NA57 experiments}
\begin{figure}[tbp]
 \centering   \includegraphics[width=0.7\textwidth]{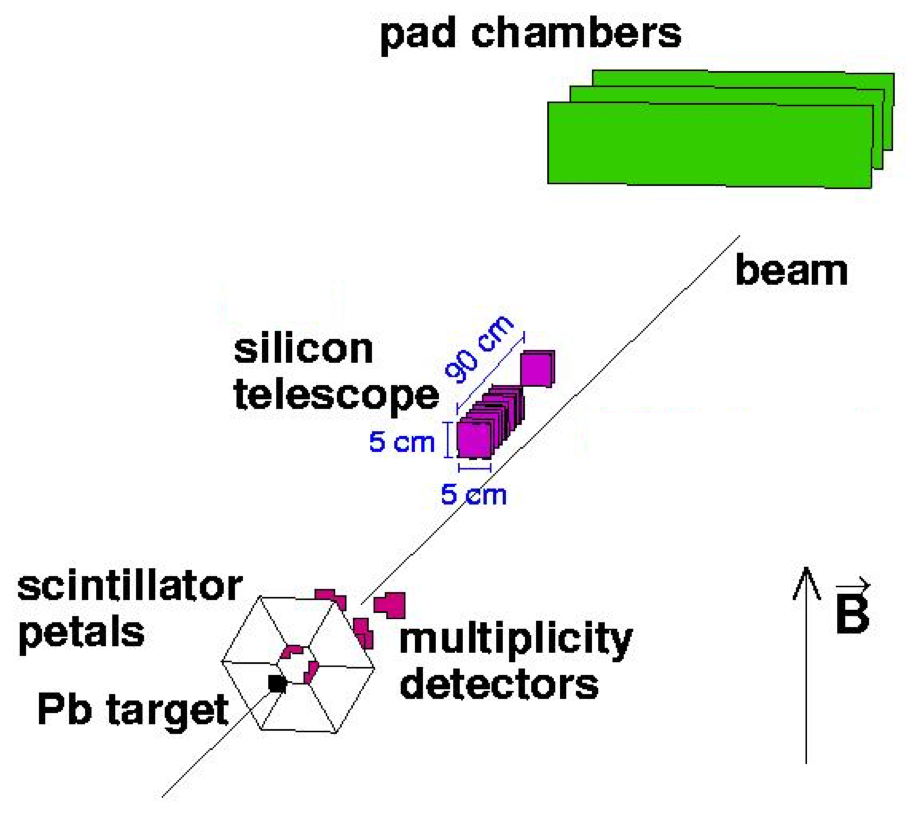}
    \caption{Layout of the WA97 experiment~\cite{WA97Proposal}
    }
    \label{AQS:WA97setup}
\end{figure}

\vspace{-0.3cm}
The layout of WA97 experiment (Spokesperson: Emanuele Quercigh) in Omega is shown in 
Fig.~\ref{AQS:WA97setup}. 
To enable larger area coverage, a second version of the Omega pixel readout chip (Omega2) was developed with modifications for integration into large systems~\cite{CAMPBELL199452}. In 1994, it was possible to construct the seven-layer, 0.5 million channel, silicon micro-pattern pixel detector telescope that allowed for electronically imaging the tracks on the lead-ion event in the WA97 experiment~\cite{WA97Proposal}.
  These new detectors could determine the space points on a track directly with a two-dimensional readout, thus avoiding the ambiguities generated from the intersections of wires or strips. In addition, because of their high granularity, they could be placed near the target, thus easing the study of the short-lived strange baryons.

The telescope, made from an array of silicon detector planes of $5\times 5$\,cm$^2$ cross section, was placed above the beam line, inclined and pointing to the target. The detectors were closely packed in a length of about 30 cm. This compact part was used for pattern recognition.
The seven pixel planes were interleaved with ten silicon microstrip planes of 50\,$\umu$m pitch, providing about half a million detecting elements.
The basic building block of the pixel telescope was the silicon detector ladder: a matrix of rectangular pixels, each one 
(size, $500 \times 75$\,$\umu$m$^2$) connected to a virtual ground via a front-end amplifier on a readout chip by a Pb-Sn solder bump. 

\begin{figure}[tbp]
\centering \includegraphics[width=0.7\textwidth]{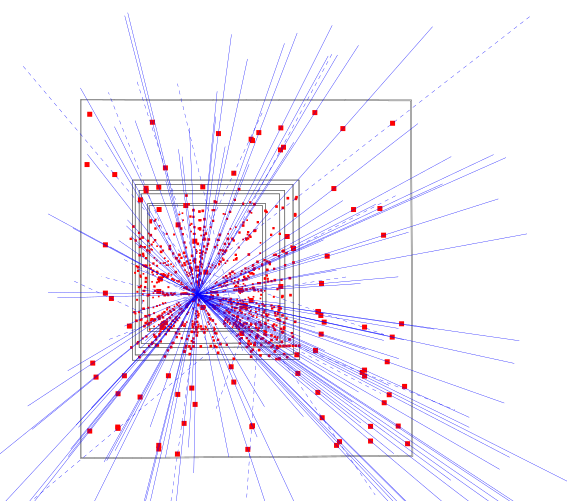}
    \caption{A fixed target Pb-Pb event reconstruction with 153 tracks, using a seven-plane pixel telescope in WA97 experiment. The dots represent hits in the silicon detectors, after which the tracks have been reconstructed in perspective view.}
    \label{AQS:WA97event}
\end{figure}

\begin{figure}[tbp]
 \centering   
\includegraphics[width=0.6\textwidth,height=0.4\textwidth]{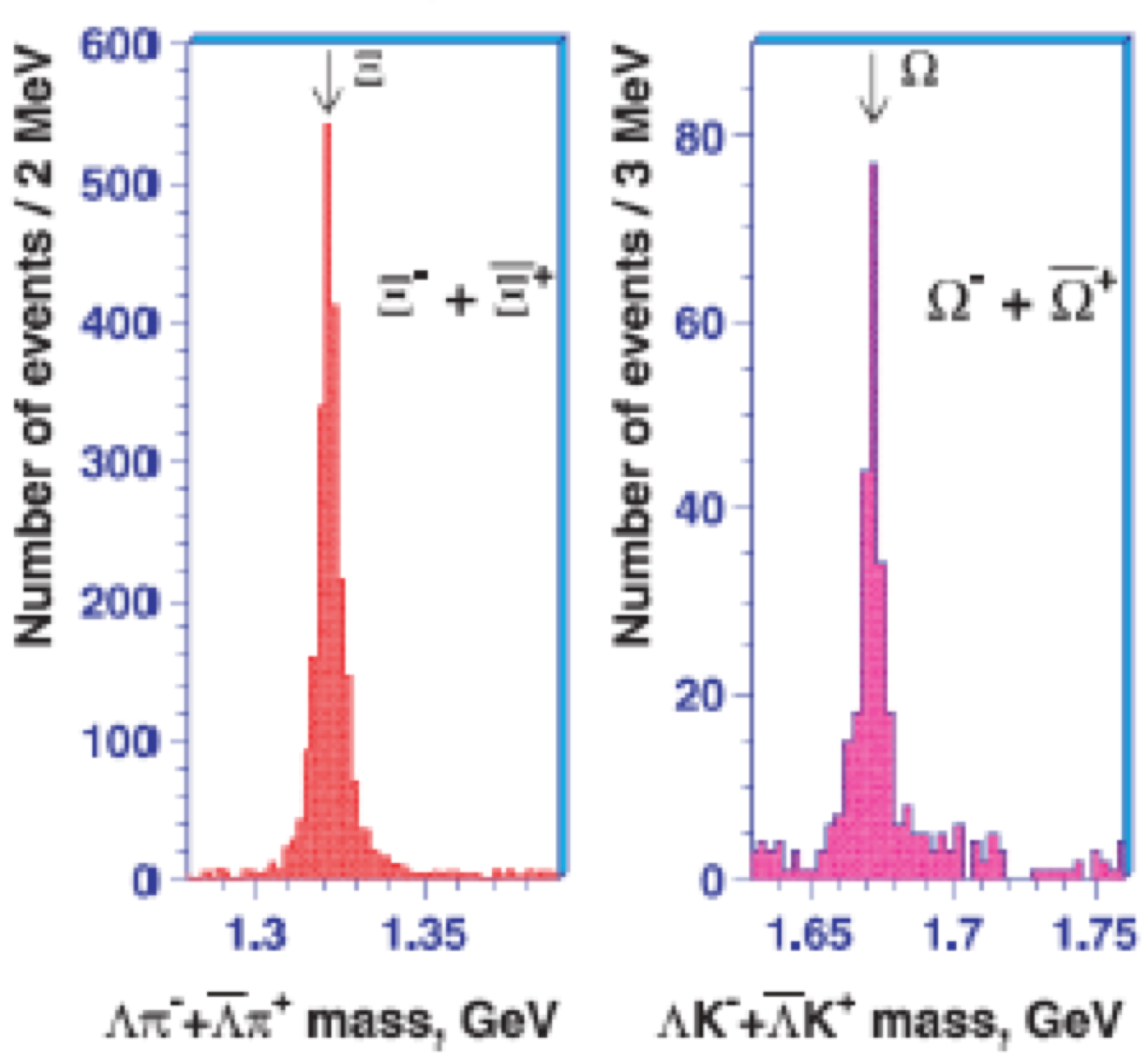}
    \caption{
    Invariant mass distributions for $\Xi$ and $\Omega$ hyperon candidates.
    }
    \label{AQS:WA97setup_omega}
\end{figure} 


\begin{figure}[btp]
\centering    \includegraphics[width=0.7\textwidth]{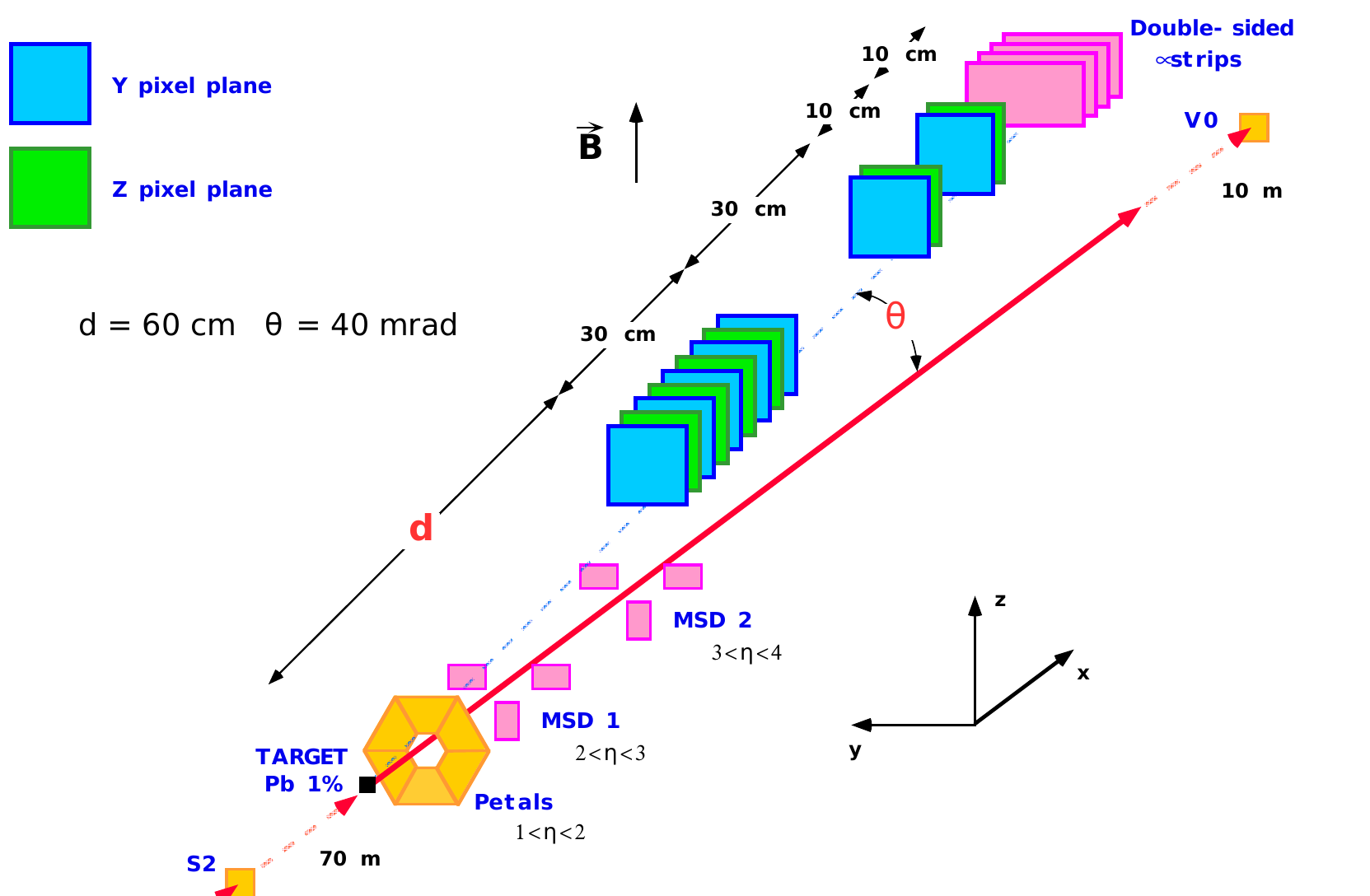}
    \caption{Schematic view of the NA57 setup in the Goliath magnet~\cite{NA57:2006aux}. 
    }
    \vspace{-3mm}
    \label{AQS:NA57layout}
\end{figure} 

Fig.~\ref{AQS:WA97event} shows the reconstructed tracks using the silicon detectors in lead ion collision at the SPS. Silicon Pixel Detectors proved to be particularly suitable for tracking in a high-multiplicity environment. 
Hyperon identification was remarkable, even in the most central Pb-Pb collisions, as illustrated in
Fig.~\ref{AQS:WA97setup_omega}. 

WA97 was the first experiment 
in high-energy physics which made use of this new technique. Later, the experience gained proved to be invaluable in the design of the LHC inner tracking detectors: the ALICE SPD sensors, in particular, employed during RUN 1 and Run 2 of the LHC, were the direct descendants of the pixel sensors of WA97 and NA57.

The Omega facility was closed at the end of 1996. This was a necessary sacrifice to free resources for LHC construction. As the programme was in full swing, it it had to be quickly relocated elsewhere.
 It was proposed to move the Goliath magnet from the West to the North Area and install it along with the WA97 detectors~\cite{NA57Proposal}, marking the beginning of the NA57 experiment (Spokesperson: Federico Antinori).

The NA57 layout (Fig.~\ref{AQS:NA57layout}) was conceptually similar to that of WA97.
While in WA97 the telescope consisted of a combination of silicon pixel and silicon microstrip detectors, in 
NA57, it was entirely made of pixel detectors for a total of about one million channel
(each pixel, 50×500 $\mu$m$^2$).
NA57 extended the physics scope of WA97 by extending centrality range and by also taking data at a lower beam energy (40\AGeV/$c$) in order to study the energy dependence of the hyperon enhancements. 


\subsection{Key findings of the Omega experiments}
\vspace{-0.3cm}

The main results of the Omega experiments are the observation of a significant increase in the production of strange particles, and especially of multi-strange an- tibaryons, in heavy-ion collisions.

One important quantitative prediction, made in 
1980-82~\cite{Rafelski:1980rk,Rafelski:1982pu}, is that the formation of QGP leads to equilibration of strange quarks on a time scale of a few fm/$c$, and thereby to enhancement of strangeness production. Moreover, multistrange baryons were predicted to be formed close to thermal and chemical hadronic equilibrium, so that the enhancement, crucially, was predicted to increase with the strangeness content of the hyperon~\cite{Koch:1986ud}. 
In order to look for such an effect, it was necessary to measure the production of baryons and antibaryons carrying one, two and three units of strangeness.

At the Quark Matter 1990 conference in Menton, WA85 reported the first ever observation of multi-strange baryons and antibaryons ($\Xi^-$ and $\overline{\Xi}^+$) in ultra-relativistic heavy-ion collisions~\cite{ABATZIS1991441}.
   
It was shown that $\Xi^-$ and $\overline{\Xi}^+$ decays can be successfully reconstructed 
with little background (see Fig.~\ref{AQS:WA85Xi} on page \pageref{AQS:WA85Xi}). 
The fact that one could observe such rare hadrons among the hundreds of particles produced on average in each collision, was one of the reasons for the enthusiasm  gathered then for this new research domain.
Then, the following year WA85 \textcolor{green}{(Physics Letters B 270 ( 1991 ) 123)}
reported the indication of an enhancement in the production of $\overline{\Xi}^+$ when going from p-W to S-W collisions, based on the evolution of the $\overline{\Xi}^+/\overline{\Lambda}$ ratio. 
A couple of years later, WA85 managed to also extract an ${\Omega}^-$ signal \textcolor{green}{Physics Letters B 316 (1993) 615–619}, and to obtain indications of enhancement also for the ${\Omega}^-$, $|s| = 3$ baryon.

The WA94 experiment then obtained results similar to those of WA85 on the $\overline{\Xi}^+$ enhancement~\cite{Abatzis1995}.


 


\begin{figure}[tbp]
\centering
\includegraphics[width=0.85\textwidth]{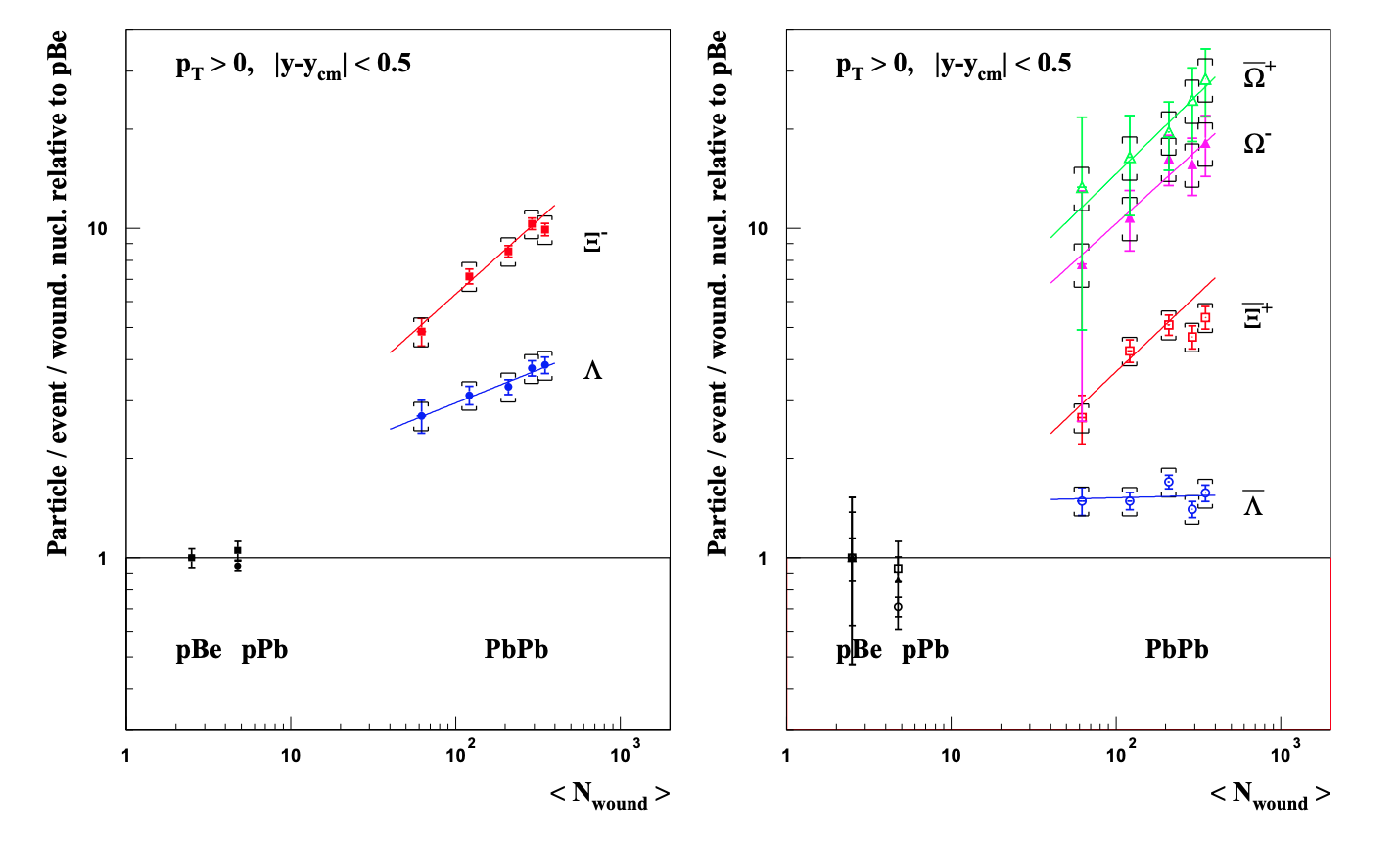}
\caption{
Hyperon enhancements as a function of the number of wounded nucleons with corresponding power-law fits (NA57)~\cite{NA57:2006aux}.
}
\label{AQS:NA57_hyperon_enhancement}
\vspace{-3mm}
\end{figure}

These first indications from the sulphur beam experiments turned into spectacular evidence with the high-precision data from Pb-Pb collisions with the silicon pixel trackers of WA97 and NA57. The legacy data from NA57 are shown in figure ~\ref{AQS:NA57_hyperon_enhancement}, where the hyperon enhancements with respect to p-Be collisions are plotted as a function of collision centrality.
The enhancements 
show a clear hierarchy, increasing with the strangeness content of the particle, for both hyperons and antihyperons, up to a value of about 20 for the triply-strange $\Omega^-$ baryon for the most central collisions, as predicted for statistical hadronisation from the QGP. These results constituted one of the main pieces of evidence for the formation of a new state of matter at CERN-SPS energies.

\section{Experiments WA80 to WA93 to WA98 - direct photons}
\label{SPS_WA98}
\vspace{-0.3cm}

Initiated by Hans Gutbrod in the mid-1980s, the WA80 experiment~\cite{CERN-SPSC-88-22}
was primarily designed for the detction of direct photons, which are messengers from the early, hot stages of nuclear collisions using oxygen and sulphur beams on heavy targets (such as gold). The experimental setup is outlined in 
Fig.~\ref{figures:SPS_2}. It measured the distribution of charged particles in a large fraction of phase space, to analyse the forward and transverse energy distributions and to investigate the target fragmentation region. Photons and neutral pions were identified near midrapidity in a high resolution lead glass calorimeter. The centrality selections were determined either by measurements of the remaining projectile energy in the forward direction by the Zero degree calorimeter (ZDC) or the charged particle multiplicity.  
For central S-Au collisions at 200\AGeV, the measured photon excess~\cite{WA80:1995xza} corresponded to 5.0\% of the total inclusive photon yield with a statistical error of 0.8\% and a systematic error of 5.8\%, yielding an upper limit on the invariant yield for direct photon production at the 90\% C.L. 

Although no definitive photon excess was observed, the WA80 experiment laid crucial groundwork in detector design and analysis techniques. Building on this, the WA93 
experiment~\cite{CERN-SPSC-90-032} (spokesperson: Hans Gutbrod) refined the photon detectors and analysis methods, pushing sensitivity further~\cite{Srivastava:1994wk}. Still, a statistically significant photon signal remained out of reach. The WA93 experiment using the India made photon multiplicity detector (PMD) discovered the azimuthal anisotropy of inclusive photons, providing early evidence of collective flow in heavy-ion collisions.

The leap into a precision measurement of direct photons came with the WA98 experiment~\cite{Angelis:295040}. 
With major detector upgrades, WA98 delivered high-precision measurements in Pb-Pb collisions. It provided the first compelling evidence of direct photon production at low \pt, hinting at thermal radiation from a quark–gluon plasma, while also offering detailed insight into hadron production,  flow and disoriented chiral condensates. In 1997, Hans Gutbrod moved to Nantes, and Terry Awes took over as the WA98 spokesperson.

\subsection{WA98 experimental setup}
\begin{figure}[tbp]
\begin{center}
\includegraphics[width=0.8\textwidth]{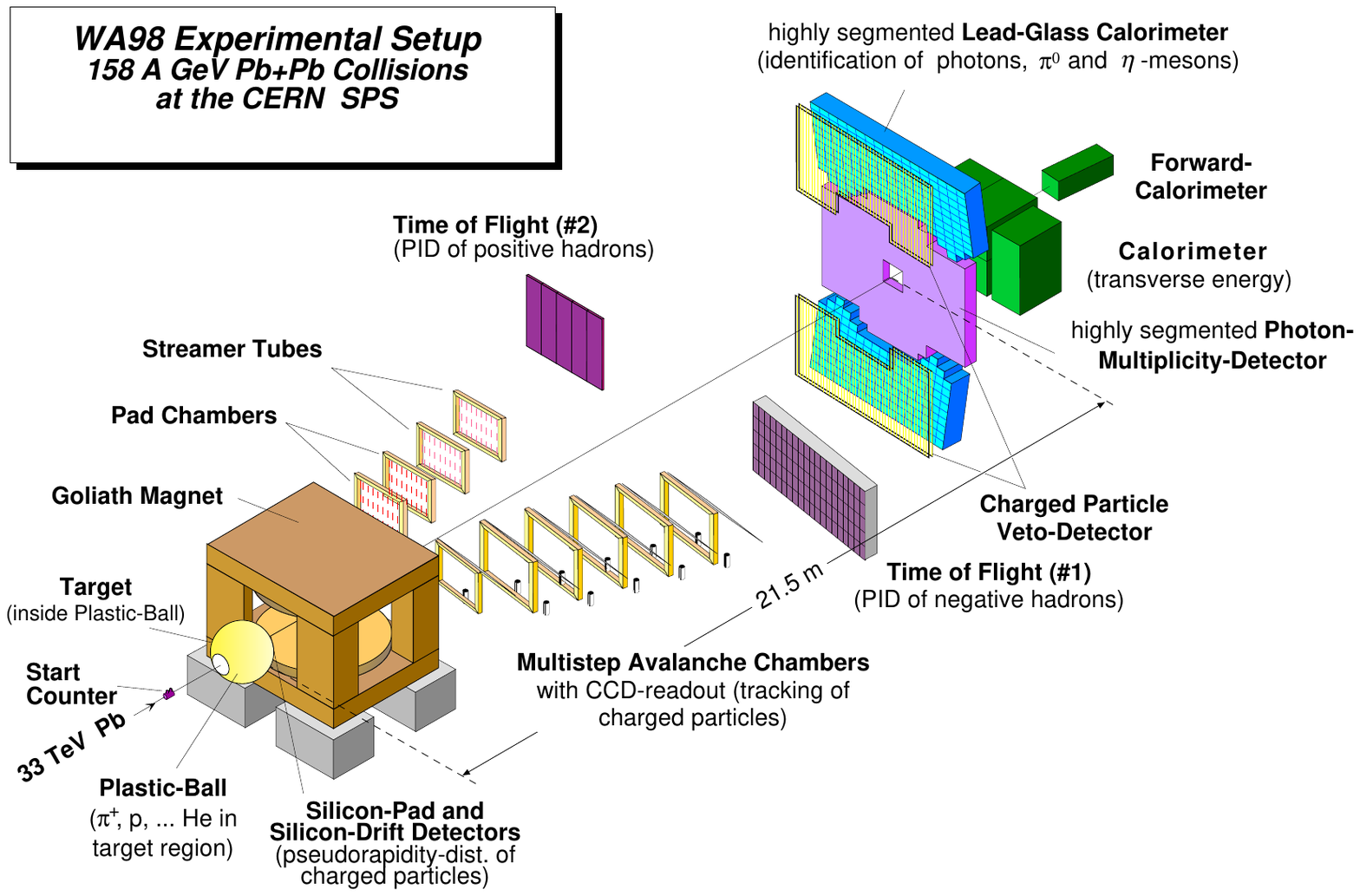}
\end{center}
\caption{The WA98 experimental setup. See text for details.}
\vspace{-6mm}
\label{fig:WA98_1}
\end{figure}

\vspace{-0.3cm}
The WA98 experiment combined photon and hadron spectrometers with several other large acceptance detectors that allow for measuring various global variables on an event-by-event basis. The experiment took data with the lead beams from SPS in 1994, 1995, and 1996. The layout of the WA98 experiment as it existed during the final WA98 run period in 1996 is shown in Fig.~\ref{fig:WA98_1}. 

The lead target was mounted in a thin target wheel at the center of a 20~cm diameter spherical, thin-walled aluminum vacuum chamber located within the Plastic Ball detector. The Plastic Ball consisted of 655 modules, which provided energy measurement and particle identification of charged pions and light particles through $\Delta E$-$E$ measurements. 
The Plastic Ball covered the interval $-1.7 < \eta < 1.3$.

The target vacuum chamber was extended downstream in a $30\deg$ conical vacuum chamber which contained the Silicon Drift Detector (SDD) and the Silicon Pad Multiplicity Detector (SPMD). These detectors provided charged particle multiplicity measurement over the intervals of $2.5 < \eta < 3.75$ and $2.3 < \eta < 3.75$, respectively.
Charged particle momentum measurement and particle identification were accomplished using two tracking spectrometer arms and a large (1.6 m) dipole magnet called Goliath, which provided 1.6 Tm bending power. Particle identification was obtained using time-of-flight measured with scintillator slat detectors in each tracking arm. 

The PMD, located at a distance of 21.5~m from the target, was a large (21 m$^2$) preshower detector consisting of 3 radiation lengths of lead converter backed by over 50000 scintillator tiles, readout via wavelength shifting optical fibers coupled to CCD cameras with image intensifiers. The PMD provided photon multiplicity measurement within $2.8 < \eta < 4.4$. 

The photon spectrometer comprised of the Lead-glass photon Detector Array, and a charged particle veto detector. It provided photon energy measurement over the interval $2.4 < \eta < 3.0$. At 24.7 m downstream, the MIRAC calorimeter provided total transverse energy measurement with varying azimuthal coverage over the interval $3.2 < \eta < 5.4$. 
Finally, the total energy of the non-interacting beam, or of the residual beam fragments and produced particles emitted near to zero degrees, was measured in the Zero Degree Calorimeter (ZDC). 

\subsection{Key findings of WA80, WA93 and WA98} 
\vspace{-0.3cm}
The WA98 experiment provided some of the earliest experimental data on charged particles, inclusive and direct photons,
suggesting the possible creation of QGP in the laboratory. Some of the key findings are:

\begin{itemize}
\item{\bf Observation of Direct Photons:}
Direct thermal photons emitted from the early stage of the QGP provide one of the most
important signals of phase transition~\cite{Kapusta:1991qp}. Since the photon emission in ultra-relativistic nuclear collisions is dominated by the background from decay photons, the measurement of thermal photons becomes very difficult. 
In the WA98 experiment, major emphasis was placed on the measurement of photons and neutral mesons (particularly the neutral pion). The neutral pions are primarily produced via hadronic interactions and decay rapidly into photon pairs. Thus, by measuring the inclusive photon spectrum and isolating the decay photons, WA98 was able to extract the spectrum of direct photons - photons not originating from hadron decays but emitted directly from the early stages of the collision. This is shown in  
Fig.~\ref{fig:WA98_2}({\it{feft}}).

\begin{figure}[tbp]
\begin{center}
\includegraphics[width=0.45\textwidth,height=0.45\textwidth]{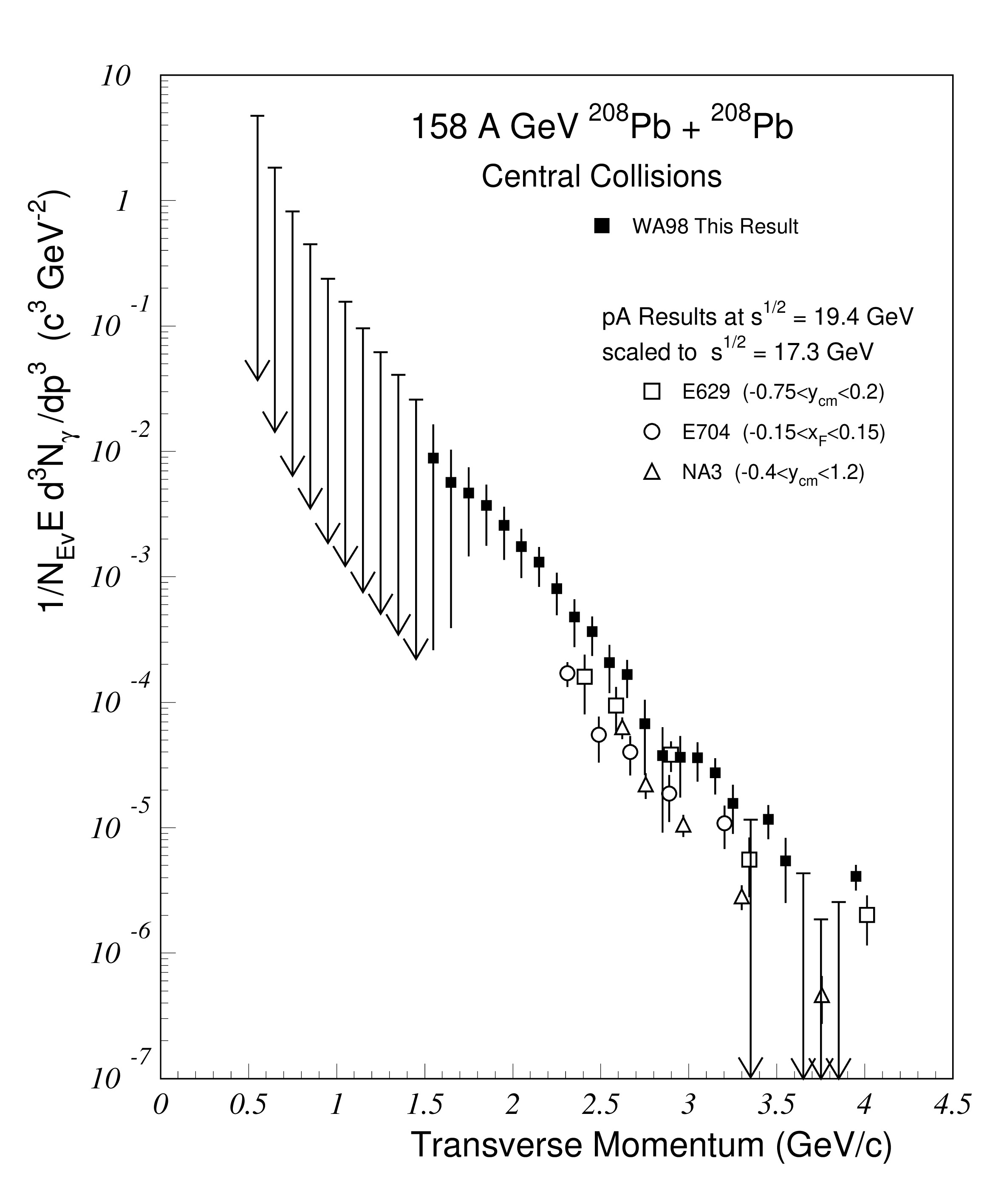}
\includegraphics[width=0.4\textwidth,height=0.43\textwidth]{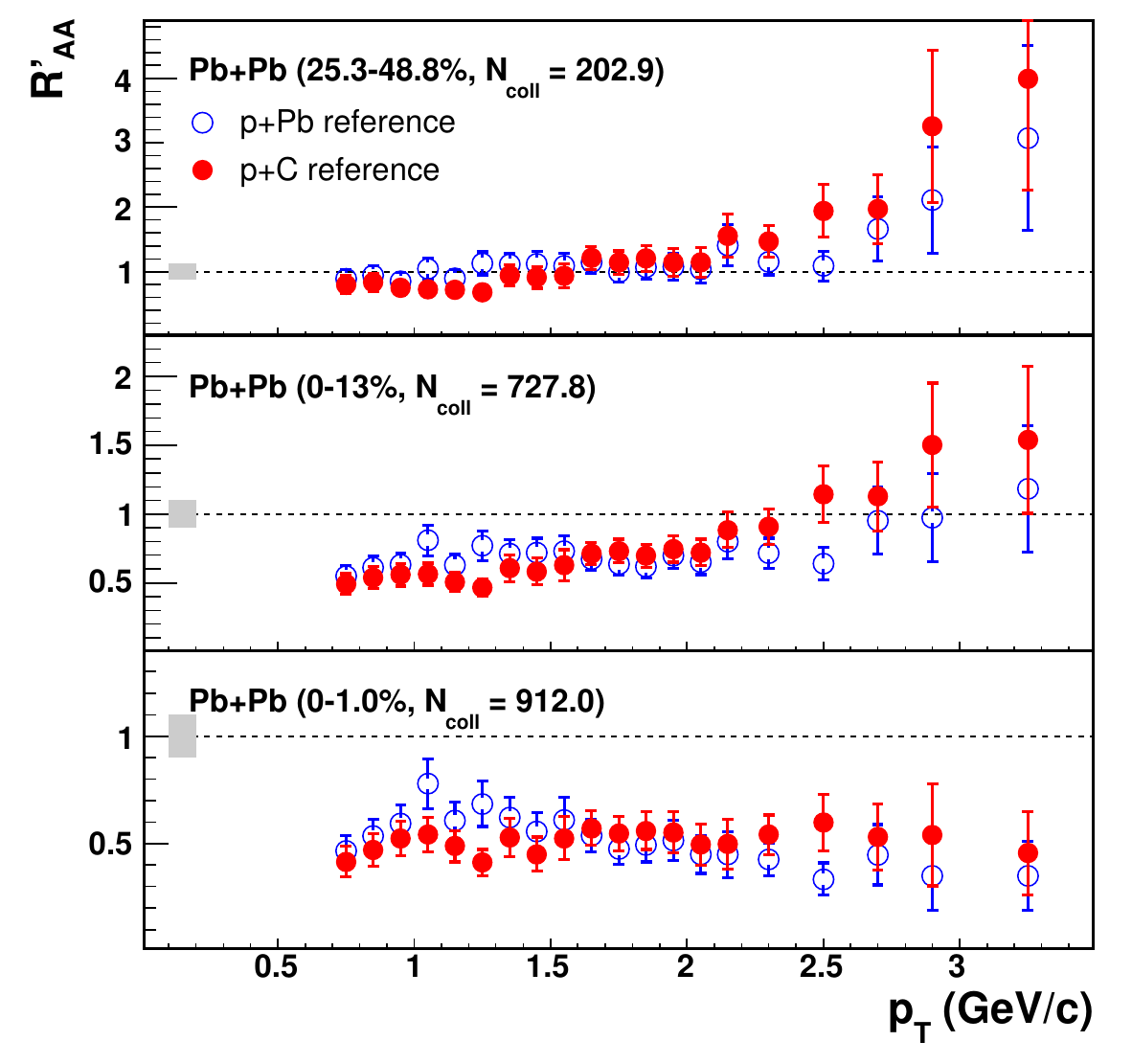}
\end{center}
\caption{(Left:) The invariant 
direct photon transverse momentum spectrum for central 
Pb-Pb collisions at SPS~\cite{WA98:2000vxl}. Results of several direct photon measurements for proton-induced reactions have been scaled to central 
collisions for comparison.
(Right:) Nuclear modification factor of neutral pions in 158\AGeV Pb on Pb collisions 
for three centrality classes 
using p-C or p-Pb spectra as
a reference. The pion suppression seen here constrain jet-quenching models~\cite{WA98:2007qib}.
}
\label{fig:WA98_2}
\vspace{-3mm}
\end{figure}

The detection of a direct photon excess above the expected background from neutral meson ($\pi^0$ and $\eta$) decays was one of the key results of WA98 experiment. The detection of the direct photons provided crucial evidence that high temperatures (up to ~200 MeV) were reached in the collisions - consistent with the formation of the QGP state of matter~\cite{WA98:2000vxl}.

\vspace{0.2cm}
\item{\bf Photon and Neutral Pion Production:}
The WA98 experiment studied neutral pion and photon spectra extensively.  In central Pb-Pb collisions at 158\AGeV, the WA98 experiment observed a suppression of high \pt $\pi^0$ mesons compared to yields expected from scaled \textit{pp} collisions, shown in  
Fig.~\ref{fig:WA98_2}~({\it{right}}). 


The observed suppression deviates from the enhancement typically associated with the Cronin effect, where multiple initial-state scatterings in proton-nucleus collisions lead to an increased yield of high \pt particles. In peripheral and semi-central Pb-Pb collisions, WA98 data exhibited this expected enhancement. However, in the most central collisions, a significant suppression was observed, suggesting that high-energy partons lose energy while traversing the dense medium created in such collisions. This is known as jet quenching. This suppression is considered a key QGP signature, indicating that the medium created in central heavy-ion collisions is dense and strongly interacting enough to significantly modify parton propagation~\cite{WA98:2007qib}.

\vspace{0.2cm}
\item{\bf Disoriented Chiral Condensates (DCC):}
 Formation of DCC domains have been predicted to form in high energy heavy-ion collisions when
the chiral symmetry is restored at 
high temperatures~\cite{Bjorken:1997re}. Anomalous event-by-event
fluctuations of the neutral to charged pions as well as neutral to charged kaons have been
predicted as signatures of the formation of DCC. 
In the WA98 experiment, fluctuations of photons to charged particle multiplicities were studied on an event-by-event basis.
The expectation was that DCC formation would produce non-statistical fluctuations, with neutral pion fractions deviating significantly from the usual 1/3 value predicted by isospin symmetry. Despite a thorough analysis, no significant evidence of DCC formation was observed. The fluctuations in neutral-to-charged pion ratios were found to be consistent with expectations from known statistical and detector-related sources. As a result, upper limits were placed on the possible DCC domain sizes and the occurrences 
of DCC~\cite{WA98:1997nnc}.
\vspace{0.2cm}
\item{\bf Interferometry of Direct Photons:}
The WA98 collaboration performed the first successful measurement of two-photon correlations providing direct evidence for direct photon emission. The obtained source size of about 6 fm is similar to that of hadrons, suggesting that photons were emitted from a large, thermalized regions~\cite{WA98:2003ukc}. This study demonstrated that photon interferometry is a powerful tool to probe the early, hot phases of heavy-ion collisions.
\end{itemize}


\section{Experiment NA61/SHINE - 
two-dimensional scan}
\label{SPS_NA61}
\vspace{-0.3cm}
NA61/SHINE is the third and most recent experiment in the family of heavy-ion experiments located at the H2 beamline of the CERN North Area at SPS. The experiment was approved by the CERN Research Board in 2007. Measurements for strong interaction physics are conducted in parallel with reference measurements for neutrino and cosmic-ray physics. The first phase of the strong interaction programme includes traditional heavy-ion physics, motivated by the results of NA35 and NA49.

The experiment conducted the world's first systematic scan in the two-dimensional space of laboratory-controlled parameters: the mass of colliding nuclei and collision energy. Results from \textit{pp} and Be-Be collisions indicate a transition from hadron production dominated by resonances to that dominated by strings at $\sqrt{s_{NN}} \approx 10$~GeV. Moreover, results at the top SPS energy reveal a transition from string-dominated to QGP-dominated production between Be-Be and Ar-Sc collisions. These findings, combined with results from RHIC and LHC, enable the establishment of the diagram of high-energy nuclear collisions. 
See Sec. \Ref{SPS_NA61_results} for details.

The need for a two-dimensional scan in collision energy and the mass of colliding nuclei at the CERN SPS was presented by Marek Gazdzicki during the NA49 Collaboration meeting in March 2003. The idea was based on results from the NA49 energy scan with Pb-Pb collisions, which revealed a markedly different collision energy dependence observed for \textit{pp} interactions and central Pb-Pb collisions; see 
Fig.~\ref{fig:na49_horn} (\textit{Left}). Importantly, the dependence for Pb-Pb collisions aligns with the predictions for the onset of deconfinement at low SPS energies~\cite{Gazdzicki:1998vd}. 

The proposal for these new measurements was supported by the NA49 spokesperson, Peter Seyboth, and a group of NA49 senior members, including Zoltan Fodor (Budapest), Wojtek Dominik (Warsaw), Rainer Renford (Frankfurt am Main), Herbert Stroebele (Frankfurt am Main), and George Vesztergombi (Budapest).

At the Research Board meeting in June 2003, Agnieszka Zalewska, the Board Chair, remarked that a detailed study of the low-energy region would be challenging at RHIC and encouraged the preparation of ideas for further studies of heavy ions at the CERN SPS beyond 2005. 
The Expression of Interest, ``A New Experimental Programme with Nuclei and Proton Beams at the CERN SPS," was submitted to the SPSC in November 2003~\cite{Gazdzicki:685283}, followed by Letter of Intent in January 2006~\cite{Gazdzicki:919966}. 

Finally, the proposal, ``Study Hadron Production in Hadron-Proton Interactions and nucleus-nucleus Collisions at the CERN SPS," was submitted to the SPSC in November 2006~\cite{Gazdzicki:995681}. It was approved by the CERN Research Board in February 2007. Marek Gazdzicki served as the spokesperson from 2007 until December 2024, when he was succeeded by Seweryn Kowalski and Eric Zimmerman.


The scan programme received significant support from the theory community. Notably, Frank Wilczek, Edward Shuryak, Krishna Rajagopal, and Misha Stephanov sent letters and emails of endorsement. For further details, see the NA61/SHINE website~\cite{NA61_history}.

\begin{figure}[tbp]
\begin{center}
\includegraphics[width=0.8\textwidth]{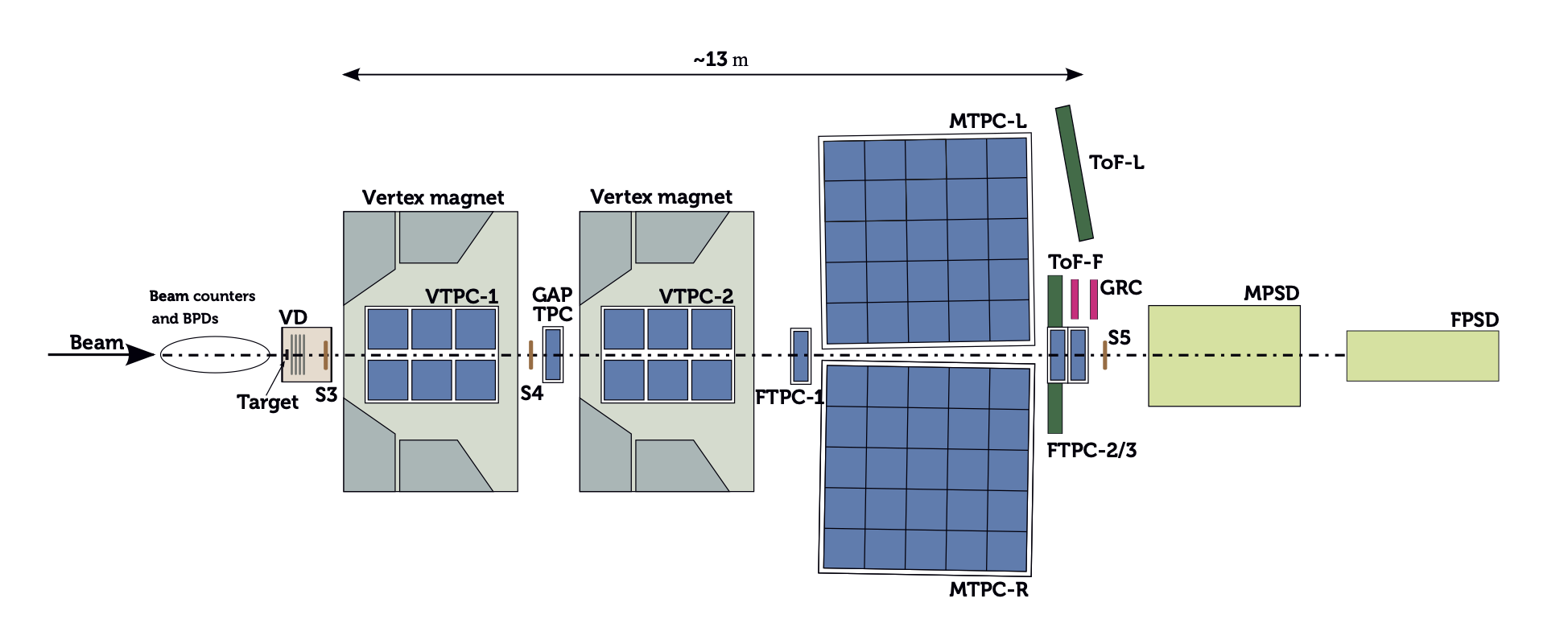}
\includegraphics[width=0.7\textwidth]{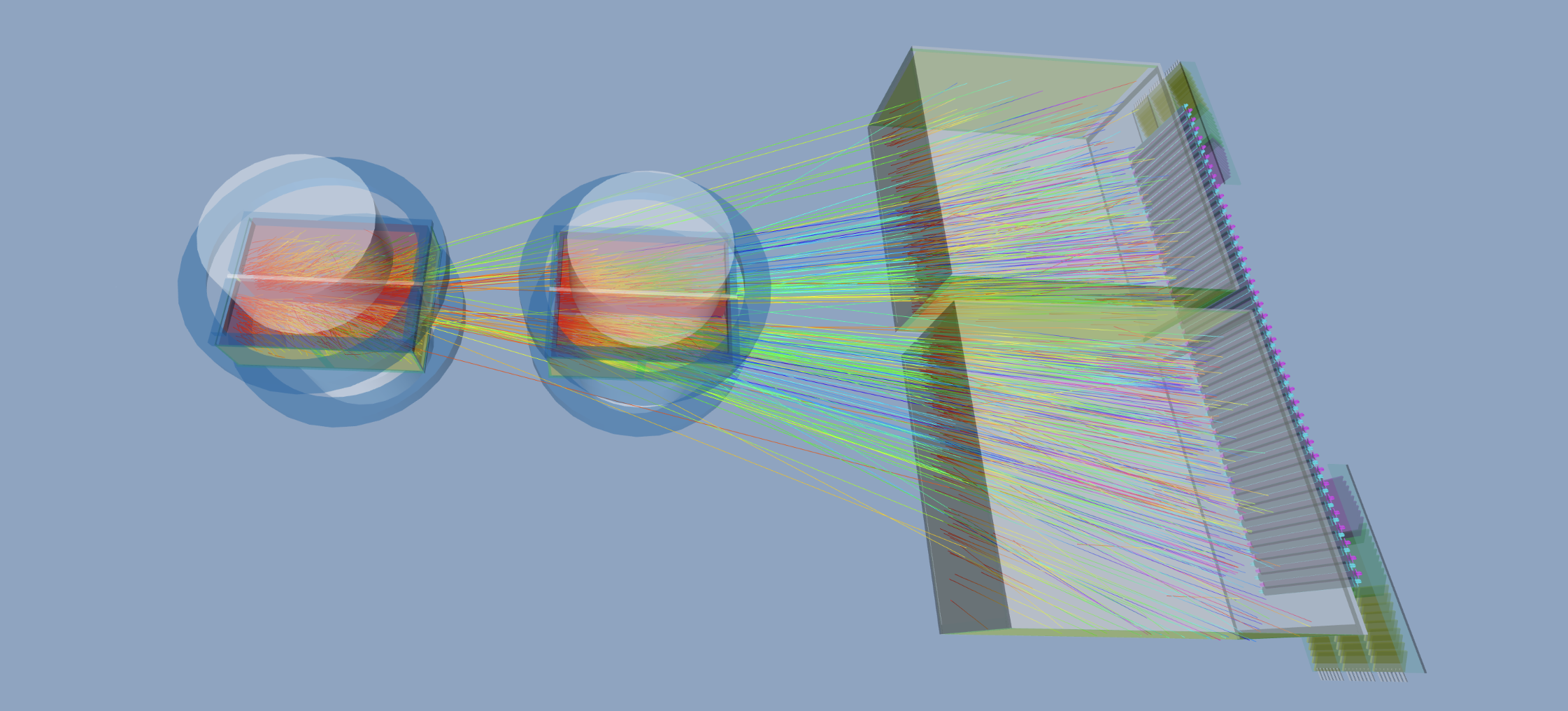}
\end{center}
\caption{
\textit{Top:} The post-LS2 NA61/SHINE detector~\cite{vonDoetinchem:2914265}.  
\textit{Bottom:} Particle tracks produced in Xe-La collision, recorded by the NA61/SHINE TPC. See text for details.
}
\label{fig:na61_postLS2}      
\end{figure}

\subsection{NA61 experimental setup}
\label{subsec:na61_exp}
\vspace{-0.3cm}

The NA49 collaboration transferred ownership of its detector to NA61/SHINE. Following a successful test in 2006, pilot physics data-taking for the long-baseline neutrino experiment, T2K at J-PARC, began in 2007~\cite{NA61SHINE:2011dsu}. Since then, the detector has undergone continuous upgrades, with major changes made during accelerator shutdowns.

Data collection for the 2D scan programme was conducted from 2009 to 2017, utilising an evolving detector setup resembling the one presented in Ref.~\cite{NA61:2014lfx}. The present setup~\cite{vonDoetinchem:2914265} is shown in Fig.~\ref{fig:na61_postLS2}~(\textit{top}). The main components of the tracking system in the large-acceptance hadron spectrometer include a silicon pixel detector (VD) located downstream of the target, followed by four large-volume TPCs. Two Vertex TPCs (VTPC-1/2) are positioned inside superconducting magnets with a maximum combined bending power of 9~Tm. The magnetic field was scaled proportionately to the beam momentum to maintain similar 
rapidity-\pt acceptance at all beam momenta. The main TPCs (MTPC-L/R) and Time-of-Flight (ToF) detectors are downstream of the magnets. The setup is completed by two forward calorimeters, the Projectile Spectator Detectors (MPSD and FPSD). An event display of particle tracks in a Xe-La collision recorded by the TPCs is presented in Fig.~\ref{fig:na61_postLS2}~(\textit{bottom}).

\begin{figure}[tbp]
\begin{center}
\includegraphics[width=0.44\textwidth]{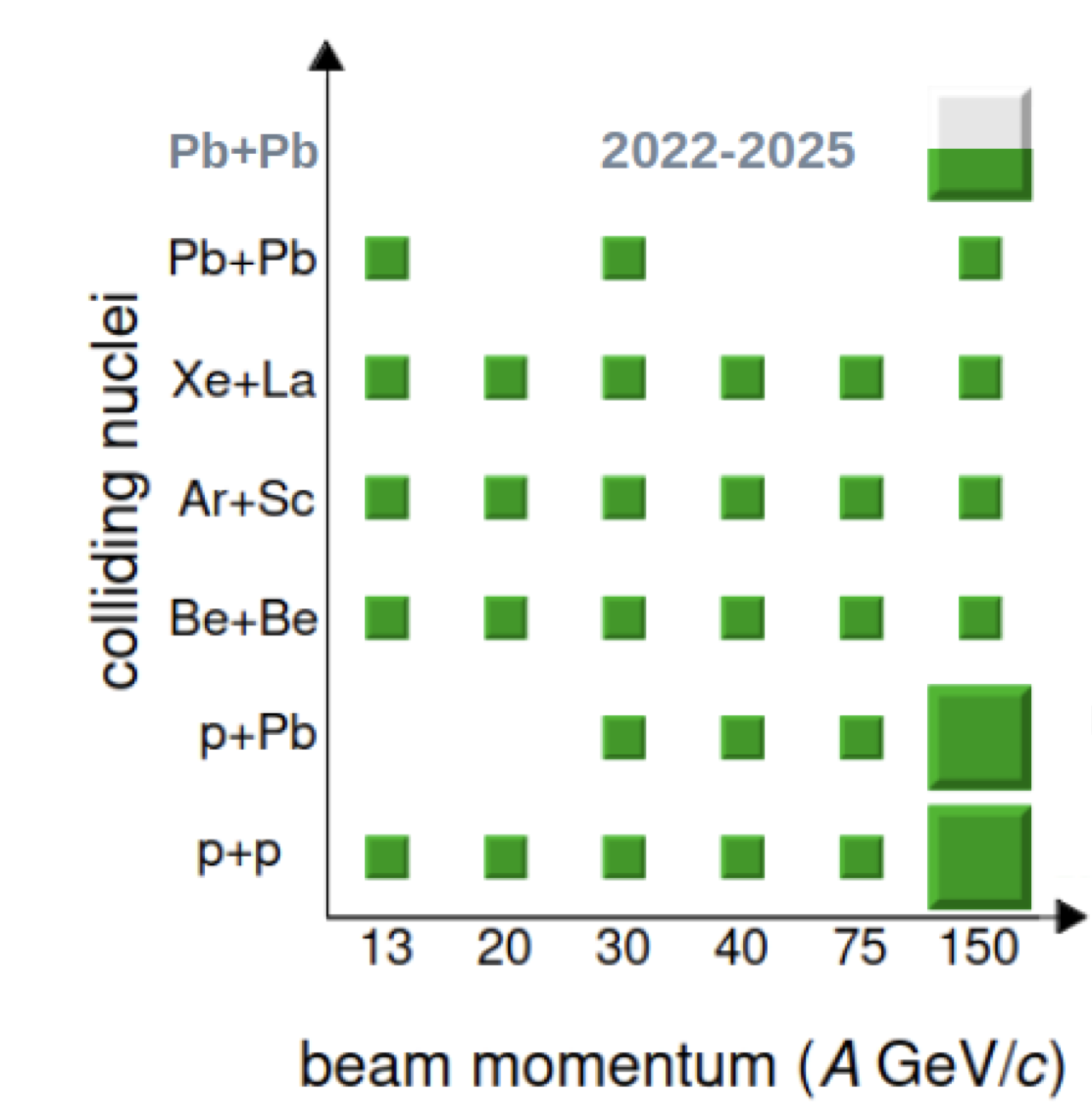}
\includegraphics[trim= 0cm 1cm 0cm 0cm, width=0.5\textwidth]{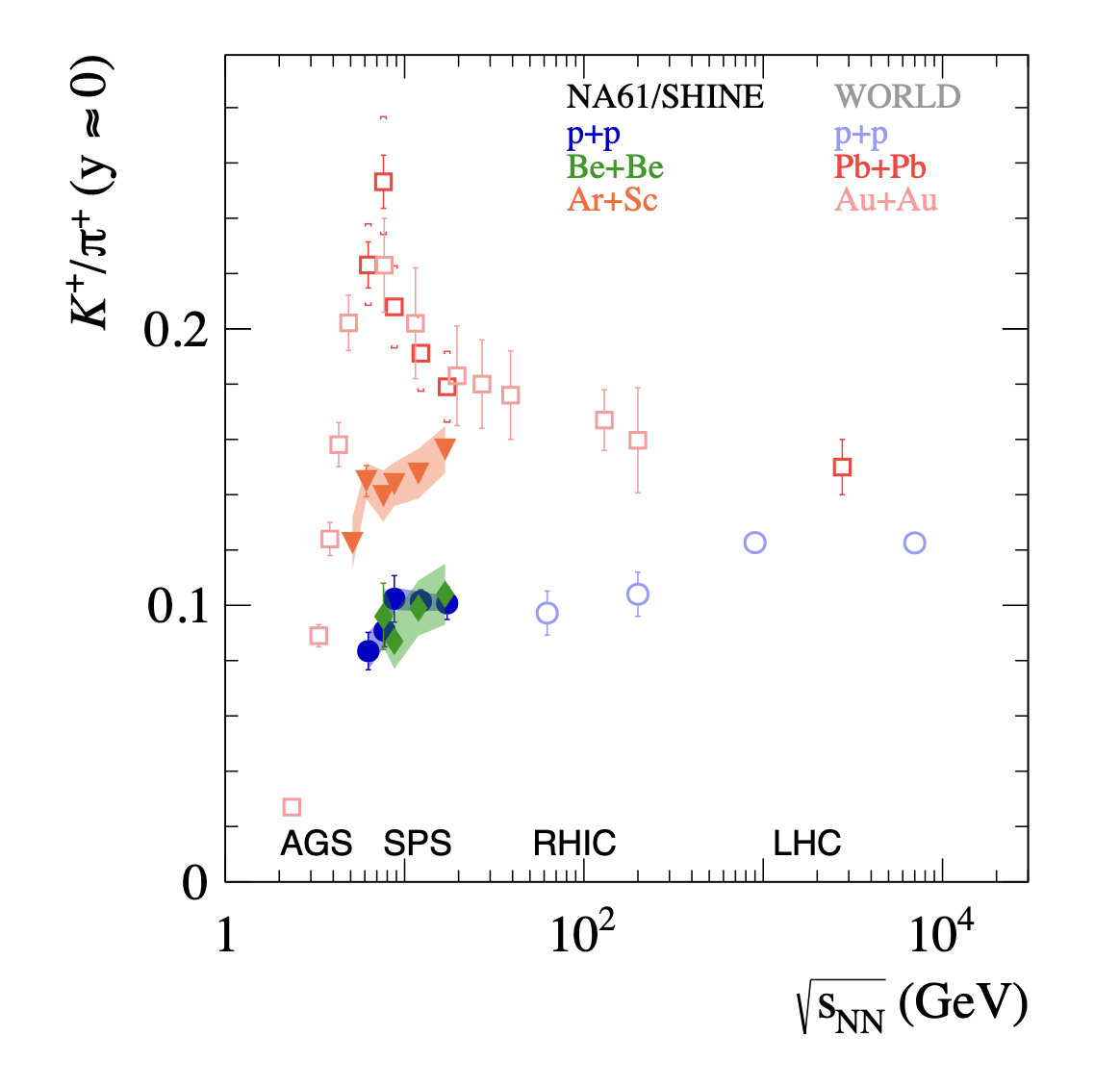}
\end{center}
\caption{
\textit{Left:}
2D scan programme of NA61/SHINE. 
\textit{Right:}
\kp/\pip at mid-rapidity as a function of collision energy~\cite{NA61SHINE:2023epu}, shown alongside a  compilation of world data. 
}
\label{fig:na61_boxplot}      
\end{figure}

Data were recorded for \textit{pp} (secondary proton beams), Be-Be (secondary $^7$Be beams), Ar-Sc, and Xe-La collisions at 13\AGeVc, 19(20)\AGeVc, 30\AGeV, 40\AGeV, 75(80)\AGeV, and 150(158)\AGeV, where the number represents the laboratory beam momentum per nucleon. Additionally, data for Pb-Pb collisions were recorded at 13\AGeV, 30\AGeV, and 150\AGeV. The datasets and years of data-taking are summarised in Fig.~\ref{fig:na61_boxplot}~(\textit{Left}).

\subsection{Key findings of NA61/SHINE}
\label{SPS_NA61_results}
\vspace{-0.3cm}
Figure~\ref{fig:na61_boxplot}~(\textit{right}) summarises the collision-energy dependence of positively charged kaon-to-pion ratio, \kp/\pip, at mid-rapidity measured 
by NA61/SHINE~\cite{NA61SHINE:2023epu} in \textit{pp}, Be-Be, Ar-Sc and Pb-Pb collisions, alongside a compilation of world data. 

First precise measurements of \textit{pp} interactions revealed a \textit{break} in the $\sqrt{s_{NN}}$ dependence at around 10~GeV~\cite{NA61SHINE:2019xkb}. This break was initially interpreted as either a transition from resonance- to string-dominated hadron production or a remnant of the
beginning of the QGP creation
observed in Pb-Pb collisions. Subsequent results for central Be-Be collisions~\cite{NA61SHINE:2020ggt,NA61SHINE:2020czq} surprisingly coincided with the \textit{pp} data, and supported the former interpretation. This confirms the prediction of the SMES model that Be-Be collisions behave similar to Pb-Pb, rather than resembling \textit{pp} 
interactions~\cite{Poberezhnyuk:2015wea}.

Thus, it was concluded that the system created in collisions of low-mass nuclei at SPS and below is far from equilibrium. Statistical models cannot approximate its properties. The concepts of phases of matter, characterised by temperature and chemical potentials, are not applicable. We had to move beyond the phase diagram of strongly interacting matter to understand the experimental results. This led us to the \textit{diagram of high-energy nuclear collisions}~\cite{Andronov:2022cna}.

\begin{figure}[tbp]
\begin{center}
\includegraphics[width=0.8\textwidth]{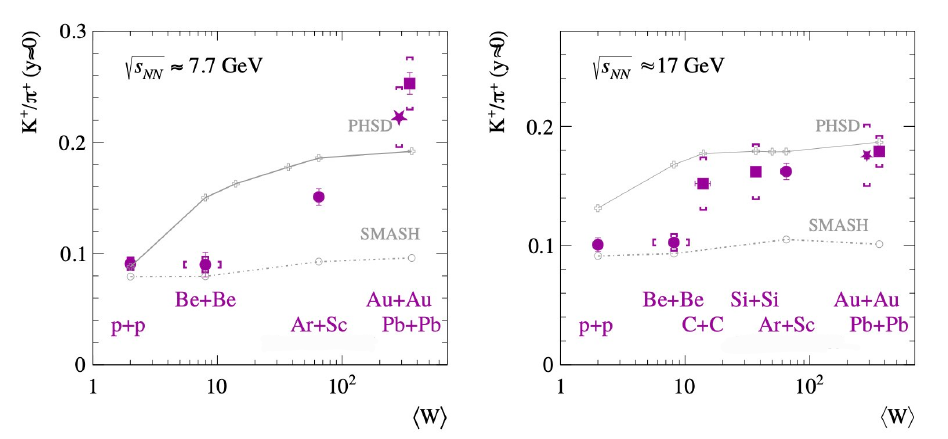}
\end{center}
\vspace{-2mm}
\caption{ $K^+/\pi^+$ ratio at mid-rapidity at  $\sqrt{s_{NN}}\approx 7.7$~GeV  and 
 17~GeV as a function of the mean number of wounded nucleons ($\langle W \rangle$) 
 for various collision systems. Results are compared to PHSD and SMASH models. See Ref.~\cite{Andronov:2022cna} for details.
}
\vspace{-2mm}
\label{fig:na61_jump}      
\end{figure}

Then, results from central Ar-Sc collisions~\cite{NA61SHINE:2021nye,NA61SHINE:2023epu} added another piece to the puzzle (see Fig.~\ref{fig:na61_jump}). At the top SPS energy, the \kp/\pip ratio matched that of Pb-Pb, but at lower SPS energies, it fell between the \textit{pp} and Pb-Pb values, showing no horn. 
The dependence of the \kp/\pip ratio at the top SPS energy exhibits a rapid change, or \textit{jump}, between collisions of small-mass nuclei (\textit{pp} and Be-Be) and those of medium- and large-mass nuclei
(Ar-Sc and Pb-Pb). In contrast, at lower energies, the dependence is continuous. Current models fail to describe these observations~\mbox{\cite{NA61SHINE:2023epu,Andronov:2022cna}}.

\begin{figure}[tbp]
\begin{center}
\includegraphics[width=0.45\textwidth]{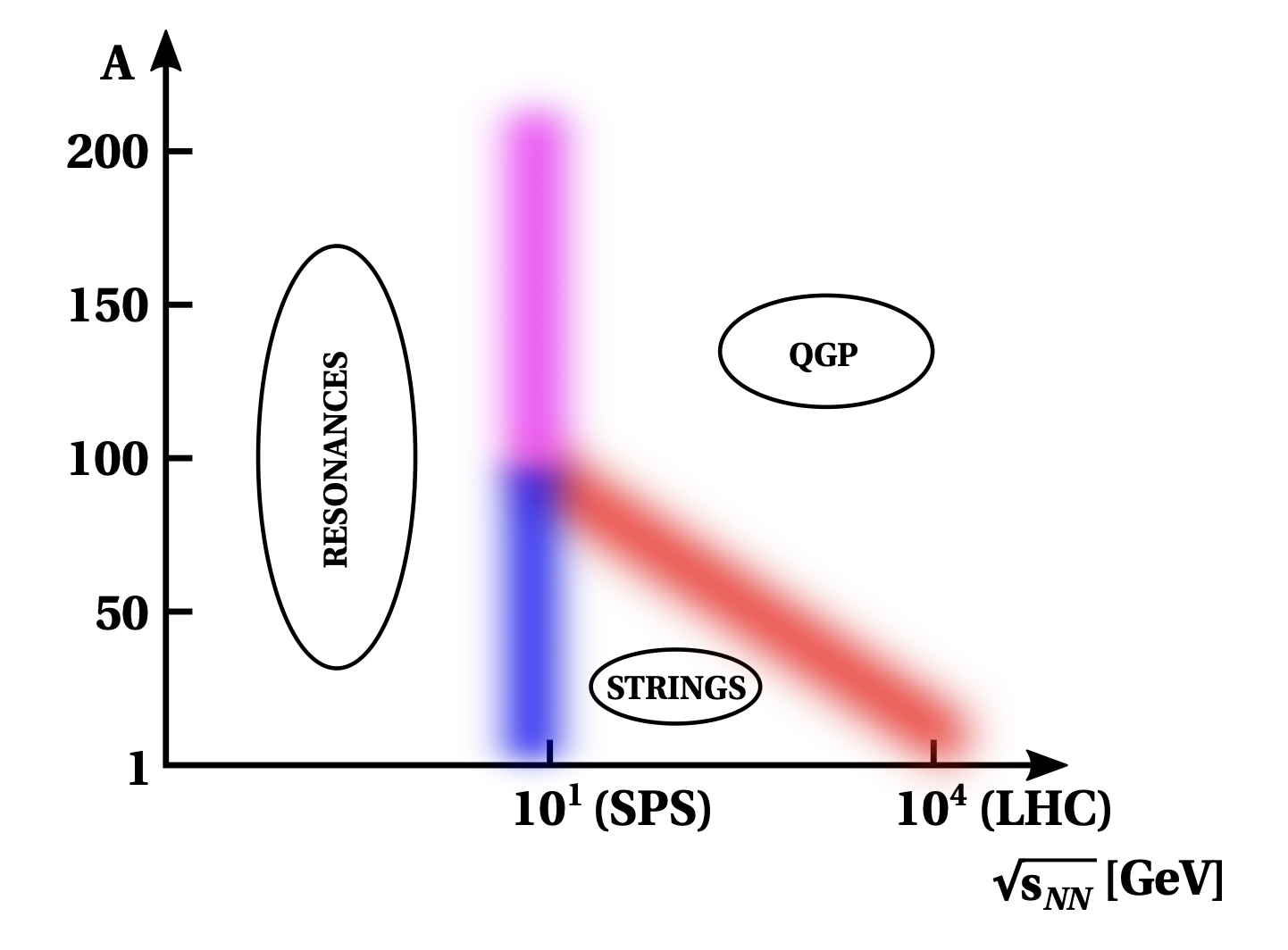}
\includegraphics[width=0.34\textwidth]{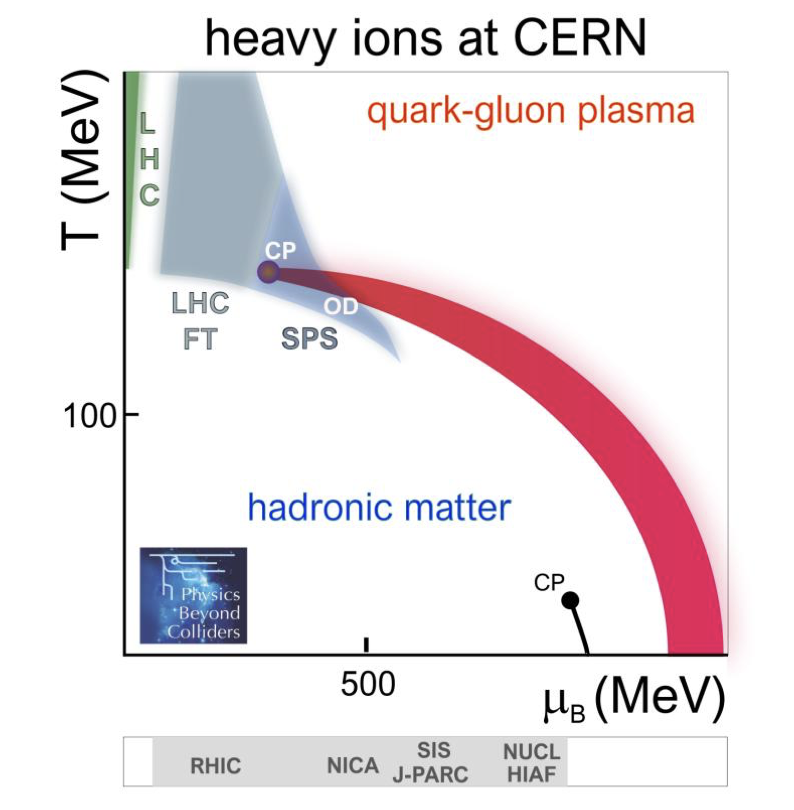}
\end{center}
\caption{
\textit{Left:} Diagram of high-energy nuclear collisions~\cite{Andronov:2022cna}, as suggested by NA61/SHINE and global data.  
\textit{Right:} 
Phase diagram of strongly interacting matter, showing regions accessible via heavy-ion collisions at 
CERN~\cite{QCDWorkingGroup:2019dyv}.
}
\vspace{-2mm}
\label{fig:na61_diagrams}      
\end{figure}

As mentioned above, the 2D scan of NA61/SHINE led to the  concept of the diagram of high-energy nuclear collisions~\cite{Andronov:2022cna}. 
This diagram depicts the domains dominated by different hadron-production processes in the space of laboratory-controlled parameters: collision energy and the nuclear mass number of the colliding nuclei. The diagram shown in Fig.~\ref{fig:na61_diagrams}~({\it{Left}}) 
gives a possible interpretation.  The lines indicate the changeover between the resonance, string, and QGP domains, with their approximate locations derived from experimental data; for details, see Ref.~\cite{Andronov:2022cna}. 

The experimental results suggest that the system created in collisions of high-mass nuclei is 
close to (local) equilibrium~\cite{Becattini:2005xt}. In this case, experiments with heavy-ion collisions provide insights into the \textit{phase diagram of strongly interacting matter}. The most widely accepted phase diagram is depicted in 
Fig.~\ref{fig:na61_diagrams}~({\it{right}}). 

The phase diagram includes a possible critical point (CP) - the endpoint of the first-order phase transition line, which has properties of a second-order phase transition. 
A significant objective of the NA61/SHINE 2D scan programme has been the search for the CP. Various experimental strategies, inspired by diverse model predictions, have been pursued; however, no signal has been observed to date. The 
CP-exclusion plots have been obtained based on the \textit{intermittency} analysis of particle multiplicity fluctuations~\cite{NA61SHINE:2023gez,NA61SHINE:2024ffp}. 

NA61/SHINE is collecting high-statistics data on Pb-Pb collisions at the top SPS energy for systematic measurements of open charm hadrons~\cite{Aduszkiewicz:2309890}.  
A detailed 2D scan, key to studying the transition from string-dominated to resonance-dominated hadron production~\cite{Mackowiak-Pawlowska:2867952}, is planned after LHC Long Shutdown~3. The idea of measuring the correlation between charm and anti-charm hadrons in central Pb-Pb collisions at the top SPS energy is also under consideration~\cite{Gazdzicki:2023niq}.

\section{A new state of matter}
\label{SPS_Summary}
\vspace{-0.3cm}
\begin{figure}[tbp]
\begin{center}
\includegraphics[width=0.94\textwidth]{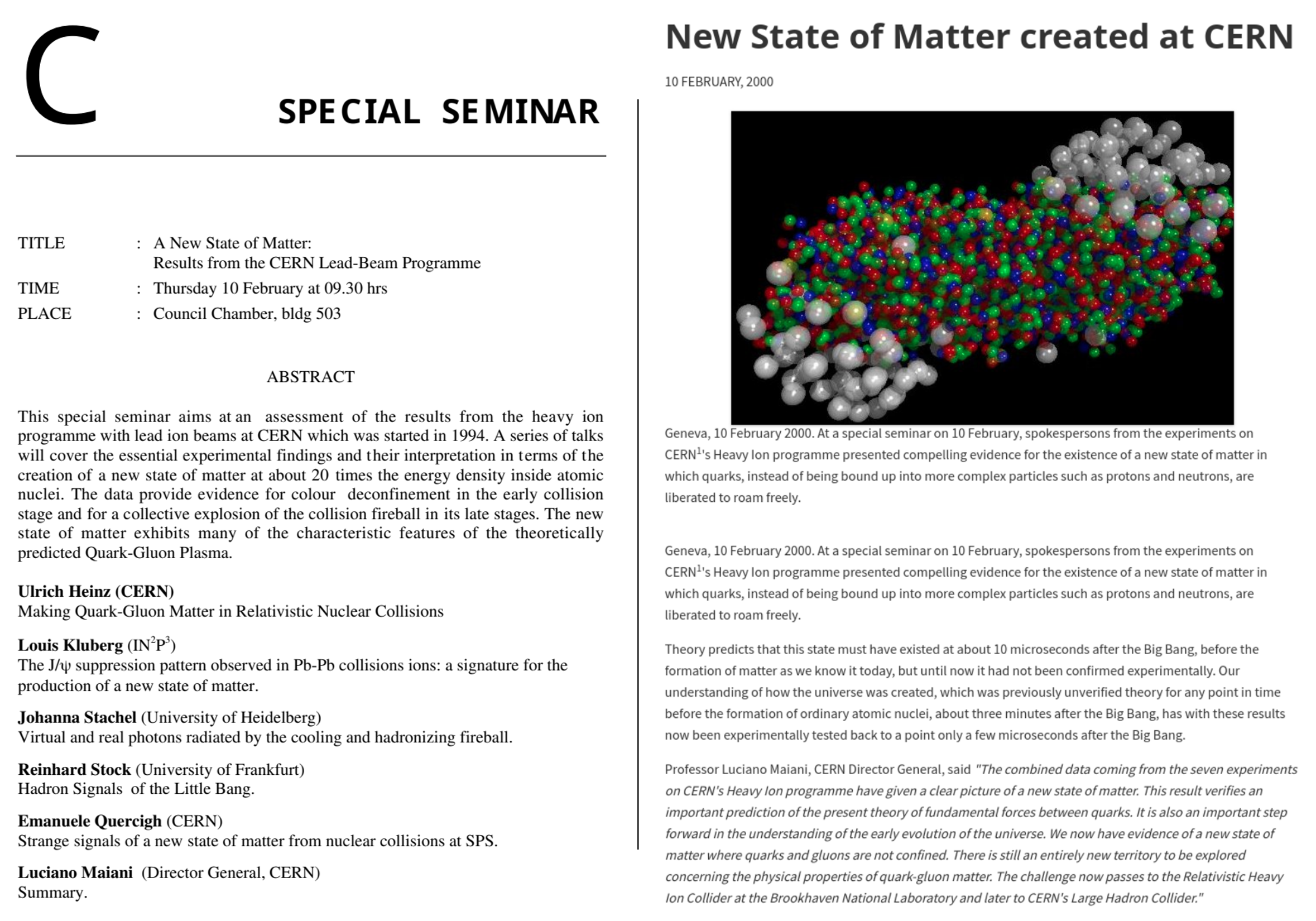}
\end{center}
\caption{February 2000: A special seminar was held at CERN, followed by a press release on the formation of a new state of matter 
at the SPS~\cite{CERN2000NewState}.}
\vspace{-3mm}
\end{figure}
On 10 February 2000, CERN hosted a special seminar titled ``A New State of Matter: Results from the CERN lead-Beam Programme", to review findings from the SPS heavy ion program.
The experimental findings
had been interpreted in terms of the creation of a new state of matter at about 20 times the energy density inside atomic nuclei. It also suggested the evidence for colour deconfinement in the early collision
stage and for a collective explosion of the collision fireball in 
its late stages~\cite{Heinz:2000bk}. 
In announcing this new state of matter~\cite{CERN2000NewState}, Luciano Maiani, CERN Director General during that time, noted that ``the challenge now passes to the Relativistic Heavy Ion Collider (RHIC) at the Brookhaven National Laboratory and later to CERN's Large Hadron Collider (LHC)".

Indeed, soon after the announcement of these first, encouraging signals of deconfinement from the SPS, the experimental focus shifted to the RHIC programme, that would soon provide full-fledged evidence for a strongly interacting QGP, and later to the LHC, where a comprehensive characterisation of the physical properties of the QGP is currently under way, while active QGP research is still being pursued at the CERN SPS.


\begin{acknowledgement}
We sincerely thank Peter Glassel for his valuable and constructive comments on this chapter.
Marek Gazdzicki's work has been supported by
the Polish National Science Centre grant 2018/30/A/ST2/00226. Tapan Nayak gratefully acknowledges the guidance of Hans Gutbrod and the support provided by GSI, Darmstadt, during his participation in the WA98 experiment as a postdoctoral fellow at CERN.

The experimental programs, detector developments, and results discussed in this chapter were made possible by the tremendous efforts of the dedicated researchers involved in the strong SPS heavy-ion program at CERN. The authors would like to thank all members of the collaborations, as well as the accelerator department at CERN, who have contributed to achieving the results presented in this chapter. The efforts of the wider scientific community are gratefully acknowledged.

\end{acknowledgement}

\bibliographystyle{atlasnote}
\bibliography{references}

@article{NA60:2006ymb,
    author = "Arnaldi, R. and others",
    collaboration = "NA60",
    title = "{First measurement of the rho spectral function in high-energy nuclear collisions}",
    eprint = "nucl-ex/0605007",
    archivePrefix = "arXiv",
    doi = "10.1103/PhysRevLett.96.162302",
    journal = "Phys. Rev. Lett.",
    volume = "96",
    pages = "162302",
    year = "2006"
}

@article{Rapp:1997fs,
    author = "Rapp, R. and Chanfray, G. and Wambach, J.",
    title = "{Rho meson propagation and dilepton enhancement in hot hadronic matter}",
    eprint = "hep-ph/9702210",
    archivePrefix = "arXiv",
    reportNumber = "SUNY-NTG-97-04",
    doi = "10.1016/S0375-9474(97)00137-1",
    journal = "Nucl. Phys. A",
    volume = "617",
    pages = "472--495",
    year = "1997"
}

@article{Heinz:2000bk,
    author = "Heinz, Ulrich W. and Jacob, Maurice",
    title = "{Evidence for a new state of matter: An Assessment of the results from the CERN lead beam program}",
    eprint = "nucl-th/0002042",
    archivePrefix = "arXiv",
    month = "1",
    year = "2000"
}

@article{Rapp:1999ej,
    author = "Rapp, R. and Wambach, J.",
    title = "{Chiral symmetry restoration and dileptons in relativistic heavy ion collisions}",
    eprint = "hep-ph/9909229",
    archivePrefix = "arXiv",
    reportNumber = "SUNY-NTG-99-29",
    doi = "10.1007/0-306-47101-9_1",
    journal = "Adv. Nucl. Phys.",
    volume = "25",
    pages = "1",
    year = "2000"
}

@article{Brown:1991kk,
    author = "Brown, G. E. and Rho, Mannque",
    title = "{Scaling effective Lagrangians in a dense medium}",
    doi = "10.1103/PhysRevLett.66.2720",
    journal = "Phys. Rev. Lett.",
    volume = "66",
    pages = "2720--2723",
    year = "1991"
}

@article{Li:1995qm,
    author = "Li, Guo-Qiang and Ko, C. M. and Brown, G. E.",
    title = "{Enhancement of low mass dileptons in heavy ion collisions}",
    eprint = "nucl-th/9504025",
    archivePrefix = "arXiv",
    doi = "10.1103/PhysRevLett.75.4007",
    journal = "Phys. Rev. Lett.",
    volume = "75",
    pages = "4007--4010",
    year = "1995"
}

@article{Brown:2001nh,
    author = "Brown, G. E. and Rho, Mannque",
    title = "{On the manifestation of chiral symmetry in nuclei and dense nuclear matter}",
    eprint = "hep-ph/0103102",
    archivePrefix = "arXiv",
    doi = "10.1016/S0370-1573(01)00084-9",
    journal = "Phys. Rept.",
    volume = "363",
    pages = "85--171",
    year = "2002"
}

@article{vanHees:2007th,
    author = "van Hees, Hendrik and Rapp, Ralf",
    title = "{Dilepton Radiation at the CERN Super Proton Synchrotron}",
    eprint = "0711.3444",
    archivePrefix = "arXiv",
    primaryClass = "hep-ph",
    doi = "10.1016/j.nuclphysa.2008.03.009",
    journal = "Nucl. Phys. A",
    volume = "806",
    pages = "339--387",
    year = "2008"
}

@article{Ruppert:2007cr,
    author = "Ruppert, Jorg and Gale, Charles and Renk, Thorsten and Lichard, Peter and Kapusta, Joseph I.",
    title = "{Low mass dimuons produced in relativistic nuclear collisions}",
    eprint = "0706.1934",
    archivePrefix = "arXiv",
    primaryClass = "hep-ph",
    doi = "10.1103/PhysRevLett.100.162301",
    journal = "Phys. Rev. Lett.",
    volume = "100",
    pages = "162301",
    year = "2008"
}

@article{Dusling:2006yv,
    author = "Dusling, K. and Teaney, D. and Zahed, I.",
    title = "{Thermal dimuon yields at NA60}",
    eprint = "nucl-th/0604071",
    archivePrefix = "arXiv",
    doi = "10.1103/PhysRevC.75.024908",
    journal = "Phys. Rev. C",
    volume = "75",
    pages = "024908",
    year = "2007"
}

@article{Kapusta:1979fh,
    author = "Kapusta, Joseph I.",
    title = "{Quantum Chromodynamics at High Temperature}",
    doi = "10.1016/0550-3213(79)90146-9",
    journal = "Nucl. Phys. B",
    volume = "148",
    pages = "461--498",
    year = "1979"
}

@article{Shuryak:1980tp,
    author = "Shuryak, Edward V.",
    title = "{Quantum Chromodynamics and the Theory of Superdense Matter}",
    doi = "10.1016/0370-1573(80)90105-2",
    journal = "Phys. Rept.",
    volume = "61",
    pages = "71--158",
    year = "1980"
}

@article{NA49:1999myq,
    author = "Afanasiev, S. and others",
    collaboration = "NA49",
    title = "{The NA49 large acceptance hadron detector}",
    reportNumber = "CERN-EP-99-001, CERN-EP-99-01, CERN-EP-99-1",
    doi = "10.1016/S0168-9002(99)00239-9",
    journal = "Nucl. Instrum. Meth. A",
    volume = "430",
    pages = "210--244",
    year = "1999"
}

@techreport{na35_roots,
      author = "Stock, Reinhard",
      title         = "{Study of relativistic nucleus-nucleus reactions induced
                       by $^{16}O$ beams of 9-13 GeV per nucleon at the CERN PS}",
      institution   = "CERN",
      reportNumber  = "CERN-PSCC-82-1, PSCC-P-53",
      address       = "Geneva",
      year          = "1982",
      url           = "https://cds.cern.ch/record/725513",
}

@article{Klapisch:1984yfi,
    author = "Klapisch, R.",
    editor = "Ludlam, T. W. and Wegner, H. E.",
    title = "{STATUS OF CERN PROGRAM IN RELATIVISTIC ION PHYSICS}",
    doi = "10.1016/0375-9474(84)90559-1",
    journal = "Nucl. Phys. A",
    volume = "418",
    pages = "347C--352C",
    year = "1984"
}

@article{Koch:1986ud,
    author = "Koch, P. and Muller, Berndt and Rafelski, Johann",
    title = "{Strangeness in Relativistic Heavy Ion Collisions}",
    doi = "10.1016/0370-1573(86)90096-7",
    journal = "Phys. Rept.",
    volume = "142",
    pages = "167--262",
    year = "1986"
}

@article{NA35:1990teq,
    author = "Bartke, J. and others",
    collaboration = "NA35",
    title = "{Neutral strange particle production in sulphur sulphur and proton sulphur collisions at 200-GeV/nucleon}",
    doi = "10.1007/BF01554465",
    journal = "Z. Phys. C",
    volume = "48",
    pages = "191--200",
    year = "1990"
}

@techreport{Panagiotou:295042,
      author        = "Panagiotou, A. D. and others",
      collaboration = "NA35",
      title         = "{Large acceptance hadron detector for an investigation of
                       Pb-induced reactions at the CERN SPS}",
      institution   = "CERN",
      reportNumber  = "CERN-SPSLC-91-31, SPSLC-P-264",
      address       = "Geneva",
      year          = "1991",
      url           = "https://cds.cern.ch/record/295042",
}

@misc{cern916772,
  author       = {{CERN PhotoLab}},
  title        = {The Omega spectrometer in the West Hall},
  year         = {1976},
  howpublished = {\url{https://cds.cern.ch/record/917722}},
  note         = {CERN photograph, original reference: 6-12-76},
}

@inproceedings{michelini1969omega,
  author       = {Aldo Michelini},
  title        = {{The Omega project}},
  booktitle    = {Proceedings of the 1969 CERN School of Physics},
  pages        = {181--206},
  year         = {1969},
  address      = {Geneva, Switzerland},
  publisher    = {CERN},
  doi          = {10.5170/CERN-1969-029.181}
}

@article{NA49:1997xvz,
    author = "Appelshauser, H and others",
    collaboration = "NA49",
    title = "{Hadronic expansion dynamics in central Pb + Pb collisions at 158-GeV per nucleon}",
    eprint = "hep-ex/9711024",
    archivePrefix = "arXiv",
    reportNumber = "IKF-HENPG-6-97",
    doi = "10.1007/s100520050168",
    journal = "Eur. Phys. J. C",
    volume = "2",
    pages = "661--670",
    year = "1998"
}

@inproceedings{Roland:1997hs,
    author = "Roland, G.",
    collaboration = "NA49",
    title = "{First results of the NA49 event-by event analysis of Pb + Pb collisions at the SPS}",
    booktitle = "{Hirschegg 97)}",
    pages = "309--318",
    year = "1997"
}

@article{Gazdzicki:1998vd,
    author = "Gazdzicki, Marek and Gorenstein, Mark I.",
    title = "{On the early stage of nucleus-nucleus collisions}",
    eprint = "hep-ph/9803462",
    archivePrefix = "arXiv",
    reportNumber = "IKF-HENPG-2-98",
    journal = "Acta Phys. Polon. B",
    volume = "30",
    pages = "2705",
    year = "1999"
}

@article{VanHove:1982vk,
    author = "Van Hove, L.",
    editor = "Giovannini, Alberto",
    title = "{Multiplicity Dependence of p(T) Spectrum as a Possible Signal for a Phase Transition in Hadronic Collisions}",
    reportNumber = "CERN-TH-3391",
    doi = "10.1016/0370-2693(82)90617-7",
    journal = "Phys. Lett. B",
    volume = "118",
    pages = "138",
    year = "1982"
}

@techreport{CPOD2024,
      author        = "Gazdzicki, M. and Seyboth, P. and Shuryak E.",
      title         = "{When Quarks and Gluons Become Free}",
      institution   = "CERN",
      reportNumber  = "CERN Courier September 2024",
      address       = "Geneva",
      year          = "2024",
      url           = "https://cerncourier.com/a/when-quarks-and-gluons-become-free/",
}

@article{NA61:2014lfx,
    author = "Abgrall, N. and others",
    collaboration = "NA61",
    title = "{NA61/SHINE facility at the CERN SPS: beams and detector system}",
    eprint = "1401.4699",
    archivePrefix = "arXiv",
    primaryClass = "physics.ins-det",
    reportNumber = "CERN-PH-EP-2014-003",
    doi = "10.1088/1748-0221/9/06/P06005",
    journal = "JINST",
    volume = "9",
    pages = "P06005",
    year = "2014"
}

@article{Sandoval:170613,
      author        = "Sandoval, A",
      collaboration = "NA35",
      title         = "{Event simulation in NA35: study of relativistic nucleus -
                       nucleus collisions}",
      reportNumber  = "CERN-EP-86-108",
      journal       = "Nucl. Phys. A",
      volume        = "461",
      pages         = "465-486",
      year          = "1987",
      url           = "https://cds.cern.ch/record/170613",
      doi           = "10.1016/0375-9474(87)90507-0",
}

@article{NA35:1988eto,
    author = "Bamberger, A. and others",
    collaboration = "NA35",
    title = "{Probing the Space-time Geometry of Ultrarelativistic Heavy Ion Collisions}",
    doi = "10.1016/0370-2693(88)90561-8",
    journal = "Phys. Lett. B",
    volume = "203",
    pages = "320--326",
    year = "1988"
}

@article{Landau:1953wku,
    author = "Landau, Lev Davidovich",
    editor = "ter Haar, D.",
    title = "{On Multiple Production of Particles during Collisions of Fast Particles}",
    doi = "10.1016/b978-0-08-010586-4.50079-1",
    journal = "Izv. Akad. Nauk Ser. Fiz.",
    volume = "15",
    year = "1953"
}

@article{Gazdzicki:2003dx,
    author = "Gazdzicki, M. and Gorenstein, Mark I. and Grassi, F. and Hama, Yogiro and Kodama, T. and Socolowski, Jr., O.",
    editor = "Navarra, F. S. and Hama, Yogiro",
    title = "{Incident energy dependence of the effective temperature in heavy ion collisions}",
    eprint = "hep-ph/0309192",
    archivePrefix = "arXiv",
    doi = "10.1590/S0103-97332004000200041",
    journal = "Braz. J. Phys.",
    volume = "34",
    pages = "322--325",
    year = "2004"
}

@techreport{NA61_history,
      author        = "NA61/SHINE",
      title         = "{History of the Collaboration}",
      institution   = "CERN",
      reportNumber  = "NA61/SHINE website: https://shine.web.cern.ch/node/10",
      address       = "Geneva",
      year          = "2024",
      url           = "https://shine.web.cern.ch/node/10"
}

@techreport{Gazdzicki:685283,
      author        = "Gazdzicki, M and Vesztergombi, G",
      collaboration = "NA61",
      title         = "{A New Experimental Programme..}",
      institution   = "CERN",
      reportNumber  = "CERN-SPSC-2003-038, SPSC-EOI-001",
      address       = "Geneva",
      year          = "2003",
      url           = "https://cds.cern.ch/record/685283",
}

@techreport{Gazdzicki:919966,
      author        = "Antonious, A. and others",
      collaboration = "NA49-future",
      title         = "{Study of Hadron Production in Collisions of Protons and Nuclei at the CERN SPS}",
      institution   = "CERN",
      reportNumber  = "CERN-SPSC-2006-001, SPSC-I-235",
      year          = "2006",
      url           = "https://cds.cern.ch/record/919966",
}

@techreport{Gazdzicki:995681,
      author        = "Gazdzicki, Marek and Fodor, Z and Vesztergombi, G",
      collaboration = "NA49-future",
      title         = "{Study of Hadron Production in Hadron-Nucleus and
                       Nucleus-Nucleus Collisions at the CERN SPS}",
      institution   = "CERN",
      reportNumber  = "CERN-SPSC-2006-034, SPSC-P-330",
      address       = "Geneva",
      year          = "2006",
      url           = "https://cds.cern.ch/record/995681",
      doi           = "10.17181/CERN.JTNE.Y0QH",
}

@techreport{vonDoetinchem:2914265,
      author        = "von Doetinchem, P",
      collaboration = "NA61/SHINE",
      title         = "{Addendum to the NA61/SHINE Proposal ..}",
      institution   = "CERN",
      reportNumber  = "CERN-SPSC-2024-028, SPSC-P-330-ADD-15",
      address       = "Geneva",
      year          = "2024",
      url           = "https://cds.cern.ch/record/2914265",
}

@article{NA61SHINE:2020czq,
    author = "Acharya, A. and others",
    collaboration = "NA61/SHINE",
    title = "{Measurements of $\pi^\pm$, $K^\pm$, $p$ and $\bar{p}$ spectra in $^7$Be+$^9$Be collisions at beam momenta from 19$A$ to 150$A$ GeV/$c$ with the NA61/SHINE spectrometer at the CERN SPS}",
    eprint = "2010.01864",
    archivePrefix = "arXiv",
    primaryClass = "hep-ex",
    reportNumber = "CERN-EP-2020-187, CERN-EP-2020-187",
    doi = "10.1140/epjc/s10052-020-08733-x",
    journal = "Eur. Phys. J. C",
    volume = "81",
    number = "1",
    pages = "73",
    year = "2021",
    note = "[Erratum: Eur.Phys.J.C 83, 90 (2023)]"
}

@article{Poberezhnyuk:2015wea,
    author = "Poberezhnyuk, R. V. and Gazdzicki, M. and Gorenstein, M. I.",
    title = "{Statistical Model of the Early Stage of nucleus-nucleus collisions with exact strangeness conservation}",
    eprint = "1502.05650",
    archivePrefix = "arXiv",
    primaryClass = "nucl-th",
    doi = "10.5506/APhysPolB.46.1991",
    journal = "Acta Phys. Polon. B",
    volume = "46",
    number = "10",
    pages = "1991",
    year = "2015"
}

@article{NA61SHINE:2024ffp,
    author = "Adhikary, H. and others",
    collaboration = "NA61/SHINE",
    title = "{Search for a critical point of strongly-interacting matter in central $^{40}$Ar~+~$^{45}$Sc collisions at 13~A\textendash{}75~A ~GeV/c beam momentum}",
    eprint = "2401.03445",
    archivePrefix = "arXiv",
    primaryClass = "nucl-ex",
    reportNumber = "FERMILAB-PUB-24-0021-AD",
    doi = "10.1140/epjc/s10052-024-13012-0",
    journal = "Eur. Phys. J. C",
    volume = "84",
    number = "7",
    pages = "741",
    year = "2024"
}

@techreport{Aduszkiewicz:2309890,
      author        = "Aduszkiewicz, A",
      collaboration = "NA61/SHINE",
      title         = "{Study of Hadron-Nucleus and Nucleus-Nucleus Collisions at
                       the CERN SPS ..}",
      institution   = "CERN",
      reportNumber  = "CERN-SPSC-2018-008, SPSC-P-330-ADD-10",
      address       = "Geneva",
      year          = "2018",
      url           = "https://cds.cern.ch/record/2309890",
}

@techreport{Mackowiak-Pawlowska:2867952,
      author        = "Mackowiak-Pawlowska,  M",
      collaboration = "NA61/SHINE",
      title         = "{Addendum to the NA61/SHINE Proposal..}",
      institution   = "CERN",
      reportNumber  = "CERN-SPSC-2023-022, SPSC-P-330-ADD-14",
      year          = "2023",
      url           = "https://cds.cern.ch/record/2867952",
}

@article{Gazdzicki:2023niq,
    author = "Gazdzicki, Marek and Kiko{\l}a, Daniel and Pidhurskyi, Ivan and Tinti, Leonardo",
    title = "{Spatial correlations of charm and anticharm quarks at hadronisation}",
    eprint = "2305.00212",
    archivePrefix = "arXiv",
    primaryClass = "hep-ph",
    doi = "10.1038/s42005-025-02213-y",
    journal = "Commun. Phys.",
    volume = "8",
    number = "1",
    pages = "304",
    year = "2025"
}

@article{Chen:1993sh,
    author = "Chen, W. and others",
    title = "{Performance of the multianode cylindrical silicon drift detector in the CERES NA45 experiment: First results}",
    doi = "10.1016/0168-9002(93)90363-M",
    journal = "Nucl. Instrum. Meth. A",
    volume = "326",
    pages = "273--278",
    year = "1993"
}

@article{CERES:2005uih,
    author = "Agakichiev, G. and others",
    collaboration = "CERES",
    title = "{e+ e- pair production in Pb - Au collisions at 158-GeV per nucleon}",
    eprint = "nucl-ex/0506002",
    archivePrefix = "arXiv",
    doi = "10.1140/epjc/s2005-02272-3",
    journal = "Eur. Phys. J. C",
    volume = "41",
    pages = "475--513",
    year = "2005"
}

@article{CERESNA45:2002gnc,
    author = "Adamova, D. and others",
    collaboration = "CERES/NA45",
    title = "{Enhanced production of low mass electron pairs in 40-AGeV Pb - Au collisions at the CERN SPS}",
    eprint = "nucl-ex/0209024",
    archivePrefix = "arXiv",
    doi = "10.1103/PhysRevLett.91.042301",
    journal = "Phys. Rev. Lett.",
    volume = "91",
    pages = "042301",
    year = "2003"
}

@article{CERES:2005wwe,
    author = "Adamova, D. and others",
    collaboration = "CERES",
    title = "{Leptonic and charged kaon decay modes of the phi meson measured in heavy-ion collisions at the CERN SPS}",
    eprint = "nucl-ex/0512007",
    archivePrefix = "arXiv",
    doi = "10.1103/PhysRevLett.96.152301",
    journal = "Phys. Rev. Lett.",
    volume = "96",
    pages = "152301",
    year = "2006"
}

@article{CERES:2009jng,
    author = "Adamova, D. and others",
    collaboration = "CERES",
    title = "{Modification of jet-like correlations in Pb-Au collisions at 158A-GeV/c}",
    eprint = "0904.2973",
    archivePrefix = "arXiv",
    primaryClass = "nucl-ex",
    doi = "10.1016/j.physletb.2009.05.048",
    journal = "Phys. Lett. B",
    volume = "678",
    pages = "259--263",
    year = "2009"
}

@article{CERESNA45:2003pat,
    author = "Agakichiev, G. and others",
    collaboration = "CERES/NA45",
    title = "{Semihard scattering unraveled from collective dynamics by two pion correlations in 158-A-GeV / c Pb + Au collisions}",
    eprint = "nucl-ex/0303014",
    archivePrefix = "arXiv",
    doi = "10.1103/PhysRevLett.92.032301",
    journal = "Phys. Rev. Lett.",
    volume = "92",
    pages = "032301",
    year = "2004"
}

@article{CERESNA45:2016rra,
    author = "Adamov\'a, D. and others",
    collaboration = "CERES NA45",
    title = "{Triangular flow of negative pions emitted in PbAu collisions at $\sqrt{s_{NN}} =$ 17.3\textasciitilde{}GeV}",
    eprint = "1604.07469",
    archivePrefix = "arXiv",
    primaryClass = "nucl-ex",
    doi = "10.1016/j.nuclphysa.2016.08.002",
    journal = "Nucl. Phys. A",
    volume = "957",
    pages = "99--108",
    year = "2017"
}

@article{CERES:2012hbp,
    author = "Adamova, D. and others",
    collaboration = "CERES",
    title = "{Elliptic flow of charged pions, protons and strange particles emitted in Pb+Au collisions at top SPS energy}",
    eprint = "1205.3692",
    archivePrefix = "arXiv",
    primaryClass = "nucl-ex",
    doi = "10.1016/j.nuclphysa.2012.08.004",
    journal = "Nucl. Phys. A",
    volume = "894",
    pages = "41--73",
    year = "2012"
}

@article{CERES:2003sap,
    author = "Adamova, D. and others",
    collaboration = "CERES",
    title = "{Event by event fluctuations of the mean transverse momentum in 40, 80 and 158 A GeV / c Pb - Au collisions}",
    eprint = "nucl-ex/0305002",
    archivePrefix = "arXiv",
    doi = "10.1016/j.nuclphysa.2003.07.018",
    journal = "Nucl. Phys. A",
    volume = "727",
    pages = "97--119",
    year = "2003"
}

@article{CERES:2002rfr,
    author = "Adamova, D. and others",
    collaboration = "CERES",
    title = "{Universal pion freezeout in heavy ion collisions}",
    eprint = "nucl-ex/0207008",
    archivePrefix = "arXiv",
    doi = "10.1103/PhysRevLett.90.022301",
    journal = "Phys. Rev. Lett.",
    volume = "90",
    pages = "022301",
    year = "2003"
}

@article{Agakichiev:1998ign,
    author = "Agakichiev, G. and others",
    title = "{Neutral meson production in p Be and p Au collisions at 450-GeV beam energy}",
    doi = "10.1007/s100529800804",
    journal = "Eur. Phys. J. C",
    volume = "4",
    pages = "249--257",
    year = "1998"
}

@article{CERES:1995rik,
    author = "Baur, R. and others",
    collaboration = "CERES",
    title = "{Search for direct photons from S - Au collisions at 200-GeV/u}",
    reportNumber = "CERN-PPE-95-116",
    doi = "10.1007/s002880050204",
    journal = "Z. Phys. C",
    volume = "71",
    pages = "571--578",
    year = "1996"
}

@article{Hohler:2013eba,
    author = "Hohler, Paul M. and Rapp, Ralf",
    title = "{Is $\rho$-Meson Melting Compatible with Chiral Restoration?}",
    eprint = "1311.2921",
    archivePrefix = "arXiv",
    primaryClass = "hep-ph",
    doi = "10.1016/j.physletb.2014.02.021",
    journal = "Phys. Lett. B",
    volume = "731",
    pages = "103--109",
    year = "2014"
}

@article{CERES:2006wcq,
    author = "Adamova, D. and others",
    collaboration = "CERES",
    title = "{Modification of the rho-meson detected by low-mass electron-positron pairs in central Pb-Au collisions at 158-A-GeV/c}",
    eprint = "nucl-ex/0611022",
    archivePrefix = "arXiv",
    doi = "10.1016/j.physletb.2008.07.104",
    journal = "Phys. Lett. B",
    volume = "666",
    pages = "425--429",
    year = "2008"
}

@article{Tserruya:2006ht,
    author = "Tserruya, Itzhak",
    editor = "Csorgo, T. and Levai, P. and David, G. and Papp, G.",
    title = "{Quark Matter 2005: Experimental conference summary}",
    eprint = "nucl-ex/0601036",
    archivePrefix = "arXiv",
    doi = "10.1016/j.nuclphysa.2006.06.063",
    journal = "Nucl. Phys. A",
    volume = "774",
    pages = "415--432",
    year = "2006"
}

@article{CMS:2012qk,
    author = "Chatrchyan, Serguei and others",
    collaboration = "CMS",
    title = "{Observation of Long-Range Near-Side Angular Correlations in Proton-Lead Collisions at the LHC}",
    eprint = "1210.5482",
    archivePrefix = "arXiv",
    primaryClass = "nucl-ex",
    reportNumber = "CMS-HIN-12-005, CERN-PH-EP-2012-320, CMS-HIN-12-015",
    doi = "10.1016/j.physletb.2012.11.025",
    journal = "Phys. Lett. B",
    volume = "718",
    pages = "795--814",
    year = "2013"
}

@article{Agakichiev:1998kip,
    author = "Agakichiev, G. and others",
    title = "{Systematic study of low mass electron pair production in p Be and p Au collisions at 450-GeV/c}",
    doi = "10.1007/PL00021659",
    journal = "Eur. Phys. J. C",
    volume = "4",
    pages = "231--247",
    year = "1998"
}

@article{CERESNA45:1997tgc,
    author = "Agakichiev, G. and others",
    collaboration = "CERES/NA45",
    title = "{Low mass e+ e- pair production in 158/A-GeV Pb - Au collisions at the CERN SPS, its dependence on multiplicity and transverse momentum}",
    eprint = "nucl-ex/9712008",
    archivePrefix = "arXiv",
    doi = "10.1016/S0370-2693(98)00083-5",
    journal = "Phys. Lett. B",
    volume = "422",
    pages = "405--412",
    year = "1998"
}

@article{CERES:1995vll,
    author = "Agakichiev, G. and others",
    collaboration = "CERES",
    title = "{Enhanced production of low mass electron pairs in 200-GeV/u S - Au collisions at the CERN SPS}",
    reportNumber = "CERN-PPE-95-026, CERN-PPE-95-26",
    doi = "10.1103/PhysRevLett.75.1272",
    journal = "Phys. Rev. Lett.",
    volume = "75",
    pages = "1272--1275",
    year = "1995"
}

@article{CERES:2008asn,
    author = "Adamova, D. and others",
    collaboration = "CERES",
    title = "{The CERES/NA45 Radial Drift Time Projection Chamber}",
    eprint = "0802.1443",
    archivePrefix = "arXiv",
    primaryClass = "nucl-ex",
    doi = "10.1016/j.nima.2008.04.056",
    journal = "Nucl. Instrum. Meth. A",
    volume = "593",
    pages = "203--231",
    year = "2008"
}

@article{CERES:1996bja,
    author = "Agakishiev, G. and others",
    collaboration = "CERES",
    title = "{Performance of the CERES electron spectrometer in the CERN SPS lead beam}",
    doi = "10.1016/0168-9002(95)01135-8",
    journal = "Nucl. Instrum. Meth. A",
    volume = "371",
    pages = "16--21",
    year = "1996"
}

@article{Faschingbauer:1995qj,
    author = "Faschingbauer, U. and others",
    editor = "Holl, P. and Lutz, G. and Richter, Rainer Helmut and Strueder, L. and Longoni, A. and Sampietro, M.",
    title = "{A Doublet of 3-inch cylindrical silicon drift detectors in the CERES / NA45 experiment}",
    reportNumber = "CERN-PPE-95-132",
    doi = "10.1016/0168-9002(95)01403-9",
    journal = "Nucl. Instrum. Meth. A",
    volume = "377",
    pages = "362--366",
    year = "1996"
}

@article{Tserruya:2020vyt,
    author = "Tserruya, I. and Aoki, K. and Woody, C.",
    title = "{Hadron blind Cherenkov counters}",
    eprint = "2003.12858",
    archivePrefix = "arXiv",
    primaryClass = "physics.ins-det",
    doi = "10.1016/j.nima.2020.163765",
    journal = "Nucl. Instrum. Meth. A",
    volume = "970",
    pages = "163765",
    year = "2020"
}

@techreport{NA44Proposal1993,
  author       = {Appelshäuser, H. and others},
  title        = {The NA44 Spectrometer: Improvements for Pb Beams},
  institution  = {CERN},
  year         = {1993},
  number       = {CERN-SPSLC-93-22, SPSLC-M-521},
  url          = {https://cds.cern.ch/record/310569/files/cm-p00043816.pdf},
}

@article{Gazdzicki:1995ze,
    author = "Gazdzicki, M.",
    title = "{Entropy in nuclear collisions}",
    doi = "10.1007/BF01579641",
    journal = "Z. Phys. C",
    volume = "66",
    pages = "659--662",
    year = "1995"
}

@article{Matsui:1986dk,
    author = "Matsui, T. and Satz, H.",
    title = "{$J/\psi$ Suppression by Quark-Gluon Plasma Formation}",
    reportNumber = "BNL-38344",
    doi = "10.1016/0370-2693(86)91404-8",
    journal = "Phys. Lett. B",
    volume = "178",
    pages = "416--422",
    year = "1986"
}

@article{Baur:1993mz,
    author = "Baur, R. and others",
    editor = "Nappi, E. and Ypsilantis, T.",
    title = "{In beam experience from the CERES UV detectors: Prohibitive spark breakdown in multistep parallel plate chambers as compared to wire chambers}",
    reportNumber = "CERN-PPE-93-169",
    doi = "10.1016/0168-9002(94)90556-8",
    journal = "Nucl. Instrum. Meth. A",
    volume = "343",
    pages = "231--240",
    year = "1994"
}

@article{Baur:1993mw,
    author = "Baur, R. and others",
    editor = "Nappi, E. and Ypsilantis, T.",
    title = "{The CERES RICH detector system}",
    reportNumber = "CERN-PPE-93-170",
    doi = "10.1016/0168-9002(94)90537-1",
    journal = "Nucl. Instrum. Meth. A",
    volume = "343",
    pages = "87--98",
    year = "1994"
}

@techreport{CERN-SPSC-88-25,
  title        = {Study of electron pair production in hadron and nuclear collisions at the CERN SPS},
  author       = {Faschingbauer, U.},
  institution  = {CERN},
  year         = {1988},
  reportNumber = {CERN-SPSC-88-25; SPSC-P-237},
  url          = {https://cds.cern.ch/record/000303389},
}

@misc{CERN2000NewState,
  author       = {{CERN}},
  title        = {New State of Matter Created at {CERN}},
  year         = {2000},
  month        = feb,
  howpublished = {\url{https://home.cern/news/press-release/cern/new-state-matter-created-cern}},
  note         = {Press release, February 10, 2000}
}

@article{Bearden:2000ex,
    author = "Bearden, I. G. and others",
    title = "{Space-time evolution of the hadronic source in peripheral to central Pb + Pb collisions}",
    doi = "10.1007/s100520000543",
    journal = "Eur. Phys. J. C",
    volume = "18",
    pages = "317--325",
    year = "2000"
}

@article{NA50:1996lag,
    author = "Gonin, Michel and others",
    editor = "Braun-Munzinger, P. and Specht, H. J. and Stock, R. and Stoecker, Horst",
    collaboration = "NA50",
    title = "{Anomalous J / psi suppression in Pb + Pb collisions at 158-A-GeV/c}",
    doi = "10.1016/S0375-9474(96)00373-9",
    journal = "Nucl. Phys. A",
    volume = "610",
    pages = "404C--417C",
    year = "1996"
}

@article{NA57:2006aux,
    author = "Antinori, F. and others",
    collaboration = "NA57",
    title = "{Enhancement of hyperon production at central rapidity in 158-A-GeV/c Pb-Pb collisions}",
    eprint = "nucl-ex/0601021",
    archivePrefix = "arXiv",
    doi = "10.1088/0954-3899/32/4/003",
    journal = "J. Phys. G",
    volume = "32",
    pages = "427--442",
    year = "2006"
}

@article{NA50:2004sgj,
    author = "Alessandro, B. and others",
    collaboration = "NA50",
    title = "{A New measurement of J/psi suppression in Pb-Pb collisions at 158-GeV per nucleon}",
    eprint = "hep-ex/0412036",
    archivePrefix = "arXiv",
    reportNumber = "CERN-PH-EP-2004-052",
    doi = "10.1140/epjc/s2004-02107-9",
    journal = "Eur. Phys. J. C",
    volume = "39",
    pages = "335--345",
    year = "2005"
}

@article{NA50:2000brc,
    author = "Abreu, M. C. and others",
    collaboration = "NA50",
    title = "{Evidence for deconfinement of quarks and gluons from the J / psi suppression pattern measured in Pb + Pb collisions at the CERN SPS}",
    reportNumber = "CERN-EP-2000-013",
    doi = "10.1016/S0370-2693(00)00237-9",
    journal = "Phys. Lett. B",
    volume = "477",
    pages = "28--36",
    year = "2000"
}

@article{NA50:2006yzz,
    author = "Alessandro, B. and others",
    collaboration = "NA50",
    title = "{psi-prime production in Pb-Pb collisions at 158-GeV/nucleon}",
    eprint = "nucl-ex/0612013",
    archivePrefix = "arXiv",
    reportNumber = "CERN-PH-EP-2006-019",
    doi = "10.1140/epjc/s10052-006-0153-y",
    journal = "Eur. Phys. J. C",
    volume = "49",
    pages = "559--567",
    year = "2007"
}

@article{NA38:2000wlp,
    author = "Abreu, M. C. and others",
    collaboration = "NA38, NA50",
    title = "{Dimuon and charm production in nucleus-nucleus collisions at the CERN SPS}",
    reportNumber = "CERN-EP-2000-012",
    doi = "10.1007/s100520000373",
    journal = "Eur. Phys. J. C",
    volume = "14",
    pages = "443--455",
    year = "2000"
}

@article{Gyulassy:1988gg,
    author = "Gyulassy, M.",
    title = "{QUARK MATTER '87: CONCLUDING REMARKS}",
    doi = "10.1007/BF01574560",
    journal = "Z. Phys. C",
    volume = "38",
    pages = "361--370",
    year = "1988"
}

@article{NA38:1988ahd,
    author = "Abreu, M. C. and others",
    collaboration = "NA 38",
    title = "{The Production of $J/\psi$ in 200-{GeV}/a Oxygen - Uranium Interactions}",
    reportNumber = "CERN-EP-88-25",
    doi = "10.1007/BF01574524",
    journal = "Z. Phys. C",
    volume = "38",
    pages = "117",
    year = "1988"
}

@article{Fermi:1950frz,
    author = "Fermi, Enrico",
    title = "{High Energy Nuclear Events}",
    doi = "10.1143/ptp/5.4.570",
    journal = "Prog. Theor. Phys.",
    volume = "5",
    number = "4",
    pages = "570--583",
    year = "1950"
}

@article{NA49:2007stj,
    author = "Alt, C. and others",
    collaboration = "NA49",
    title = "{Pion and kaon production in central Pb + Pb collisions at 20-A and 30-A-GeV: Evidence for the onset of deconfinement}",
    eprint = "0710.0118",
    archivePrefix = "arXiv",
    primaryClass = "nucl-ex",
    doi = "10.1103/PhysRevC.77.024903",
    journal = "Phys. Rev. C",
    volume = "77",
    pages = "024903",
    year = "2008"
}

@article{Gorenstein:2003cu,
    author = "Gorenstein, Mark I. and Gazdzicki, M. and Bugaev, K. A.",
    title = "{Transverse activity of kaons and the deconfinement phase transition in nucleus-nucleus collisions}",
    eprint = "hep-ph/0303041",
    archivePrefix = "arXiv",
    doi = "10.1016/j.physletb.2003.06.043",
    journal = "Phys. Lett. B",
    volume = "567",
    pages = "175--178",
    year = "2003"
}

@article{CBM:2016kpk,
    author = "Ablyazimov, T. and others",
    collaboration = "CBM",
    title = "{Challenges in QCD matter physics --The scientific programme of the Compressed Baryonic Matter experiment at FAIR}",
    eprint = "1607.01487",
    archivePrefix = "arXiv",
    primaryClass = "nucl-ex",
    doi = "10.1140/epja/i2017-12248-y",
    journal = "Eur. Phys. J. A",
    volume = "53",
    number = "3",
    pages = "60",
    year = "2017"
}

@techreport{Bächler:356588,
      author        = "Bächler, J",
      collaboration = "NA49",
      title         = "{Status and future programme of the NA49 Experiment:
                       addendum-2 to proposal SPSLC/P264}",
      institution   = "CERN",
      reportNumber  = "CERN-SPSC-98-4, SPSLC-P-264-Add-2",
      address       = "Geneva",
      year          = "1998",
      url           = "https://cds.cern.ch/record/356588",
}

@article{NA49:1994hfj,
    author = "Alber, T. and others",
    collaboration = "NA49",
    title = "{Transverse energy production in Pb-208 + Pb collisions at 158-GeV per nucleon}",
    reportNumber = "LBL-37450",
    doi = "10.1103/PhysRevLett.75.3814",
    journal = "Phys. Rev. Lett.",
    volume = "75",
    pages = "3814--3817",
    year = "1995"
}

@article{NA49:1997qey,
    author = "Appelshauser, H. and others",
    collaboration = "NA49",
    title = "{Directed and elliptic flow in 158-GeV / nucleon Pb + Pb collisions}",
    eprint = "nucl-ex/9711001",
    archivePrefix = "arXiv",
    reportNumber = "LBL-41016, LBNL-41016",
    doi = "10.1103/PhysRevLett.80.4136",
    journal = "Phys. Rev. Lett.",
    volume = "80",
    pages = "4136--4140",
    year = "1998"
}

@article{Gazdzicki:1996pk,
    author = "Gazdzicki, Marek and Rohrich, Dieter",
    title = "{Strangeness in nuclear collisions}",
    eprint = "hep-ex/9607004",
    archivePrefix = "arXiv",
    reportNumber = "IKF-HENPG-8-95",
    doi = "10.1007/s002880050147",
    journal = "Z. Phys. C",
    volume = "71",
    pages = "55--64",
    year = "1996"
}

@article{Gazdzicki:1995zs,
    author = "Gazdzicki, M. and Roehrich, D.",
    title = "{Pion multiplicity in nuclear collisions}",
    doi = "10.1007/BF01571878",
    journal = "Z. Phys. C",
    volume = "65",
    pages = "215--223",
    year = "1995"
}

@article{NA49:2002pzu,
    author = "Afanasiev, S. V. and others",
    collaboration = "NA49",
    title = "{Energy dependence of pion and kaon production in central Pb + Pb collisions}",
    eprint = "nucl-ex/0205002",
    archivePrefix = "arXiv",
    doi = "10.1103/PhysRevC.66.054902",
    journal = "Phys. Rev. C",
    volume = "66",
    pages = "054902",
    year = "2002"
}

@article{STAR:2010vob,
    author = "Aggarwal, M. M. and others",
    collaboration = "STAR",
    title = "{An Experimental Exploration of the QCD Phase Diagram: The Search for the Critical Point and the Onset of De-confinement}",
    eprint = "1007.2613",
    archivePrefix = "arXiv",
    primaryClass = "nucl-ex",
    note = "arXiv:1007.2613 [nucl-ex]",
    year = "2010"
}

@article{MPD:2022qhn,
    author = "Abgaryan, V. and others",
    collaboration = "MPD",
    title = "{Status and initial physics performance studies of the MPD experiment at NICA}",
    eprint = "2202.08970",
    archivePrefix = "arXiv",
    primaryClass = "physics.ins-det",
    doi = "10.1140/epja/s10050-022-00750-6",
    journal = "Eur. Phys. J. A",
    volume = "58",
    number = "7",
    pages = "140",
    year = "2022"
}

@article{NA61SHINE:2011dsu,
    author = "Abgrall, N and others",
    collaboration = "NA61/SHINE",
    title = "{Measurements of Cross Sections and Charged Pion Spectra in Proton-Carbon Interactions at 31 GeV/c}",
    eprint = "1102.0983",
    archivePrefix = "arXiv",
    primaryClass = "hep-ex",
    reportNumber = "CERN-PH-EP-2011-005",
    doi = "10.1103/PhysRevC.84.034604",
    journal = "Phys. Rev. C",
    volume = "84",
    pages = "034604",
    year = "2011"
}

@article{NA61SHINE:2019xkb,
    author = "Aduszkiewicz, A. and others",
    collaboration = "NA61/SHINE",
    title = "{Proton-Proton Interactions and Onset of Deconfinement}",
    eprint = "1912.10871",
    archivePrefix = "arXiv",
    primaryClass = "hep-ex",
    reportNumber = "FERMILAB-PUB-19-664-AD-SCD",
    doi = "10.1103/PhysRevC.102.011901",
    journal = "Phys. Rev. C",
    volume = "102",
    number = "1",
    pages = "011901",
    year = "2020"
}

@article{NA61SHINE:2020ggt,
    author = "Acharya, A. and others",
    collaboration = "NA61/SHINE",
    title = "{Measurements of $\pi^-$ production in $^7$Be+$^9$Be collisions ..}",
    eprint = "2008.06277",
    archivePrefix = "arXiv",
    primaryClass = "nucl-ex",
    reportNumber = "CERN-EP-2020-150",
    doi = "10.1140/epjc/s10052-020-08514-6",
    journal = "Eur. Phys. J. C",
    volume = "80",
    number = "10",
    pages = "961",
    year = "2020",
    note = "[Erratum: Eur.Phys.J.C 81, 144 (2021)]"
}

@article{NA61SHINE:2021nye,
    author = "Acharya, A. and others",
    collaboration = "NA61/SHINE",
    title = "{Spectra and mean multiplicities of $\pi ^{-}$ in central${}^{40}$Ar+${}^{45}$Sc collisions at 13A, 19A, 30A, 40A, 75A and 150$A\,\text{ Ge }\text{ V }\!/\!\textit{c}$ beam momenta measured by the NA61/SHINE spectrometer at the CERN SPS}",
    eprint = "2101.08494",
    archivePrefix = "arXiv",
    primaryClass = "hep-ex",
    reportNumber = "CERN-EP-2021-010",
    doi = "10.1140/epjc/s10052-021-09135-3",
    journal = "Eur. Phys. J. C",
    volume = "81",
    number = "5",
    pages = "397",
    year = "2021"
}

@article{NA61SHINE:2023epu,
    author = "Adhikary, H. and others",
    collaboration = "NA61/SHINE",
    title = "{Measurements of $\pi ^\pm $, $K^\pm $, p and $\bar{p}$ spectra in $^{40}\hbox {Ar+}^{45}\hbox {Sc}$ collisions at 13A to 150A~$\text{ Ge }\hspace{-1.00006pt}\text{ V }\!/\!c$}",
    eprint = "2308.16683",
    archivePrefix = "arXiv",
    primaryClass = "nucl-ex",
    reportNumber = "CERN-EP-2023-179, FERMILAB-PUB-23-563-AD",
    doi = "10.1140/epjc/s10052-024-12602-2",
    journal = "Eur. Phys. J. C",
    volume = "84",
    number = "4",
    pages = "416",
    year = "2024"
}

@article{Andronov:2022cna,
    author = "Andronov, Evgeny and Kuich, Magdalena and Ga\'zdzicki, Marek",
    title = "{Diagram of High-Energy Nuclear Collisions \textdagger{}}",
    eprint = "2205.06726",
    archivePrefix = "arXiv",
    primaryClass = "hep-ph",
    doi = "10.3390/universe9020106",
    journal = "Universe",
    volume = "9",
    number = "2",
    pages = "106",
    year = "2023"
}

@article{Becattini:2005xt,
    author = "Becattini, F. and Manninen, J. and Gazdzicki, M.",
    title = "{Energy and system size dependence of chemical freeze-out in relativistic nuclear collisions}",
    eprint = "hep-ph/0511092",
    archivePrefix = "arXiv",
    doi = "10.1103/PhysRevC.73.044905",
    journal = "Phys. Rev. C",
    volume = "73",
    pages = "044905",
    year = "2006"
}

@techreport{omega1968,
  author       = {Baker, W. F. et al.},
  title        = {Proposal for a Large Magnet and Spark Chamber System},
  institution  = {CERN},
  year         = {1968},
  reportnumber = {NP-68-11},
  url          = {https://cds.cern.ch/record/299898/files/cer-000221645.pdf},
  note         = {OMEGA Project Working Group report}
}

@inproceedings{Angert:1982rel,
  author       = {Angert, N. and others},
  title        = {Study of relativistic nucleus reactions induced by $^{16}$O beams of 9--13 GeV per nucleon at the CERN PS},
  booktitle    = {Workshop on Quark Matter Formation and Heavy-ion Collisions, Bielfeld},
  year         = {1982},
  url          = {https://cds.cern.ch/record/904115}
}

@techreport{RD19Proposal,
  author       = {Beusch, W. and others},
  title        = {R\&D Proposal: Development of Hybrid and Monolithic Silicon Micropattern Detectors},
  institution  = {CERN},
  year         = {1990},
  number       = {CERN-DRDC-90-81; DRDC-P-22},
  url          = {https://cds.cern.ch/record/292598},
}

@article{Srivastava:1994wk,
    author = "Srivastava, D. K. and Sinha, B.",
    title = "{Single photons from S + Au collisions at the CERN Super Proton Synchrotron and the quark - hadron phase transition}",
    doi = "10.1103/PhysRevLett.73.2421",
    journal = "Phys. Rev. Lett.",
    volume = "73",
    pages = "2421--2424",
    year = "1994"
}

@article{WA80:1995xza,
    author = "Albrecht, R. and others",
    collaboration = "WA80",
    title = "{Limits on the production of direct photons in 200-A/GeV S-32 + Au collisions}",
    reportNumber = "CERN-PPE-95-186",
    doi = "10.1103/PhysRevLett.76.3506",
    journal = "Phys. Rev. Lett.",
    volume = "76",
    pages = "3506--3509",
    year = "1996"
}

@techreport{Wa97Proposal,
  author       = {Armenise, N. and others},
  title        = {Proposal: Study of baryon and antibaryon spectra in lead--lead interactions at 160 GeV/c per nucleon},
  institution  = {CERN},
  year         = {1991},
  reportnumber = {CERN-SPSLC-91-29; SPSLC-P-263},
  url          = {https://cds.cern.ch/record/294873},
}

@techreport{NA57Proposal,
  author       = {Calindro, R. and others},
  title        = {Study of strange and multistrange particles in ultrarelativistic nucleus--nucleus collisions},
  institution  = {CERN},
  year         = {1996},
  reportnumber = {CERN-SPSLC-96-40; SPSLC-P-300},
  url          = {https://cds.cern.ch/record/309075},
  note         = {Submitted to the CERN SPS-LEAR Experiments Committee}
}

@article{NA61SHINE:2023gez,
    author = "Adhikary, H. and others",
    collaboration = "NA61/SHINE",
    title = "{Search for the critical point of strongly-interacting matter in $^{40}$Ar ~+~$^{45}$Sc collisions at 150A ~Ge V /c using scaled factorial moments of protons}",
    eprint = "2305.07557",
    archivePrefix = "arXiv",
    primaryClass = "nucl-ex",
    reportNumber = "CERN-EP-2023-082",
    doi = "10.1140/epjc/s10052-023-11942-9",
    journal = "Eur. Phys. J. C",
    volume = "83",
    number = "9",
    pages = "881",
    year = "2023"
}

@article{QCDWorkingGroup:2019dyv,
    author = "Dainese, A. and others",
    collaboration = "QCD Working Group",
    title = "{Physics Beyond Colliders: QCD Working Group Report}",
    eprint = "1901.04482",
    archivePrefix = "arXiv",
    primaryClass = "hep-ex",
    reportNumber = "CERN-PBC-REPORT-2018-008",
    year = "2019"
}

@article{NA44:1995lds,
    author = "Simon-Gillo, Jehanne and others",
    editor = "Poskanzer, Arthur M. and Harris, J. W. and Schroeder, L. S.",
    collaboration = "NA44",
    title = "{Deuteron and anti-deuteron production in CERN experiment NA44}",
    doi = "10.1016/0375-9474(95)00259-4",
    journal = "Nucl. Phys. A",
    volume = "590",
    pages = "483C--486C",
    year = "1995"
}

@article{WA98:2003ukc,
    author = "Aggarwal, M. M. and others",
    collaboration = "WA98",
    title = "{Interferometry of direct photons in central Pb-208+Pb-208 collisions at 158-A-GeV}",
    eprint = "nucl-ex/0310022",
    archivePrefix = "arXiv",
    doi = "10.1103/PhysRevLett.93.022301",
    journal = "Phys. Rev. Lett.",
    volume = "93",
    pages = "022301",
    year = "2004"
}

@article{WA98:2007qib,
    author = "Aggarwal, M. M. and others",
    collaboration = "WA98",
    title = "{Suppression of High-$p_T$ Neutral Pions in Central Pb + Pb Collisions at $\sqrt {s_{NN}}$ = 17.3 GeV Relative to $p$ + C and $p$ + Pb Collisions}",
    eprint = "0708.2630",
    archivePrefix = "arXiv",
    primaryClass = "nucl-ex",
    doi = "10.1103/PhysRevLett.100.242301",
    journal = "Phys. Rev. Lett.",
    volume = "100",
    pages = "242301",
    year = "2008"
}

@article{WA98:2000vxl,
    author = "Aggarwal, M. M. and others",
    collaboration = "WA98",
    title = "{Observation of direct photons in central 158-A-GeV Pb-208 + Pb-208 collisions}",
    eprint = "nucl-ex/0006008",
    archivePrefix = "arXiv",
    doi = "10.1103/PhysRevLett.85.3595",
    journal = "Phys. Rev. Lett.",
    volume = "85",
    pages = "3595--3599",
    year = "2000"
}

@article{Bjorken:1997re,
    author = "Bjorken, J. D.",
    editor = "Czyzewski, J. and Wosiek, J.",
    title = "{Disoriented chiral condensate: Theory and phenomenology}",
    eprint = "hep-ph/9712434",
    archivePrefix = "arXiv",
    reportNumber = "SLAC-PUB-7720",
    journal = "Acta Phys. Polon. B",
    volume = "28",
    pages = "2773--2791",
    year = "1997"
}

@article{WA98:1997nnc,
    author = "Nayak, Tapan K. and others",
    editor = "Hatsuda, T. and Miake, Y. and Yagi, K. and Nagamiya, S.",
    collaboration = "WA98",
    title = "{Present status and future of DCC analysis}",
    eprint = "hep-ex/9802019",
    archivePrefix = "arXiv",
    reportNumber = "VECC-NEX-01",
    doi = "10.1016/S0375-9474(98)00358-3",
    journal = "Nucl. Phys. A",
    volume = "638",
    pages = "249C--260C",
    year = "1998"
}

@article{Kapusta:1991qp,
    author = "Kapusta, Joseph I. and Lichard, P. and Seibert, D.",
    title = "{High-energy photons from quark - gluon plasma versus hot hadronic gas}",
    doi = "10.1103/PhysRevD.47.4171",
    journal = "Phys. Rev. D",
    volume = "44",
    pages = "2774--2788",
    year = "1991",
    note = "[Erratum: Phys.Rev.D 47, 4171 (1993)]"
}

@techreport{Angelis:295040,
      author        = "Angelis, Aris L S and others",
      title         = "{Proposal for a large acceptance hadron and photon
                       spectrometer}",
      institution   = "CERN",
      reportNumber  = "CERN-SPSLC-91-17, SPSLC-P-260",
      year          = "1991",
      url           = "https://cds.cern.ch/record/295040",
}

@article{Abatzis1995,
  author  = {S. Abatzis and others},
  title   = {Results on the Production of Baryons with |S| = 1, 2, 3 and Strange Mesons in S-W Collisions at 200 GeV/c per Nucleon},
  journal = {Nuclear Physics A},
  volume  = {590},
  year    = {1995},
  pages   = {307c--316c},
  doi     = {10.1016/0375-9474(95)00243-T},
  url     = {https://www.sciencedirect.com/science/article/pii/037594749500243T}
}

@article{NA44:1996xlh,
    author = "Bearden, I. G. and others",
    collaboration = "NA44",
    title = "{Collective expansion in high-energy heavy ion collisions}",
    reportNumber = "CERN-PPE-96-163",
    doi = "10.1103/PhysRevLett.78.2080",
    journal = "Phys. Rev. Lett.",
    volume = "78",
    pages = "2080--2083",
    year = "1997"
}

@TechReport{CERN-SPSC-88-22,
  author = "Albrecht, R and others",
  title       = "{Study of relativistic nucleus-nucleus collisions at the CERN SPS: Addendum to memorandum M 406}",
  institution = {CERN},
  number      = {CERN-SPSC-88-22, SPSC-P-236},
  year        = {1988},
  url         = {https://cds.cern.ch/record/303388},
}

@Article{Bearden2000Antideuteron,
  author = "Bearden, I. G. and others",
  title        = {Antideuteron Production in 158 A {GeV}/c Pb+Pb Collisions},
  doi          = {10.1103/PhysRevLett.85.2681},
  journal = "Phys. Rev. Lett.",
  year         = "2000",
  volume       = "85",
  number       = {13},
  pages        = "2681--2684",
}

@inproceedings{Rafelski:1980rk,
    author = "Rafelski, Johann and Hagedorn, R.",
    title = "{From Hadron Gas to Quark Matter. 2.}",
    booktitle = "{International Symposium on Statistical Mechanics of Quarks and Hadrons}",
    reportNumber = "CERN-TH-2969",
    year = "1980"
}

@article{Rafelski:1982pu,
    author = "Rafelski, Johann and M{\"u}ller, Berndt",
    title = "{Strangeness Production in the Quark - Gluon Plasma}",
    doi = "10.1103/PhysRevLett.48.1066",
    journal = "Phys. Rev. Lett.",
    volume = "48",
    pages = "1066--1069",
    year = "1982",
    note = "[Erratum: Phys. Rev. Lett. {\bf 56}, 2334 (1986)]"
}

@techreport{Beusch:345021,
      author        = "Beusch, W. and others",
      title         = "{{\it OMEGA PRIME, A Project of Improving the Omega                        Particle Detector System}}",
      institution   = "CERN",
      year          = "1977",
url           = "https://cds.cern.ch/record/345021",
}

@techreport{WA72proposal,
      author        = "Armstrong, T and others",
      title         = "{{\it Proposal: A Study of Fast Proton Production ..}}",
      institution   = "CERN",
      year          = "1981",
url           = "http://cds.cern.ch/record/601374",
}

@article{BEUSCH1986391,
title = "A fast on-line trigger for events with multiple high transverse momentum tracks",
journal = "Nucl. Instrum. Methods Phys. Res., Sect. A",
volume = "249",
pages = "391-398",
year = "1986",
issn = "0168-9002",
doi = "https://doi.org/10.1016/0168-9002(86)90693-5",
IGNOREurl = "http://www.sciencedirect.com/science/article/pii/0168900286906935",
author = "Beusch, W and others",
}

@article{HEIJNE20011,
title = {Semiconductor micropattern pixel detectors: a review of the beginnings},
journal = {Nucl. Instrum. Methods Phys. Res., Sect. A},
volume = {465},
pages = {1-26},
year = {2001},
doi = {https://doi.org/10.1016/S0168-9002(01)00340-0},
IGNOREurl = {https://www.sciencedirect.com/science/article/pii/S0168900201003400},
author = {Erik H.M. Heijne},
}

@article{NA60:2008dcb,
    author = "Arnaldi, R and others",
    collaboration = "NA60",
    title = "{Evidence for the production of thermal-like muon pairs with masses above 1-GeV/c**2 in 158-A-GeV Indium-Indium Collisions}",
    eprint = "0810.3204",
    archivePrefix = "arXiv",
    primaryClass = "nucl-ex",
    doi = "10.1140/epjc/s10052-008-0857-2",
    journal = "Eur. Phys. J. C",
    volume = "59",
    pages = "607--623",
    year = "2009"
}

@article{NA60:2007lzy,
    author = "Arnaldi, R. and others",
    collaboration = "NA60",
    title = "{Evidence for radial flow of thermal dileptons in high-energy nuclear collisions}",
    eprint = "0711.1816",
    archivePrefix = "arXiv",
    primaryClass = "nucl-ex",
    doi = "10.1103/PhysRevLett.100.022302",
    journal = "Phys. Rev. Lett.",
    volume = "100",
    pages = "022302",
    year = "2008"
}

@article{NA60:2016nad,
    author = "Arnaldi, R. and others",
    collaboration = "NA60",
    title = "{Precision study of the $\eta\to\mu^+\mu^-\gamma$ and $\omega\to\mu^+\mu^-\pi^0$ electromagnetic transition form-factors and of the $\rho\to\mu^+\mu^-$ line shape in NA60}",
    eprint = "1608.07898",
    archivePrefix = "arXiv",
    primaryClass = "hep-ex",
    doi = "10.1016/j.physletb.2016.04.013",
    journal = "Phys. Lett. B",
    volume = "757",
    pages = "437--444",
    year = "2016"
}

@article{NA60:2019tfy,
    author = "Arnaldi, R. and others",
    collaboration = "NA60",
    title = "{Nuclear dependence of light neutral meson production in p-A collisions at 400 GeV with NA60}",
    eprint = "2012.14385",
    archivePrefix = "arXiv",
    primaryClass = "hep-ex",
    doi = "10.1140/epjc/s10052-019-6848-7",
    journal = "Eur. Phys. J. C",
    volume = "79",
    number = "5",
    pages = "443",
    year = "2019"
}

@article{NA60:2009una,
    author = "Arnaldi, R. and others",
    collaboration = "NA60",
    title = "{Study of the electromagnetic transition form-factors in eta ---\ensuremath{>} mu+ mu- gamma and omega ---\ensuremath{>} mu+ mu- pi0 decays with NA60}",
    eprint = "0902.2547",
    archivePrefix = "arXiv",
    primaryClass = "hep-ph",
    doi = "10.1016/j.physletb.2009.05.029",
    journal = "Phys. Lett. B",
    volume = "677",
    pages = "260--266",
    year = "2009"
}

@article{NA60:2011zkq,
    author = "Arnaldi, R. and others",
    collaboration = "NA60",
    title = "{A Comparative measurement of $\phi\rightarrow K^+K^-$ and $\phi\rightarrow \mu^+\mu^-$ in In-In collisions at the CERN SPS}",
    eprint = "1104.4060",
    archivePrefix = "arXiv",
    primaryClass = "nucl-ex",
    doi = "10.1016/j.physletb.2011.04.028",
    journal = "Phys. Lett. B",
    volume = "699",
    pages = "325--329",
    year = "2011"
}

@article{NA60:2009eih,
    author = "Banicz, K and others",
    collaboration = "NA60",
    title = "{phi Production in In-In Collisions at 158-A-GeV}",
    eprint = "0906.1102",
    archivePrefix = "arXiv",
    primaryClass = "hep-ex",
    doi = "10.1140/epjc/s10052-009-1137-5",
    journal = "Eur. Phys. J. C",
    volume = "64",
    pages = "1--18",
    year = "2009"
}

@article{NA60:2010wey,
    author = "Arnaldi, R and others",
    collaboration = "NA60",
    title = "{J/psi production in proton-nucleus collisions at 158 and 400 GeV}",
    eprint = "1004.5523",
    archivePrefix = "arXiv",
    primaryClass = "nucl-ex",
    doi = "10.1016/j.physletb.2011.11.042",
    journal = "Phys. Lett. B",
    volume = "706",
    pages = "263--267",
    year = "2012"
}

@article{NA60:2006ncq,
    author = "Arnaldi, R. and others",
    editor = "Sissakian, Alexey and Kozlov, Gennady and Kolganova, Elena",
    collaboration = "NA60",
    title = "{$J/\psi$ production in Indium-Indium collisions at 158- GeV/nucleon}",
    eprint = "0706.4361",
    archivePrefix = "arXiv",
    primaryClass = "nucl-ex",
    doi = "10.1103/PhysRevLett.99.132302",
    journal = "Conf. Proc. C",
    volume = "060726",
    pages = "430--434",
    year = "2006"
}

@article{NA60:2008iqj,
    author = "Arnaldi, R. and others",
    collaboration = "NA60",
    title = "{First results on angular distributions of thermal dileptons in nuclear collisions}",
    eprint = "0812.3100",
    archivePrefix = "arXiv",
    primaryClass = "nucl-ex",
    doi = "10.1103/PhysRevLett.102.222301",
    journal = "Phys. Rev. Lett.",
    volume = "102",
    pages = "222301",
    year = "2009"
}

@article{Specht:2010xu,
    author = "Specht, Hans J.",
    editor = "Nieves, Juan M. and Oset, Eulogio and Vicente Vacas, Manuel J.",
    collaboration = "NA60",
    title = "{Thermal Dileptons from Hot and Dense Strongly Interacting Matter}",
    eprint = "1011.0615",
    archivePrefix = "arXiv",
    primaryClass = "nucl-ex",
    doi = "10.1063/1.3541982",
    journal = "AIP Conf. Proc.",
    volume = "1322",
    number = "1",
    pages = "1--10",
    year = "2010"
}

@techreport{Cambell:240426,
      author        = "Campbell, M and others",
      title         = "{\it Design and Performance of the Omega--Ion Hybrid Silicon Pixel Detector}",
      institution   = "CERN",
      collaboration = "RD19",
      year          = "1992",
      url           = "http://cds.cern.ch/record/240426",
}

@article{CAMPBELL199452,
title = {Development of a pixel readout chip compatible with large area coverage},
journal = {Nucl. Instrum. Methods Phys. Res., Sect.},
volume = {342},
number = {1},
pages = {52--58},
year = {1994},
issn = {0168-9002},
doi = {https://doi.org/10.1016/0168-9002(94)91410-9},
IGNOREurl = {https://www.sciencedirect.com/science/article/pii/0168900294914109},
author = {M. Campbell and others}
}

@techreport{Abatzis:195599,
  author       = {Abatzis, S. and others},
  title        = {Nucleus--nucleus interactions using the CERN OMEGA spectrometer and a multiparticle high-$p_T$ detector},
  institution  = {CERN},
  year         = {1988},
  reportnumber = {CERN-SPSC-88-20; SPSC-P-206-Add-2},
  url          = {https://cds.cern.ch/record/195599},
}

@TechReport{CERN-SPSC-90-032,
  author       = "Dönni, P. and others",
  title        = "{Proposal for a Light Universal .. }",
  institution  = {CERN},
  number       = {CERN-SPSC-90-032, SPSC-P-252-Add-1},
  year         = {1990},
  url          = {https://cds.cern.ch/record/000493141},
}

@article{CERESNA45:2024pkm,
    author = "Adamov{\'a}, D. and others",
    collaboration = "CERES NA45",
    title = "{The $v^{1/3}_{3}/v^{1/2}_{2}$ ratio in PbAu collisions at $\sqrt{s_{\textrm{NN}}} = $ 17.3~GeV: a hint of a hydrodynamic behavior}",
    eprint = "2402.10895",
    archivePrefix = "arXiv",
    primaryClass = "nucl-ex",
    doi = "10.1140/epjc/s10052-024-13416-y",
    journal = "Eur. Phys. J. C",
    volume = "84",
    number = "10",
    pages = "1090",
    year = "2024"
}

@techreport{WA94proposal,
      author        = "Vasileiadis, G and others",
      title         = "{\it Proposal Study of Baryon and Antibaryon Spectra ..}",
      institution   = "CERN",
      address       = "Geneva",
      number        = "CERN-SPSLC-91-5, SPSLC-P-257",
      year          = "1991",
url           = "http://cds.cern.ch/record/294863",
}

@article{ABATZIS1991441,
title = {Multi-strange baryon and antibaryon production in sulphur-tungsten and proton-tungsten interactions at 200 GeV/c per nucleon},
journal = {Nucl. Phys. A},
volume = {525},
pages = {441c-444c},
year = {1991},
issn = {0375-9474},
doi = {https://doi.org/10.1016/0375-9474(91)90360-I},
IGNOREurl = {https://www.sciencedirect.com/science/article/pii/037594749190360I},
author = {S. Abatzis and others}
}
\end{document}